\documentclass[twocolumn]{aastex63}
\usepackage{amsmath}
\usepackage{comment}

\usepackage{enumitem}

\usepackage{physics}
\usepackage{CJK}

\newcommand{\be}{\begin{equation}}
\newcommand{\ee}{\end{equation}}

\shorttitle{Evolution of TDE Envelopes}
\shortauthors{Tuna et al.}

\graphicspath{{./}}

\begin{document}
\begin{CJK*}{UTF8}{gbsn}

\title{Super-Eddington Chimneys: On the Cooling Evolution of Tidal Disruption Event Envelopes}

\author[0000-0002-2002-6860]{Semih Tuna}
\affil{Department of Physics and Columbia Astrophysics Laboratory, Columbia University, New York, NY 10027, USA}

\author[0000-0002-4670-7509]{Brian D.~Metzger}
\affil{Department of Physics and Columbia Astrophysics Laboratory, Columbia University, New York, NY 10027, USA}
\affil{Center for Computational Astrophysics, Flatiron Institute, 162 5th Ave, New York, NY 10010, USA}

\author[0000-0002-2624-3399]{Yan-fei Jiang (姜燕飞)}
\affil{Center for Computational Astrophysics, Flatiron Institute, 162 5th Ave, New York, NY 10010, USA}

\author[0000-0001-7448-4253]{Andrea Antoni}
\affil{Center for Computational Astrophysics, Flatiron Institute, 162 5th Ave, New York, NY 10010, USA}

\correspondingauthor{Semih Tuna}
\email{semih.tuna@columbia.edu}

\begin{abstract}

The formation of a compact accretion disk following a tidal disruption event (TDE) requires that the shocked stellar debris cool efficiently as it settles toward the black hole. While recent simulations suggest that stream dissipation occurs rapidly, how the weakly bound debris subsequently loses its thermal energy to assemble a compact disk near the circularization radius remains uncertain. We investigate this cooling process using axisymmetric radiation-hydrodynamic simulations of quasi-hydrostatic “TDE envelopes,” initialized with the total mass, angular momentum, and binding energy expected from a complete stellar disruption. The envelopes, supported by radiation pressure on large scales and rotation near the circularization radius, evolve through a combination of radiative diffusion, turbulent mixing, and polar outflows. In our fiducial model, a quasi-steady state is achieved in which a polar outflow radiates and expels matter at several times the Eddington luminosity. This enables the envelope to cool and contract, forming a dense, rotationally supported ring near the circularization radius, but on a timescale roughly ten times shorter than the naive photon-diffusion (Kelvin-Helmholtz) timescale.  Comparative models without radiation transport confirm that cooling, not purely adiabatic evolution, is essential to driving this rapid inflow. Nevertheless, across a range of envelope masses, the effective envelope cooling time scales only weakly with its optical depth, implying that advective and wind-driven energy transport dominate over diffusion. Our results demonstrate the cooling-induced contraction, even absent viscosity and associated black hole accretion, can produce luminosities and large photosphere radii consistent with early UV/optical TDE emission. However, more quantitative light-curve predictions must incorporate self-consistent formation and feeding of the envelope by fall-back accretion.

\end{abstract}


\section{Introduction}
\label{sec:intro}

Stars orbiting supermassive black holes (SMBH) in galactic nuclei occasionally stray close enough to be torn apart by the black hole's strong tidal field. Following this encounter, part of the stellar debris falls back to the SMBH, powering a transient luminous across the electromagnetic spectrum, dubbed a ``Tidal Disruption Event'' (TDE; \citealt{Hills75, Rees88}). 

TDEs were historically of interest as tracers of SMBH demographics at the low end of the mass distribution 
$M_{\bullet} \lesssim 10^8\,M_\odot$, which reside at the centers of otherwise quiescent galaxies \citep{Hills75, Frank&Rees76, Lidskii&Ozernoi79, Lacy+82, Nolthenius&Katz82}. A basic physical picture was formulated in these early works, elucidating the stellar disruption process into streams of gas \citep{Carter&Luminet83}, the canonical mass fallback rate $\dot{M} \sim t^{-5/3}$ \citep{Rees88, Phinney89}, circularization of the streams by shocks \citep{Kochanek94}, and the subsequent formation of
a compact accretion disk of radius $r \sim R_{\rm t} \sim 10-100\,R_{\rm g}$, where $R_{\rm t}$ is the tidal radius and $R_{\rm g} \equiv GM_{\bullet}/c^{2}$. The disk emission was assumed to follow the mass fallback rate, generating a multicolor thermal spectrum peaking at temperatures $kT \sim 10^5 - 10^6\,$eV, bounded by a Wien tail detectable at soft X-ray energies and a Rayleigh-Jeans tail extending into the UV/optical band. 

The first TDEs discovered in X-ray surveys \citep{Bade+96, Komossa&Bade99} appeared to support this theoretical picture. However, this victory was short-lived, as the large abundance of TDE flares subsequently discovered in UV/optical surveys \citep{Stern+04, Gezari+06, vanVelzen+11, Cenko+12, Arcavi+14}, now challenge the view that the earliest and brightest emission phase is powered by standard disk accretion. The UV/optical spectra reveal thermal emission peaking at luminosities $L \sim 10^{43}\,{\rm erg}\,{\rm s}^{-1}$, but with temperatures $T_\text{eff} \sim 3-5\,\times\,10^4\,$K an order of magnitude below the predicted values for a thin disk, with correspondingly large implied photosphere radii $R_{\rm ph} \sim 10^{15}\,{\rm cm}$ (e.g., \citealt{Gezari21}), greatly exceeding the tidal radius.  Spectroscopic measurements further indicate typical line widths $v \lesssim 10^4\,{\rm km}\,{\rm s}^{-1}$ (e.g., \citealt{Charalampopoulos+21}), much lower than the Keplerian motion of a compact disk. Interestingly, a power-law decay $L \sim t^{-5/3}$ nevertheless provides a reasonable fit to the early evolution of some optical TDE light curves \citep{Gezari21,Hammerstein+23a}.

UV/optical TDE flares rise on a timescale of roughly a month, comparable to when the most tightly bound debris returns to the black hole; however, the emission at other wavelengths is frequently delayed.  UV/optically-discovered TDEs usually only peak in the X-rays several months or longer after the UV/optical peak \citep{Gezari+17,Kajava+20,Liu+22,Yao+22}; conversely, most X-ray selected TDEs are optically faint \citep{Sazonov+21, Eyles-Ferris+25}. Synchrotron radio emission due to jets or outflows from the SMBH \citep{Giannios&Metzger11} is similarly delayed from the disruption by several months to years (e.g., \citealt{Horesh+21a, Horesh+21b, Sfaradi+22, Cendes+22};  see \citealt{Alexander+20} for a review)
with a few notable exceptions \citep{Bloom+11, Burrows+11}.  Neutrinos, which likely originate from the hot inner accretion flow (e.g., \citealt{Murase+20}), were observed in coincidence with a handful of TDEs \citep{vanVelzen+24,Reusch+22}, but again delayed by many months from the optical/UV peak \citep{Stein+21}. Although other explanations have been proposed (e.g., \citealt{Teboul&Metzger23}), collectively this behavior may indicate that the SMBH accretion rate peaks long after the disruption, even though a large gaseous structure must come to encase the system much faster.

Simulating the canonical disruption of a solar-type star of mass $M_{\star} \sim M_{\odot}$ on a parabolic orbit is numerically challenging due to
the high eccentricity ($e > 0.99$, e.g., \citealt{Stone+13}) and correspondingly 
large dynamical range $\gtrsim a_{0}/R_{\rm t} \sim (M_\bullet/M_{\star})^{1/3} \sim 100$ of the debris stream orbits, where $a_{0}$ is the apocenter distance of the most tightly bound debris (e.g., \citealt{Ayal+00}). Despite this obstacle, simulations of the disruption process have confirmed and refined analytical expectations for the specific energy distribution of debris, the mass fallback rate, and how these depend on the stellar structure \citep{Lodato+09}, orbit \citep{Guillochon&RR13} and the adapted numerical technique (\citealt{Mainetti+17,Ryu+20a,Ryu+20b}). How the debris goes on to form a compact accretion flow remains, however, hotly debated (\citealt{Bonnerot&Stone21}). 

Most TDEs arise from stars orbiting a SMBH on nearly zero-energy, parabolic orbits (e.g., \citealt{Stone&Metzger16}).  However, to reduce the scale separation $a_{0}/R_{\rm t}$, initial numerical studies explored disruptions of stars on tightly bound elliptical orbits \citep{Hayasaki+13,Hayasaki+16,Bonnerot+16, Sadowski+16, Liptai+19,Hu+24}, or parabolic orbits but around intermediate mass black holes ($M_\bullet \sim 10^3\,M_\odot$; \citealt{Rosswog+09,Guillochon+14,Shiokawa+15}). An alternative strategy to overcome the limitations of dynamical range is to extrapolate results of local simulations of self-crossing shocks \citep{Jiang+16,Lu&Bonnerot20,Huang+23} to global simulations by injecting gas with prescribed energy and momentum into the computational grid and following the evolution of the collision debris \citep{Bonnerot&Lu20,Bonnerot+21,Meza+25}. Shocks, either through self-intersection of debris streams due to relativistic precession effects \citep{Rees88,Hayasaki+13,Cheng+14,Bonnerot+16}, or strong compression of the debris at the pericenter \citep{Kochanek94,Evans&Kochanek89,RR&Rosswog09, Bonnerot&Lu22}, are likely mechanisms by which the debris streams dissipate their energy. 

Another approach to the TDE problem takes the formation of a torus as given and explores the subsequent accretion and emission using GR(R)MHD simulations \citep{Dai+18,Curd&Narayan19,Curd21,Thomsen+22,Zhang+25}. These works provide insight into the properties of the super-Eddington accretion flow close to the SMBH, such as jets and winds, and may offer a unification scheme for X-ray and optically selected TDEs based on observer viewing angle (see also \citealt{Metzger&Stone16}).
However, torii simulated thus far are generally endowed with either too much angular momentum (\citealt{Dai+18}; their Section 2.1) or greater initial binding energy (\citealt{Curd&Narayan19}) than the TDE debris possesses. This questions the predicted evolution of the SMBH accretion rate and TDE light-curve, if mass is not promptly supplied to the inner disk from the weakly bound debris feeding it from larger scales.  

Recent simulations may finally have begun to crack the canonical problem of a solar-type star disrupted by a $\sim 10^{6}M_{\odot}$ SMBH. GRHD simulations of deeply penetrating (high $\beta$) disruption by \citet{Andalman+22} found efficient circularization via self-intersection shocks over a simulation time of around a week, corresponding to about 20\% of the fall-back time of the most tightly bound debris, $t_{\rm fb}$. \citet{Steinberg&Stone24}
simulated a canonical TDE for $\sim 60\,{\rm days}$ ($\sim 1.5\,t_{\rm fb}$), finding efficient stream  dissipation through shocks leading to the creation of a quasi-spherical structure. \citet{Price+24} present adiabatic hydrodynamic simulations of a canonical TDE extending to $\sim {\rm year}$ after disruption; they find the formation of a radiation-dominated ``envelope'' of size $\sim 10^{15}$ cm, which by the end of their simulation has become largely unbound in a quasi-spherical outflow. Although this envelope would radiate near the Eddington limit, by neglecting radiative cooling their adiabatic simulations likely over-estimate the importance of outflows.\footnote{Crudely modeling the effects of efficient cooling with an isentropic equation of state, \citet{Price+24} found that a compact accretion ring forms near the tidal radius within a few $t_{\rm fb}$.}
 
 The early UV/optical emission of TDE flares is typically attributed to either (1) direct powering by shock dissipation \citep{Shiokawa+15,Piran+15, Ryu+20}, usually under the assumption of inefficient dissipation; or (2) ``reprocessing" of X-rays from the inner accretion disk by extended gas, which could take the form of a quasi-hydrostatic envelope \citep{Loeb&Ulmer97,Coughlin&Begelman14,Roth+16}, precessing debris streams (e.g., \citealt{Guillochon&RR15,Calderon+24}), a fast disk wind \citep{Strubbe&Quataert09, Strubbe&Quataert11,Guillochon+14,Miller15} or slow outflow associated with the circularization process \citep{Metzger&Stone16}. However, both scenarios leave open questions. As already mentioned, recent global simulations find efficient stream dissipation \citep{Steinberg&Stone24,Price+24}, contrary to original motivations for the shock-powered emission scenario. Furthermore, as the shock-heated debris sinks into the potential well of the SMBH, the associated gravitational energy release far exceeds that dissipated by stream-stream or stream-disk collisions.  Where does this energy go, if not into radiation and outflows?  On the other hand, reprocessing scenarios rely on prompt feeding of the inner SMBH accretion flow and hence do not naturally account for delayed X-ray/radio emission.

 Fortunately, at very late times, this complex and unsettled picture yields to something simpler. TDEs observed years to decades after disruption reveal slowly decaying X-ray \citep{Jonker+20} and plateau-like UV emission \citep{vanVelzen+19}, in remarkable agreement with thermal emission from a thin accretion disk viscously spreading outward from the tidal radius carrying a sizable fraction of the bound stellar mass \citep{Cannizzo+90,Shen&Matzner14,Mummery&VanVelzen24}.  Independent of the details, {\it significant energy must have been released to form such massive compact disks seen at late times from the initially weakly bound debris.}   

Building on earlier ideas by \citet{Loeb&Ulmer97} and \citet{Coughlin&Begelman14}, \citet{Metzger22} proposed the \emph{cooling envelope} model for TDE emission as a middle ground solution to the above issues (see also \citealt{Sarin&Metzger24}). This model posits that the early UV/optical flare in TDEs is powered primarily by the gravitational energy released via accretion, but not directly through a compact disk feeding the SMBH.  Instead, most of the UV/opical emission arises from a quasi-spherical envelope formed through rapid stream dissipation, which releases gravitational energy as it undergoes slow, quasi-hydrostatic contraction from its initial outer radius ($a_{0} \sim 100R_{\rm t}$) down to the circularization radius ($\sim R_{\rm t}$) over several months to a year.  Although energy liberated from the inner accretion disk onto the SMBH can also contribute to the envelope luminosity through reprocessing, this is not an essential aspect of the model, which relies on quasi-spherical inflow and requires no viscosity.  As the rate limiting step to feeding the SMBH is not the fall-back time of the debris, but rather than cooling time of the envelope, this scenario could in principle lead to the delayed peak in SMBH activity implied by X-ray, radio and neutrino observations.

Here we use 2D radiation hydrodynamics simulations to investigate the large-scale ($r \sim 10^3\,R_{\rm g}$), long-term ($\gtrsim t_{\rm fb}$ after formation) evolution of quasi-hydrostatic TDE envelopes 
\citep{Loeb&Ulmer97, Metzger22}.
Our methodology is similar to \citet{Dai+18} and \citet{Curd&Narayan19}, in that we initialize our simulations with rotating, hydrostatic tori on circular orbits (hence we implicitly assume efficient stream dissipation); however, the initial total energy and angular momentum of our tori are chosen to be consistent with the tidal disruption process, leading to a more tenuously bound structure initially spanning a wide range of radii $10\,R_{\rm g}\,\lesssim r \lesssim \text{few}\times \,10^3\,R_{\rm g}$, with the bulk of the mass residing at $r \sim 10^3\,R_{\rm g}$, but rotational support only becoming relevant inside $r \lesssim 50\,R_{\rm g}$. 

In part because of computational limitations related to the large spatial and temporal range of the system, we restrict our simulations to axisymmetry and neglect magnetic fields, the latter precluding self-consistent angular momentum transport and extension of the accretion down to SMBH horizon. The evolution is therefore, entirely controlled by cooling of the envelope (i.e., loss of pressure support), instead of viscous transport of angular momentum. 
A primary aim is to quantify the  \emph{effective cooling timescale} of the envelope through
winds and radiative losses, and their effects on 
(i) the time-dependent mass supply rate to the TDE accretion disk (neglecting strong feedback from the inner disk), and (ii) the fraction of the available mass unbound to winds.  Because the long-term evolution of the envelope may be sensitive to the initial conditions, we explore how the initial mass and radial distribution of the envelope affects the global evolution. A follow-up study 
(Tuna et al., in prep), will explore the role of energy feedback from the inner TDE accretion disk through the inclusion of an explicit viscosity, as well as the impact of continuous fall-back of mass and energy, on the envelope evolution.

\begin{figure*}
    \centering
    \includegraphics[width= \textwidth]{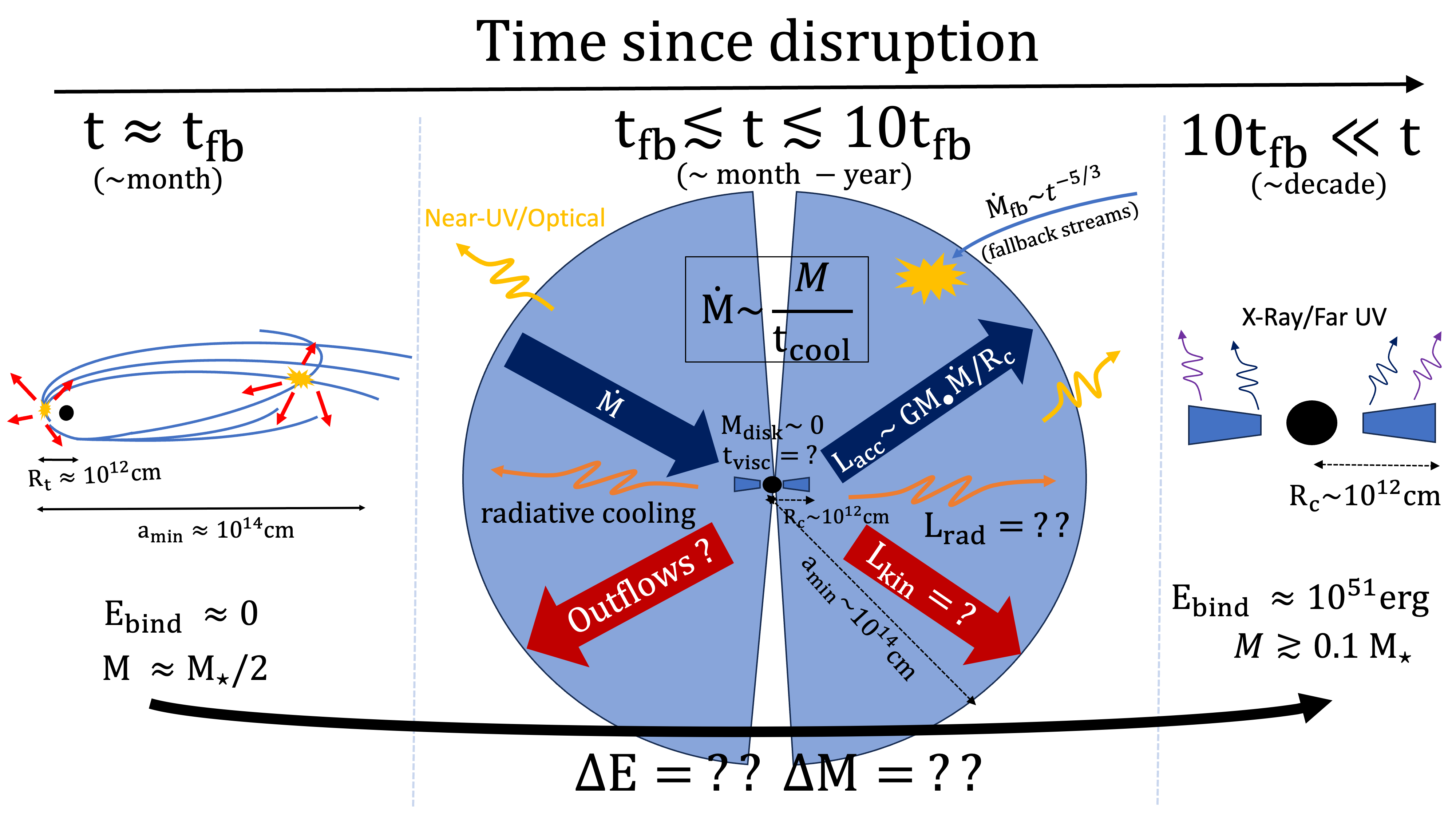}
    \caption{Schematic diagram describing the three stages of a tidal disruption event: after the debris streams pass through pericenter ($\sim \rm month$, left panel), their kinetic energy is dissipated into heat through a combination of compression-induced shock heating inside the tidal radius and self-crossing shocks at larger radii.  On the other hand, at very late times after the disruption ($\sim \rm year - decade$), X-ray and far-UV observations reveal a geometrically thin disk of size $\sim R_c \sim 10^{12}\,\rm cm$, containing significant mass $\gtrsim 0.1\,M_\star$ (e.g., \citealt{vanVelzen+19, Jonker+20,Mummery&VanVelzen24}).  The intermediate phase, $t_{\rm fb} \lesssim t \lesssim 10\,t_{\rm fb}$, which corresponds to the epoch of peak UV/optical light, is subject to larger uncertainties compared to the early and late phases. The qualitative evolution in this phase depends on how the timescales for shock dissipation ($t_{\rm diss}$), cooling of the debris ($t_{\rm cool}$), and viscous accretion
    ($t_{\rm visc}$), compare to the fallback timescale of the debris ($t_{\rm fb}$)
    (Sec.~\ref{sec:analytical_estimates}; Fig.~\ref{fig:TDEtree}). The present paper explores a scenario in which shock dissipation is fast, but cooling is slow, i.e., $t_{\rm cool} \gtrsim t_{\rm fb} \sim t_{\rm diss}$. In this case, the stellar debris forms of a large quasi-hydrostatic envelope of size $\sim a_{0}$ (middle panel) as defined by the weak initial binding energy of the debris (Eq.~\eqref{eq:epsilon_t}), as seen in several recent simulations (e.g., \citealt{Steinberg&Stone24,Price+24}). Even in the likely event that some mass during this process is placed into a Keplerian disk extending interior to the tidal radius $\lesssim R_t \lesssim 10^{13}\,\rm cm$, its mass is limited to that supplied by the large scale, highly sub-Keplerian envelope on a timescale $t_{\rm cool}$ set by its cooling rate (rather than viscous transport rate of angular momentum). An envelope cooling through a combination of radiation and outflow could power the optical/near-UV emission near peak by releasing the $\epsilon_{\rm c} \sim 10^{52}\,\rm erg\,M_\odot^{-1}$ of energy necessary to form a disk at $r \sim R_c$ (Eq.~\eqref{eq:epsilon_c}) through quasi-spherical accretion over $\sim \rm months$, which goes on to power UV and X-ray emission for $\sim \text{decades}$ 
    (right-panel).}
    \label{fig:cartoon}
\end{figure*}

The remainder of this paper is organized as follows. In Sec.~\ref{sec:analytical_estimates}, we introduce the relevant model parameters in the context of basic TDE theory, and estimate the basic properties of quasi-hydrostatic TDE envelopes. In Sec.~\ref{sec:numerical_setup}, we describe our numerical setup, initial conditions, and diagnostics used throughout the paper. 
In Sec.~\ref{sec:results}, we present our simulation results. In Sec.~\ref{sec:discussion}, we discuss the implications of our results for TDEs. In Sec.~\ref{sec:conclusion}, we summarize our conclusions.

\section{Analytical estimates}
\label{sec:analytical_estimates}

\subsection{Brief review of TDE theory}
\label{sec:review_TDE_theory}

We consider a star of mass $M_\star = m_\star\,M_\odot$ and radius $R_\star = r_\star\,R_\odot$ approaching a SMBH of mass $M_\bullet = M_{\bullet,6}\,10^6\,M_\odot$ on a nearly parabolic orbit with a pericenter radius $R_{\rm p}$ inside the tidal radius \citep{Hills75},
\begin{align}
    R_t &\equiv R_\star\left(\frac{M_\bullet}{M_\star}\right)^{1/3}
    = 46\,R_{\rm g}
     r_\star\,m_\star^{-1/3}
     M_{\bullet,6}^{-2/3}.
     \label{eq:Rt}
\end{align} 
The penetration factor is defined as
\begin{align}
    \beta \equiv \frac{R_{\rm t}}{R_p} \label{eq:beta},
\end{align}
with $\beta \gtrsim 1$ required for a complete disruption. 

As the star enters the tidal sphere, gravitational forces squeeze and elongate the stellar material into streams on the plane of the orbit.  This disruption process is often treated in the ``frozen-in'' approximation \citep{Lodato+09,Stone+13}, whereby the stellar streams are placed onto ballistic orbits with specific energy (about the zero-energy initial orbit) set by the relative depth of the stream within the SMBH potential. This energy spread is roughly given by the work done by the tidal force across the star:
\begin{align}
    \epsilon_t = k\frac{GM_\bullet\,R_\star}{R_t^2},
    \label{eq:epsilon_t}
\end{align}
where $k \sim 1$ is a dimensionless parameter that depends on the stellar structure and, to a lesser extent, the orbital properties. Following the encounter, the debris acquires specific energies  $-\epsilon_t \leq \epsilon \leq \epsilon_t$: 
 roughly half of the star closest to the BH remains bound, while the other half is unbound. The most tightly bound debris are placed onto highly eccentric orbits of semi-major axis
\begin{align}
    a_{0} \equiv 
    \frac{GM_\bullet}{2\Delta\epsilon}
    \approx 2.3\times 10^3 \,R_{\rm g}\,k^{-1}\,
    M_{\bullet, 6}^{-1/3}
    m_\star^{-2/3}\label{eq:a_min}
\end{align}
and return to pericenter after a ``fallback time'',
\begin{align}
    t_{\rm fb} &= 2\pi\,(GM_\bullet)^{-1/2}\,
    a_{0}^{3/2}\nonumber\\
    &= 41\,{\rm day}\,k^{-3/2}\,
    M_{\bullet, 6}^{1/2}
    m_\star^{-1}
    r_\star^{3/2}.
    \label{eq:tfb}
\end{align}
For debris with mass uniformly distributed around $\epsilon = 0$, the fall-back rate obeys \citep{Rees88}
\begin{align}
     \dot{M}_{\rm fb} &= \frac{dM}{d\epsilon}\frac{d\epsilon}{dt}
     = \dot{M}_{\rm pk}
     \left(\frac{t}{t_{\rm fb}}\right)^{-5/3}, \label{eq:mdot_fb}
\end{align}
where 
\begin{align}
    \dot{M}_{\rm pk} \equiv \frac{2M_{\rm bnd}}{3t_{\rm fb}} \approx
    3\,M_\odot\,{\rm yr}^{-1} \left(\frac{M_{\rm bnd}}{M_\star/2}\right)k^{3/2}
    M_{\bullet, 6}^{-1/2}
    m_\star^{2}
    r_\star^{-3/2} ,
\end{align}
and $M_{\rm bnd} \approx M_\star/2$ is the total bound mass. 
The peak mass fallback rate is highly super-Eddington, i.e.,
\begin{align}
    \frac{\dot{M}_{\rm pk}}{\dot{M_{\rm Edd}}} \approx
    135
    \left(\frac{M_{\rm bnd}}{M_\star/2}\right)
    \,k^{3/2}\eta_{0.1}
    \kappa_{\rm es}
    M_{\bullet, 6}^{-3/2}
    m_\star^{2}
    r_\star^{-3/2},
    \label{eq:mdot_edd_rat}
\end{align}
where $\dot{M}_{\rm edd} \equiv L_{\rm edd}/\eta c^2$, $L_{\rm edd} \equiv 4\pi GM_{\bullet}c/\kappa_{\rm es}$, $\kappa_{\rm es} = 0.38\,{\rm cm}^2\,{\rm g}^{-1}$, and $\eta = 0.1\eta_{0.1}$ is the radiative efficiency. 

The fallback streams carry the same angular momentum of the disrupted orbit near pericenter. Were the bound debris to eventually form a rotationally-supported disk, its (``circularization'') radius is given by
\begin{align}
        R_c = 2\,R_{\rm p} = 2R_{\rm t}/\beta =  92\,R_{\rm g}\,\beta^{-1}
     r_\star\,m_\star^{-1/3}
     M_{\bullet,6}^{-2/3},
    \label{eq:Rc}
\end{align}
at which location the dynamical timescale is
\be
\Omega_{\rm c}^{-1} \equiv \left(\frac{R_{\rm c}^{3}}{GM_{\bullet}}\right)^{1/2} \approx 75\,{\rm min}\, \beta^{-3/2}
m_{\star}^{-1/2} r_{\star}^{3/2} .
\label{eq:Omega_c}
\ee
The viscous accretion timescale at the circularization radius, $t_{\rm visc} = R_c^2/\nu$, where $\nu_{\rm c} \equiv \alpha R_{\rm c}^{2}\Omega(R_{\rm c})(h/R_{\rm c})^{2}$ is the kinematic viscosity and $h$ the disk vertical scale-height, is typically shorter than the fall-back time:
\begin{align}   
\frac{t_{\rm visc}}{t_{\rm fb}} \approx 
    0.1 \,
    \left(\frac{\alpha}{0.1}\right)^{-1}
    \left(\frac{h}{R_c}\right)^{-2}
    k^{3/2}\,\beta^{-3/2}
    m_\star^{1/2}
    M_{\bullet, 6}^{-1/2}
\label{eq:tvisc}
\end{align}
The fact that $t_{\rm visc} \lesssim t_{\rm fb}$ for a thick-disk $h/R_{\rm c} \sim 1$ shows that disk accretion rate can in principle follow the mass fallback rate, as many early works assumed \citep{Rees88,Ulmer99}.  At late times, when the fall-back rate drops and the disk becomes thinner, then $t_{\rm visc} > t_{\rm fb}$ and the accretion rate will no longer match the fall-back rate (e.g., \citealt{Mummery&VanVelzen24,Mummery+24}).

However, forming a thin, Keplerian disk of radius $R_c$ requires dissipating a specific energy
\begin{align}
    \epsilon_c \equiv \frac{GM_{\bullet}}{2R_c} - \epsilon_t \approx \frac{GM_{\bullet}}{2R_c} \approx 10^{52}\,\frac{{\rm erg}}{\,M_\odot}\, 
    \beta\,M_{\bullet,6}^{2/3} m_{\star}^{1/3} r_{\star}^{-1} 
    \label{eq:epsilon_c}.
\end{align}
This scale separation $\epsilon_c/\epsilon_t \sim (M_{\bullet}/M_{\star})^{1/3} \sim 100$ (equivalently, $a_0 \gg R_{\rm c}$) is the hallmark of TDE accretion flows, and highlights the role of two processes other than viscous evolution, which potentially limit accretion if they occur on timescales longer than $t_{\rm fb}$ (\citealt{Shiokawa+15}): 
(1) Dissipation of stream orbital kinetic energy via shocks 
(occurring on timescale $t_{\rm diss}$); (2) cooling of the debris to form a disk (on a timescale $t_{\rm cool}$).  The energy which must be radiated to form a rotationally-supported disk (Eq.~\eqref{eq:epsilon_c}) can be lost from the system through radiation, or carried in outflows. Our use of the term \emph{cooling} includes both channels.

\begin{figure}
    \centering
    \includegraphics[width= 0.5\textwidth]{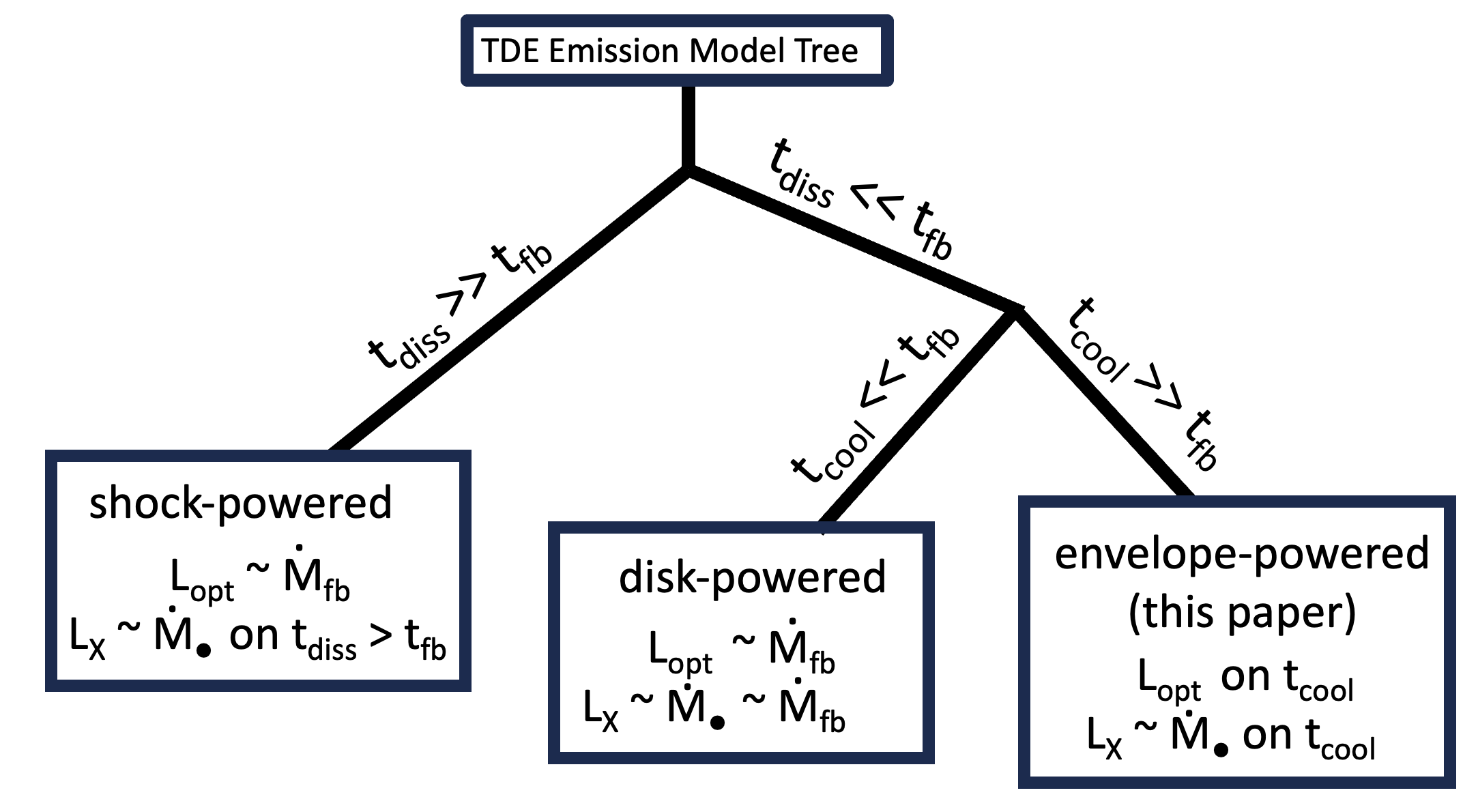}
    \caption{Schematic decision tree classifying models for early-time TDE emission according to the relative ordering of three timescales: (1) $t_{\rm fb}$, fall-back time over which debris returns to the black hole; (2) $t_{\rm diss}$, dissipation time over which streams kinetic energy is converted to debris thermal energy via shocks; (3) $t_{\rm cool}$, cooling time over which debris energy is lost to radiation and outflows. The dependence of $t_{\rm diss}$ and $t_{\rm cool}$ on TDE parameters is highly uncertain, and focus of extensive research. If dissipation is slow ($t_{\rm diss} \gg t_{\rm fb}$), shocks can power optical emission which tracks the fall-back rate, but accretion onto the SMBH is delayed.  However, even if dissipation is fast, this does not necessarily lead to rapid circularization and SMBH accretion.  If the cooling time of the debris is slow, the debris forms a large quasi-spherical envelope, as studied in this paper, whose slow contraction powers UV/optical emission and can also lead to the delayed feeding of the SMBH.   
    }
    \label{fig:TDEtree}
\end{figure}

As described in Sec.~\ref{sec:intro} and illustrated schematically in Fig.~\ref{fig:TDEtree}, various TDE scenarios can be classified based on how the timescales $\{ t_{\rm diss}$, $t_{\rm cool}$, $t_{\rm visc} \}$ compare to themselves and to the fall-back time of the debris $t_{\rm fb}$.  The finding of slow dissipation of the stream energy ($t_{\rm diss} \gg t_{\rm fb}$), which precludes rapid formation of a Keplerian disk, motivated the first shock-powered light-curve models \citep{Shiokawa+15,Guillochon&RR15,Piran+15}. However, recent global TDE simulations find that dissipation is rapid \citep{Andalman+22,Steinberg&Stone24,Price+24}.  When dissipation is rapid, $t_{\rm diss} \lesssim t_{\rm fb}$, the system's evolution instead depends on how efficiently the shocked debris can cool through radiation and outflows.  If cooling is rapid ($t_{\rm cool} \lesssim t_{\rm fb}$), then this supports slim/thick disk scenarios, in which even the earliest UV/optical TDE emission is powered by accretion onto the SMBH \citep{Rees88,Cannizzo+90,Ulmer99,Guillochon&RR13, Strubbe&Quataert09,Metzger&Stone16,Dai+18}, and the light-curve is expected to track the mass fall-back rate.  

On the other hand, if the shocked debris cools slowly compared to the fall-back time $t_{\rm cool} \gtrsim t_{\rm fb}$, then the returning mass cannot immediately form a compact disk near $R_{\rm c}$, but instead will build up in a loosely bound envelope extending to large scales $r \sim a_0 \gg R_{\rm c}$ (Eq.~\eqref{eq:a_min}), as dictated by the binding energy of the debris (Eq.~\eqref{eq:epsilon_t}; \citealt{Metzger22}). In this regime, the cooling time $t_{\rm cool}$ controls the mass inflow rate from large to small scales $a_{0} \rightarrow R_c$. The gradual, spherical inflow of mass allowed by cooling could in principle power the optical/UV light curve \citep{Metzger22}, even while the SMBH accretion rate (and, potentially, the X-ray and radio light curves) lag the mass-fall back rate.  We estimate the properties of such an envelope in the next section.  

\subsection{Hydrostatic TDE envelopes}
\label{sec:hs_TDE_envelopes}
The structure of the hypothesized TDE envelope following its formation
by $t \sim t_{\rm diss} \sim t_{\rm fb}$ is highly constrained by the assumption $t_{\rm cool} \gtrsim t_{\rm fb}$: Total mass, energy and angular momentum of the bound debris would be mostly conserved since 
the time of disruption. Since $t_{\rm fb}$ is roughly the dynamical
time at $r \sim a_{0} \gg R_c$, the envelope could also 
find approximate force equilibrium. Hence, this
would be a quasi-hydrostatic envelope of mass 
$M_{\rm env} \lesssim M_\star/2$, 
specific binding energy $\epsilon_t$ (Eq.~\eqref{eq:epsilon_t}) 
and specific angular momentum $l \approx \sqrt{GM\,R_c}$ (Eq.~\eqref{eq:Rc}). 

The weak binding energy (Eq.~\eqref{eq:epsilon_c}) requires most of the envelope mass to reside near its outer radius $r \sim a_0 \gg R_c$. Rotational support against gravity is negligible at these radii, rendering the envelope largely spherical in its geometry, except within a narrow polar funnel along the rotation axis. We focus on this outer, quasi-spherical portion of the envelope, in the estimates to follow.

The radial distribution of the envelope mass is less constrained, so we allow for a general density profile of the form
\begin{align}
    \rho(r) \propto r^{-n},
    \quad R_c \lesssim r \lesssim a_{0}\label{eq:rho_sph}.
\end{align}
This profile satisfies radial hydrostatic balance for pressure $P \propto \rho^{1 + 1/n}$, where $n$ is the polytropic index $n$. The ZEBRA models of \citet{Coughlin&Begelman14} predict $1/2 < n < 3$, while values of $n \approx 1-3$ are measured at the final state of numerical simulations of the disruption process that find efficient dissipation \citep{Andalman+22,Steinberg&Stone24,Price+24}.  To order of magnitude, the dynamical time near the outer radius, $\Omega_0^{-1} \equiv (a_0^{3}/GM_{\bullet})^{1/2}$ matches the fall-back time, $t_{\rm fb}$ (Eq.~\eqref{eq:tfb}).

For $n < 3$, a characteristic
density at radius $r \sim a_{0}$ can be estimated as
\begin{align}
    &\rho(a_{0}) \approx 
    \frac{3M_{\rm env}}{4\pi\,a_{0}^3}\nonumber\\
    & \approx 5.7\,\times\,10^{-12}\,
    {\rm g\,\rm cm^{-3}}
    \left(\frac{M_{\rm env}}{0.5\,M_\star}\right)
    k^{3}\,
    M_{\bullet, 6}^{-2}
    m_\star^{3}
    r_\star^{-3}
    \label{eq:rho_char}
\end{align}
Taking $c_s \approx (GM_\bullet/a_{0})^{1/2}$ as an estimate of the sound speed in the radiation-dominated envelope, hydrostatic balance, $P_{\rm rad}/r \sim GM_\bullet\rho/r^2$, gives a characteristic temperature:
\begin{eqnarray}
    T(a_0) &\approx& \left(\frac{3}{a}\frac{GM_\bullet \rho(a_0)}{a_{0}}\right)^{1/4} \nonumber \\
    &\approx& 2\,\times\,10^5\,K\,
    \left(\frac{M_{\rm env}}{0.5\,M_\star}\right)^{1/4}
    k\,
    M_{\bullet, 6}^{-5/12}
    m_\star^{11/12}
    r_\star^{-1}. \label{eq:temp_char}
\end{eqnarray}
Electron scattering dominates other sources of opacity at these temperatures and
densities, and the envelope is highly optically thick:
\begin{align}
    \tau(a_{0}) \approx \kappa_{\rm es}\rho(a_0)\,a_{0}
    \approx 789\,
    \left(\frac{M_{\rm env}}{0.5\,M_\star}\right)
    k^{2}\,
    M_{\bullet, 6}^{-4/3}
    m_\star^{7/3} 
    r_\star^{-2} \label{eq:tau_amin},
\end{align} 
where again we take $\kappa_{\rm es} \approx 0.38\,\rm cm^2\,\rm g^{-1}$.  Photons diffuse through the envelope on the timescale,
\begin{align}
    t_{\rm diff}(a_{0}) &\equiv \tau(a_{0})\frac{a_{0}}{c}
     \nonumber\\
    &\approx 106\,{\rm day\,}
    \left(\frac{M_{\rm env}}{0.5\,M_\star}\right)
    k\,
    M_{\bullet, 6}^{-2/3}
    m_\star^{5/3} 
    r_\star^{-1}, \label{eq:t_diff}
\end{align}
which moderately exceeds the dynamical timescale at $r \sim a_{0}$ (i.e., $t_{\rm diff} \gtrsim t_{\rm fb})$, and this remains true at smaller radii for $n > 1/2$,
since $t_{\rm diff}/t_{\rm dyn} \propto \kappa\rho{r}^2/r^{3/2} \propto r^{1/2 - n}$. 

A quasi-static, radiation-dominated envelope would cool and contract on its diffusion timescale, similar to the Kelvin-Helmholtz contraction of a proto-star \citep{Metzger22}. However, as is true also in proto-stars, the envelope can be unstable to convection, depending on its density profile (Eq.~\eqref{eq:rho_sph}). Radiation-dominated envelopes are also prone to instabilities driven by the temperature/density dependence of the opacity 
\citep{Kiriakidis+93, Blaes&Socrates03}, or inhomogeneities causing a reduction in the effective radiation force (\citealt{Shaviv98, Shaviv00, Shaviv01, vanMarle+08}, see \citealt{Owocki15} for a review).

The stability of a rotating fluid against (adiabatic) convection is assessed using the Solberg-Hoiland criteria (e.g., \citealt{Blandford&Begelman04}; their Sec. 3.1). At large radii, where rotation is subdominant and geometry is spherical, 
this reduces to the usual Schwarzschild stability criterion $ds/dr \geq 0$ on the specific entropy \citep{Schwarzschild58}.  For a radiation dominated envelope $P \propto T^4 \propto \rho^{(1+n)/n}$, and hence
\begin{align}
    s \propto T^3/\rho \propto \rho^{(3 - n)/4n} \propto r^{n - 3}.
\end{align}
TDE envelopes with $n < 3$ are thus Schwarzschild unstable and expected to develop convection, whose growth timescale can be as short as the local dynamical timescale \citep{Kippenhahn&Weigert13}.  Even if the envelope is stable according to the Hoiland/Schwarzschild criterion, diffusive losses at optically thin surface layers can generate super-adiabatic temperature gradients, driving convection.

A convective envelope can maintain its quasi-hydrostatic structure, if convection can efficiently transport energy \citep{Joss+73}. This is likely to be the case in the TDE envelopes of interest. Energy transport by convective eddies will be inefficient if the required eddy velocity, $v_{\rm c}$, becomes supersonic, or if radiative losses within a characteristic turnover time become significant (i.e., the convection becomes ``lossy'').  Assuming the envelope carries a convective energy flux $L_{\rm c} \approx 4\pi a_0^{2}(\rho(a_0) c_{\rm s}^{2}v_{\rm c})$, the convective Mach number can be estimated as
\begin{align}
\mathcal{M}_{\rm c} &= \frac{v_{\rm c}}{c_{\rm s}} \approx
\frac{L_{\rm c}}{3 M_{\rm env} a_0^2\Omega^3(a_0)} \nonumber \\
\approx& 0.06\,\left(\frac{L_{\rm c}}{L_{\rm edd}}\right)
\left(\frac{M_{\rm env}}{0.5\,M_\star}\right)^{-1}
    k^{-5/2}\,
    M_{\bullet, 6}^{7/6}
    m_\star^{-8/3} 
    r_\star^{5/2},
\end{align}
where we have used Eq.~\eqref{eq:rho_char} for $\rho(a_0)$.  We see that $\mathcal{M}_{\rm c} < 1$ for $L_{\rm c} \lesssim 20\, L_{\rm edd}$ for fiducial parameters.

Convection becomes lossy below a critical optical depth, $\tau_c$, which can be estimated by equating the local diffusive flux to convective flux of thermal energy $c\,aT^4/\tau_c \sim v_{\rm c}\,aT^4 \lesssim c_s\,aT^4$ 
\citep{Jiang+15}:
\begin{align}
    \tau_c(r) \equiv \frac{c}{c_s} 
    \underset{c_s \approx r\Omega(r)}{\sim} 
    \left(\frac{r}{R_g}\right)^{1/2} 
\end{align}
The outer layers of the TDE envelope at $r \lesssim a_{0} \sim 10^3\,R_g$, obey $\tau_c \sim 30 \ll \tau(a_{0})$ (Eqs.~\eqref{eq:a_min}, \eqref{eq:tau_amin}), and hence should be capable of efficient convection; since $\tau/\tau_c \propto r^{1/2 - n}$, this likely remains true also deeper in the envelope $r \ll a_0$ for $n > 1/2$.

Perturbations around hydrostatic equilibrium occur on the diffusion timescale. A layer of gas that moves radially inward on the local diffusion timescale  generates a local accretion power close to the Eddington luminosity:
\begin{align}
    L_{\rm grav} \equiv \frac{GM_\bullet\,\dot{M}}{r}
    \underset{\dot{M} \sim M/t_{\rm diff}}{\sim} 
    \frac{GM_\bullet\,4\pi\,r^3\,\rho}{r\,\kappa\rho r^2/c}
     = L_{\rm edd}.
\end{align}
Faster perturbations will generate accretion power
at super-Eddington rates, trapping radiation. 
If convection cannot redistribute the energy efficiently, a natural outcome is the development of winds or outflows that carries it away from the system
\citep{Quataert+16}. 

Regardless of whether by means of convection, 
a wind or radiative diffusion, the envelope is expected to evolve on the 
\emph{effective cooling timescale} $t_{\rm cool}$ (Fig.~\ref{fig:cartoon}). This cooling time can be in principle be (much) shorter than a diffusion time (leading to super-Eddington energy fluxes across the envelope), if a combination 
of convective energy transport and winds can redistribute energy
in the system efficiently. Nevertheless, even then, the evolution can maintain quasi-hydrostatic balance, if $t_{\rm diff} \gtrsim t_{\rm cool} \gg t_{\rm dyn}$, as we shall demonstrate in the following sections. 

\section{Numerical setup}
\label{sec:numerical_setup}

We employ the explicit radiation transfer module \citep{Jiang+14b, Jiang21} of \texttt{Athena++} \citep{Stone+20} to solve the equations of radiation hydrodynamics in axisymmetry.  We summarize the numerical set-up and initial conditions of the envelope in Sec.~\ref{sec:basic_numerical_setup}, with greater details provided in Appendices~\ref{app:numerical_details} and \ref{app:initial_conditions}.  In Sec.~\ref{sec:diagnostics} we define key diagnostics used to analyze our simulation results.

\subsection{Initial conditions and numerical procedure}
\label{sec:basic_numerical_setup}

We initialize our simulations with analytical solutions of rotating, adiabatic, axisymmetric tori in mechanical equilibrium (\citealt{Papaloizou&Pringle84}, see Appendix~\ref{app:initial_conditions} for details). 
These solutions assume force-balance with
a prescribed angular momentum profile, and a polytropic relationship between pressure and density $P \propto \rho^{1 + 1/n}$. The density and pressure distributions are given by
\begin{align}
    \rho(r, \theta) &= \rho_0\,\left(\frac{\text{Be}_0 - \Psi}{\text{Be}_0 - \Psi_0}\right)^n,\nonumber\\
    P(r, \theta) &= \frac{\rho}{1 + n}
    \left(\text{Be}_0 - \Psi 
   \right)
    \label{eq:init_rho_press}
\end{align}
where $\rho_0$ is a constant set by the total mass of the torus, $\text{Be}_0$ is the Bernoulli parameter, and
\begin{align}
    \Psi(r, \theta) \equiv 
        \Phi
    + \frac{GM_\bullet\,R_c}{2\,(r\sin\theta)^2}
    \label{eq:eff_potential}
\end{align}
is the effective potential.  We employ a (static) Newtonian gravitational potential
\begin{align}
    \Phi = -\frac{GM_\bullet}{r}.   
    \label{eq:Phi_BH}
\end{align}
General relativistic effects are potentially important in the tidal disruption
process and the formation of the envelope from debris streams (e.g., \citealt{Kesden12,Cheng+14,Bonnerot&Stone21}), and our
numerical grid extends down to radii $R_{\rm in} = 6\,R_{\rm g}$ close to the horizon. A Newtonian 
potential nevertheless suffices because our focus is on the evolution of the system at large radii $r \gtrsim R_{\rm c} \gtrsim  30\,R_{\rm g}$ (Eq.~\eqref{eq:Rc}) where relativistic
effects are relatively minor. Eq.~\eqref{eq:eff_potential} assumes a constant specific initial angular momentum distribution $l_z = \sqrt{GM_\bullet\,R_c}$ (Eq.~\eqref{eq:Rc}), as adopted in all our simulations, so the velocity field is initialized to be purely azimuthal:
\begin{align}
    v_\phi\,(r,\theta) = 
    \frac{\sqrt{GM_\bullet\,R_c}}{r\sin\theta},   
\end{align}
with all poloidal velocity components set to zero, i.e. $v_r = 0$, $v_\theta= 0$. 

Bound tori with $\rm Be_0 <0$ possess a well-defined surface $\Psi = \rm Be_0$, beyond which $\rho = P = 0$. Outside of this surface, we impose density/temperature floor values, and set $v_\phi = 0$. We initialize the gas temperature $T_g(r,\theta)$ by solving
\begin{align}
    \frac{aT_g^4}{3} = P\,(r,\theta) - 
    \frac{k_B\rho\,T_g}{\mu m_p},
\end{align}
where $k_B$ is the Boltzmann constant and $m_p$ is the proton mass,
and $\mu = 1$ is mean molecular mass. The radiation intensity field is initialized to be locally isotropic,
\begin{align}
    I(t=0, r, \theta) = 
    \frac{a\,T_g^4\,(r, \theta)}{4\pi} \label{eq:intensity_init},  
\end{align}
  Electron scattering dominates other sources of opacity throughout the envelope (Sec.~\ref{sec:analytical_estimates}, see also Fig.~\ref{fig:initial_conditions}). We employ a simple grey prescription for the Rosseland mean opacity:
\begin{align}
    \kappa = \begin{cases}
        0.38\,\rm cm^{2}\,g^{-1}&, \quad T_g > T_{\rm rec}\\
        0&, \quad \rm otherwise\\
    \end{cases},  \label{eq:opacity}
\end{align}
where $T_{\rm rec} = 10^{4}$ K is the recombination temperature, below which the gas becomes largely neutral and the scattering opacity will drop.

We impose \emph{selective boundary conditions} (e.g., \citealt{Caproni+23})
on all of the hydrodynamic variables and the intensity field: the values of all primitive variables (density, pressure, velocity) in the ghost zones are copied from the boundary zone, except for the radial velocity $v_r$. In the inner and outer ghost zones, we set $v_r = 0$ when $v_r > 0$ and $v_r < 0$, respectively. This minimizes the flow of gas into the grid from outside. We apply a similar boundary condition on the radiation intensity: rays leave the grid with zero radial gradient, but are not allowed to enter.

\begin{deluxetable*}{ccc}
\tablecolumns{3}
\tablewidth{0pt}
 \tablecaption{Parameters of the \texttt{FIDUCIAL} model.
 \label{tab:fiducial_param_table}}
 \tablehead{
 \colhead{Symbol} & \colhead{Description} & \colhead{Value}}
 \startdata 
    \hline 
    & Disruption  & \\
   \hline
 $M_\bullet$ & Mass of SMBH & $10^6\,M_\odot$ \\
 $M_\star$ & Mass of disrupted star & $M_\odot$ \\
 $R_\star$ & Radius of disrupted star & $R_\odot$ \\
 $M_{\rm env}$  & Bound mass after disruption $\sim$ initial envelope mass 
 & $0.5\,M_\odot$ \\
 $k$ & Prefactor entering debris binding energy (Eq.~\eqref{eq:epsilon_t}) & $0.8$ \\
 $\beta$ & Orbital penetration factor (Eq.~\eqref{eq:beta}) & $2.5$\\
 \hline 
    & Initial Conditions & \\
   \hline
   $n$ & Polytropic/power-law index of radial density profile (Eq.~\eqref{eq:rho_sph}) & $1.6$ \\
   $\text{Be}_0$ & Bernoulli parameter of the initial gas profile & $-5\,\times\,10^{50}\,\rm erg\,M_\odot^{-1}$\\
    $p$ & power-law index of specific angular momentum profile (Eq.~\eqref{eq:lprofile})
   & $0$\\
 \hline 
    & Derived & \\
   \hline
   $R_{\rm g}$ & Gravitational radius of the SMBH &
   $1.5\,\times\,10^{11}\,\rm cm$\\
   $R_c$ & Circularization radius (Eq.~\eqref{eq:Rc}) & $37\,R_{\rm g}$\\
   $a_{0}$ & Semi-major axis of most tightly bound debris (Eq.~\eqref{eq:a_min})
   & $2935\,R_{\rm g}$\\
   $t_{\rm fb}$ & Fallback time (Eq.~\eqref{eq:tfb}) & 
   $57\,{\rm day}$\\
   $\Omega_c^{-1}$ & Dynamical timescale at $R_c$ (Eq.~\eqref{eq:Omega_c}) & $19\,\rm min$\\
$\Omega_{0}^{-1}$ & Initial dynamical time of outer envelope ($r \sim a_0$) & $13\,{\rm day}$\\
   $t_{\rm diff}$ & Initial diffusion time through envelope ($r\sim a_0$; Eq.~\eqref{eq:t_diff})
   & $85\, \rm day$\\
    \hline 
    & Simulation & \\
   \hline
   $R_{\rm in}-R_{\rm out}$ & Radial range of the 
   computational domain & $6\,R_{\rm g}-10^{5}R_{\rm g}$\\
 \enddata
 \end{deluxetable*}

\subsection{Diagnostics}
\label{sec:diagnostics}

A cell on the grid is considered \emph{bound},
if its total specific energy obeys:
\begin{align}
    u_{\rm tot} &= -\frac{GM}{r} + \frac{1}{2}v^2 + u_{\rm rad} + 
    \frac{2}{3}P_{\rm gas} < 0. \label{eq:u_tot_def}
\end{align}
We include radiation energy in this accounting scheme if the local optical depth $\rho \kappa r$ exceeds unity, i.e.,
\begin{align}
    \rho u_{\rm rad} &= \begin{cases} 
    \int\,d\Omega\,I, &\quad \rho\kappa{r} > 1\\
    0 &\quad \text{otherwise.}
    \end{cases} \label{eq:u_rad_def}
\end{align}
Typically, if the cell is not in the floor region (Sec.~\ref{sec:basic_numerical_setup}), the condition $\kappa\rho\,{r} \gg 1$ is satisfied. The Bernoulli parameter is likewise defined as
\begin{align}
    \text{Be}(t, r, \theta) \equiv  -\frac{GM}{r} + \frac{1}{2}v^2 + \frac{4}{3}\,u_{\rm rad} + 
    \frac{5}{2}P_{\rm gas}.  \label{eq:bernoulli_def}
\end{align}
The flows we simulate are highly turbulent, hence we found it useful to introduce several space/time averages. \emph{Angular averages}
are taken as
\begin{align}
    \langle\dots\rangle = \frac{1}{2}\int\,\dots \sin\theta d\theta
\end{align}
Bounds of time averages are reported when used. 

We analyze radial transport of
energy and mass through \emph{isotropic fluxes}:
\begin{align}
    \dot{m} &\equiv
    4\pi{r}^2\,\rho\,{v_r} \nonumber\\
    l_{\rm rad} &\equiv 4\pi\,r^2\,F_{\rm rad, r} \nonumber\\
     l_{\rm kin} &\equiv 4\pi\,r^2\,v_r\,
    \frac{\rho\,v^2}{2} \nonumber\\
    l_{\rm{grav}} &\equiv -\frac{GM_\bullet\,\dot{m}}{r}\nonumber\\
    l_{\rm tot} &\equiv l_{\rm rad} + l_{\rm gas}
    + l_{\rm grav}
    \label{eq:iso_fluxes}
\end{align}
Gas pressure and its contribution to total energy is found to be negligible
compared to kinetic and radiative contributions at all times in our simulations. 

We reserve capital letters for angular averages, e.g.,
\begin{align}
    \dot{M}(t, r) &\equiv \langle\dot{m}\rangle_\theta \nonumber\\
    L_{\rm tot}(t, r) &\equiv \langle l_{\rm tot} \rangle_\theta 
    \label{eq:shell_fluxes}
\end{align}
Flat, radial profiles of $L_{\rm tot}$ and $\dot{M}$ 
indicate that a steady state has been achieved locally.

We shall find it convenient to separate some properties of the envelope into distinct bound and unbound parts. For example, the instantaneous \emph{bound mass density} on the grid is defined as
\begin{align}
    \rho_{\rm bnd} \equiv \begin{cases}
        \rho(t, r, \theta), &\quad u_{\rm tot} < 0\\
        0 &\quad \text{otherwise.}
    \end{cases}  \label{eq:rho_bnd}
\end{align}
When showing angle-averaged quantifies, we define an ``average optical depth'' according to,
\begin{align}
    \tau(r, t) &\equiv \frac{1}{2}\int_r^\infty\,dr\,\int_0^\pi\,d\theta\,\sin\theta\, 
    \kappa\,\rho(t, r, \theta) \label{eq:tau_av}
\end{align}
and identify an average photosphere radius as $\tau(t, R_{\rm ph}(t)) = 1$. Radii of enclosed mass are defined as
\begin{align}
    m = \int_{R_{\rm in}}^{r(t, m)}\,dr\,r^2
    \int_0^\pi\,d\theta\sin\theta\,\rho(t, r, \theta).   
\end{align}
\section{Results}
\label{sec:results}
\begin{figure*}
    \centering
    \includegraphics[width= \textwidth]{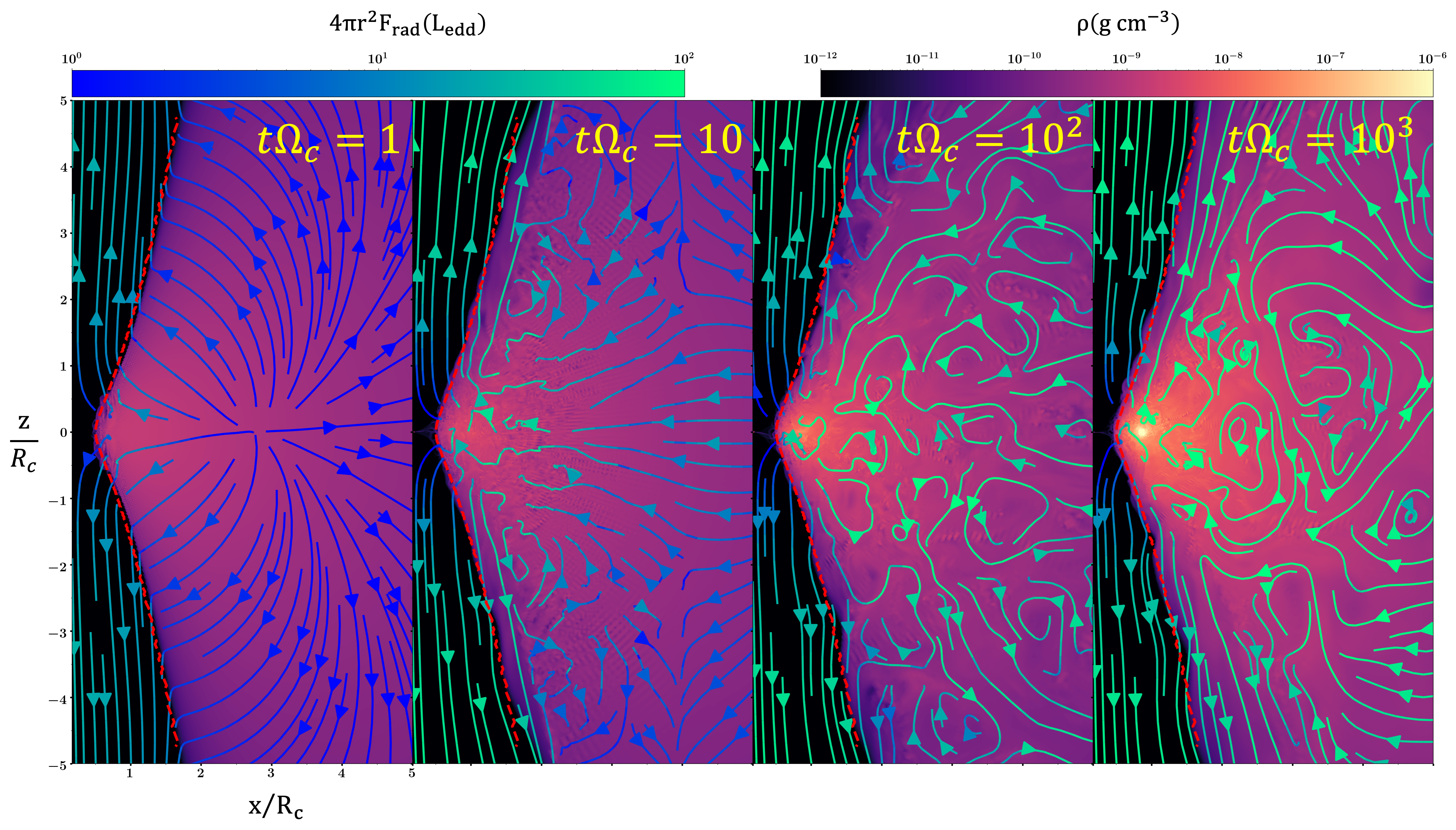}
    \caption{Snapshots of mass density $\rho$ and streamlines of radiation flux $\mathbf{F}_{\rm rad}$ at times $t\Omega_c \in \{1, 10, 10^2, 10^3\}$, and radii $r < 5\,R_c$. Flux streamlines are colored by the equivalent isotropic luminosity $4\pi r^2 F_{\rm rad}$. The polar funnel region is populated by the density floor (Eq.~\eqref{eq:floor_values}) and hence is optically thin to radiation at all times; its initial boundary at $t = 0$ is marked by a red dashed curve (see Eq.~\eqref{eq:init_rho_press}). 
    The initial equilibrium torus first becomes destabilized at the funnel boundary near the base of the envelope at $r \approx R_{\rm c}$ already by $t\Omega_c \approx 1$, while the entire flow is turbulent by $t\Omega_c = 100$. Radiative cooling through the funnel allows mass to sink into the potential well, as evidenced by the over-dense ``ring'', or disk, accumulating at $r = R_c$, that appears already by $t\Omega_c \approx 10$.
    The region encompassed by this plot encloses $3.6\%$, $3.6\%$, $5.2\%$, $19\%$ of the total mass at these times, requiring an effective cooling/accretion times shorter than the local diffusion timescale (see also Figs.~~\ref{fig:mass_radius_time}, \ref{fig:initial_conditions}).
    Radiation is transported at (locally) super-Eddington rates $\gtrsim 10\,L_{\rm edd}$ by advection within the envelope, and $\gtrsim\,100\,L_{\rm edd}$ in the funnel. The integrated luminosity leaving the funnel region is found to be $\approx 3-4\,L_{\rm edd}$ throughout the evolution (see Fig.~\ref{fig:wind_cooling}).
    } 
    \label{fig:fiducial_early}
\end{figure*}

\subsection{Brief overview of simulation suite}
 \label{sec:results_overview}
 
Our suite of numerical simulations consists of 5 models as summarized in Table~\ref{tab:sim_suite}. The fiducial model \texttt{FIDUCIAL} considers the canonical disruption of a sun-like star ($M_\star = M_\odot$, $R_\star = R_\odot$) by a SMBH of mass $M_\bullet = 10^6\,M_\odot$ on a mildly penetrating $\beta = 2.5$ parabolic orbit.  The model parameters for \texttt{FIDUCIAL} are listed in Table~\ref{tab:fiducial_param_table}. The next section (Sec.~\ref{sec:fiducial}) will describe in detail the evolution of \texttt{FIDUCIAL}. We shall show that a quasi-hydrostatic envelope is established, similar to that described in Sec.~\ref{sec:analytical_estimates}, but which cools faster than one estimates by radiative diffusion alone, due to turbulent energy transport and polar outflows. 

Given what appears to be a pronounced role of advective energy transport in model \texttt{FIDUCIAL}, we next explore the dynamical impact of radiative cooling on the evolution by performing (an otherwise equivalent) model \texttt{ADIABATIC}, for which radiation transport is artificially turned off (Sec.~\ref{sec:ADIABATIC}). The fact that the resulting evolution differs significantly from the \texttt{FIDUCIAL} indicates that radiation plays a key role in the dynamics of the envelope, even if radiative diffusion doesn't directly dominate the envelope cooling. To explore further the role of radiative diffusion on the envelope cooling time, in models \texttt{HIGH$\_$MASS} and \texttt{LOW$\_$MASS}, we increase and decrease, respectively, the envelope mass relative to \texttt{FIDUCIAL} in order to explore the impact of larger or smaller radiative diffusion timescale through the envelope (Sec.~\ref{sec:envelope_mass}). Motivated by the envelope structure found by \citet{Steinberg&Stone24}, we take $n = 1.6$ for the initial density profile $\rho \propto r^{-n}$ for most of our models, corresponding to a convectively unstable envelope.  We therefore ran a separate model \texttt{STEEP}, with the same envelope mass as \texttt{FIDUCIAL} but with a steeper initial density profile ($n = 3$) which is marginally stable to convection in the adiabatic limit. The qualitative long-term evolution of \texttt{STEEP} is similar to \texttt{FIDUCIAL}, indicating that our results are not contingent upon the envelope being unstable to convection in the initial state.

\begin{deluxetable}{cccc}
\tablecolumns{3}
\tablewidth{0pt}
 \tablecaption{Simulation suite}
 \label{tab:sim_suite}
 \tablehead{
 \colhead{Model} 
 & \colhead{Radiative transfer}
 & \colhead{n}
 & \colhead{$M_{\rm env}$ ($M_\odot$)}}
 \startdata 
 \hline
 $\texttt{FIDUCIAL}$ & ON & $1.6$ & $0.5$\\
 $\texttt{ADIABATIC}$ & OFF & $1.6$ & $0.5$\\
 $\texttt{STEEP}$ & ON & $3.0$ & $0.5$\\
 $\texttt{HIGH$\_$MASS}$ & ON & $1.6$ & $5$\\
 $\texttt{LOW$\_$MASS}$ & ON & $1.6$ & $0.05$  \\
 \enddata
 \end{deluxetable}

\subsection{\texttt{FIDUCIAL}}
\label{sec:fiducial}
Our fiducial model is initialized as an equilibrium torus with total mass, energy and angular momentum dictated by the disruption process (Sec.~\ref{sec:analytical_estimates}). Fig.~\ref{fig:initial_conditions} in Appendix \ref{app:initial_conditions} shows angle-averaged radial profiles of the initial conditions, from which we highlight three features: (1) the total mass and thermal energy of the envelope are concentrated at large radii $r \approx 40\,R_c$ (panel (iv)); (2) rotational support is relevant only on small scales $r \sim R_c$ (panel (v)), i.e. the bulk of the envelope is supported by radial pressure gradients, more similar to a ``star'' than a standard disk; (3) the characteristic photon diffusion timescale through envelope is $\gtrsim 10^3\,\Omega_c^{-1}$ (panel (vii)), i.e., many inner dynamical times. 

After the start of the simulation, we almost immediately witness the development of turbulence near the inner edge of the torus.  Fig.~\ref{fig:fiducial_early} shows maps of the mass density and streamlines of
(lab frame) radiation flux at radii $r < 5\,R_c$, taken at several snapshots in time from $t\Omega_c = 1$ until $t\Omega_c = 10^3$.  A funnel region, which is devoid of mass and optically-thin to radiation, is present at all times; its initial boundary at $t = 0$ is shown with a red dashed line (Eq.~\eqref{eq:init_rho_press}). 
The inner envelope becomes turbulent by $t\Omega_c \approx 100$, as evident from irregularity of the streamlines and density contours.  This turbulence is initiated by an instability that develops at the torus-funnel interface, starting as early as $t\Omega_c \approx \text{a few}$. 

An over-dense ring near $r = R_c$, a nascent rotationally-supported disk, becomes noticeable
by $t\Omega_c = 100$, as matter flows to small radius through the envelope.  Indeed, between $t\Omega_c = 10$ and 
$100$, the mass enclosed within $r = 5\,R_c$ is found to increase
from $\approx 3.6\%$ to $5.2\%$ of the total envelope mass. This inward flow of mass requires the removal 
of pressure support against gravity by cooling, since in the initial state $\lesssim 20\%$ of the gravitational force at $r\approx 5-7\,R_c$ is balanced by rotation (panel (v) of Fig.~\ref{fig:initial_conditions}).

One explanation for this loss of pressure support is radiative diffusion to the optically-thin funnel; however, the inflow of gas occurs on a timescale much shorter than the local photon diffusion timescale of $\kappa\rho r^2/c \gtrsim 10^3\Omega_c^{-1}$ at the radii $5 - 7\,R_c$ of interest (Fig.~\ref{fig:initial_conditions}, panel (vii)).  The inner envelope is cooling and settling into a disk faster than allowed by photon diffusion only, a point we return to in Sec.~\ref{sec:envelope_mass}.

\begin{figure*}
    \centering
    \includegraphics[width= \textwidth]{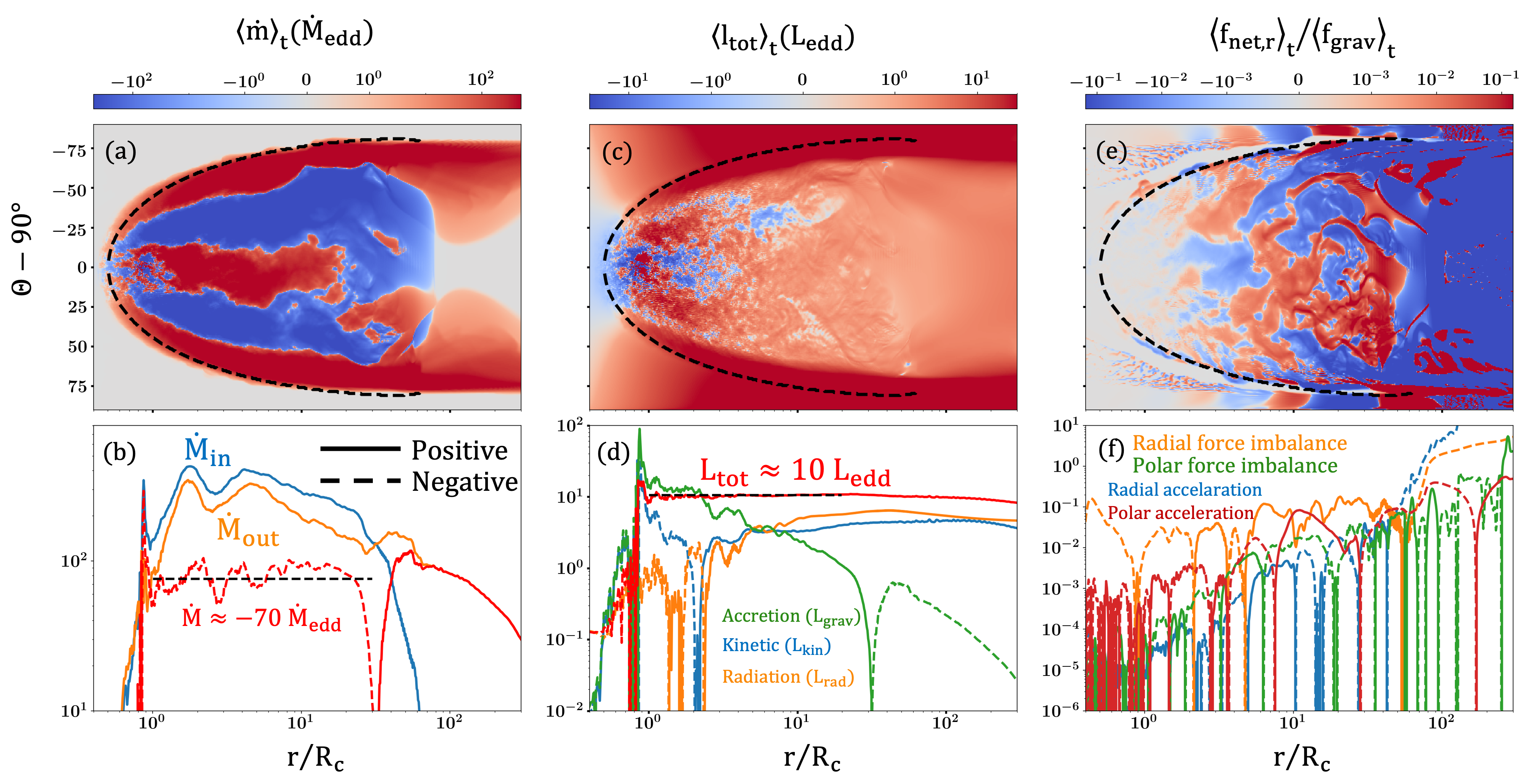}
    \caption{Quasi-steady-state properties of envelope
    mass, energy and momentum on $r-\theta$ plane (top panels), and their angular averages (bottom panels) 
    averaged over the time interval $t\Omega_c\,\in\,[500, 1000]$. In the bottom panels, dashed and solid lines indicate negative and positive values of the quantity plotted, respectively (e.g., the red line in panel (b) shows $\dot{M} \approx -70\,\dot{M}_{\rm edd}$ within $r \lesssim 10\,R_c$). Isotropic fluxes and angle averages
    are defined in Eq.~\eqref{eq:iso_fluxes}-\eqref{eq:shell_fluxes}.
    (a): Isotropic mass accretion rate $\dot{m}$ in units of
    Eddington accretion rate $\dot{M}_{\rm edd} \equiv 10\,L_{\rm edd}/c^2$ (b) the angular average
    $\dot{M}$ with its breakdown into inflow and outflow rates
    $\dot{M}_{\rm in}$, $\dot{M}_{\rm out}$
    (c) total isotropic 
    luminosity $l_{\rm tot}$, (d) breakdown of 
    total luminosity $L_{\rm tot}$ to local 
    accretion power $L_{\rm grav}$,
    (lab-frame) radiation flux
    $L_{\rm rad}$, and
    kinetic energy flux
    $L_{\rm kin}$,
    (e) radial force imbalance defined as 
    $\langle f_{\rm rad, r} + f_{\rm grav}\rangle_t/\langle f_{\rm grav}\rangle_t$ where $f_{\rm rad, r}$ is the radial component
    of the radiation force, $f_{\rm grav} = -GM_\bullet/r^2$,
    (f) shell averages $\langle f_{\rm rad, r} + f_{\rm grav}\rangle_{t,\theta}/\langle f_{\rm grav}\rangle_{t,
    \theta}$ (similarly for polar 
    component of the momentum equation), as well as average
    rate of change for momentum components $\langle\partial_t\rho\mathbf{v}\rangle_{t,\theta}/
    \langle f_{\rm grav}\rangle_{t,\theta}$. The flow
    interior to $r \lesssim 10\,R_c$ is in a quasi-steady state
    with radially constant net mass accretion rate $\dot{M} \approx -70\dot{M}_{\rm edd}$ (panel b), 
    and outwards flux of energy
    $L_{\rm tot} \approx 10\,L_{\rm edd}$ (panel d). 
    Force
    imbalance is $\lesssim 10\%$ of gravitational force 
    in the radial direction
    (panel f, orange line) 
    $\lesssim \text{a few} \%$ in the polar direction
    (panel f, green line). 
    }
    \label{fig:fiducial_steady}
\end{figure*}

Despite this rapid cooling and associated inflow, by $t\Omega_c \approx 5-10\,\times\,10^2$ ($1-2\,\rm weeks$),
the envelope has found a quasi-steady state across radii
$r \lesssim 10\,R_c$. Fig.~\ref{fig:fiducial_steady}
shows time averages of the mass/energy
transport rates (left and mid panels), as well as force 
imbalance (right panel). Most of the mass inflow occurs along intermediate latitudes at isotropic accretion
rates (Eq.~\eqref{eq:iso_fluxes}) of 
$|\dot{m}| \approx 200-400\,\dot{M}_{\rm edd}$ 
(panel (a), blue regions). An outflow of gas carrying mass at a similar rate is directed along the mid-plane and polar latitudes (red regions), largely balancing
the inflow at every shell within $r \lesssim 10\,R_c$ 
(blue and orange lines in panel (b)).  The net accretion rate across this steady-state region $r \lesssim 10\,R_c$ exhibits large fluctuations around a mean inward value $\dot{M} \approx -70\,\dot{M}_{\rm edd}$ 
(red dashed line, panel(b)). On the other hand, mass accretion interior
to the angular momentum barrier near $r \approx R_c$
is negligible, indicating that angular momentum 
transport is inefficient in the absence of magnetic fields, or explicit viscosity.

The net radial flux of energy, $l_{\rm tot}$ (Eq.~\eqref{eq:iso_fluxes}), is outwards throughout the 
majority of the envelope (panel (c), red regions),
dominated by accretion power $l_{\rm grav}$ at small radii and radiation $l_{\rm rad}$ at large radii (panel (d), green and orange lines).  The angle-integrated accretion power $L_{\rm grav}$ (panel (d), green line, see Eq.~\eqref{eq:iso_fluxes}-\eqref{eq:shell_fluxes}) 
increases moving radially inwards from $L_{\rm grav} \approx 2\,L_{\rm edd}$ at $r \approx 10\,R_c$ to $L_{\rm grav} \approx 20\,L_{\rm edd}$ at 
$r \approx R_c$, consistent with the radially constant inflow rate $\dot{M} \approx -70\,\dot{M}_{\rm edd}$. 
Radiation and gas kinetic energy are advected inwards 
at $r \lesssim 2\,R_c$ with the flow, and 
outwards at $2\,R_c \lesssim r \lesssim 10\,R_c$ 
(panel (d), orange and blue lines), such that the net luminosity is radially constant 
$L_{\rm tot} \approx 10\,L_{\rm edd}$ (red line).
Outside the steady-state zone ($r \approx 20\,R_c$), energy is carried by advection of photons at
a rate $L_{\rm rad} \approx 5-6\,L_{\rm edd}$
(orange), and kinetic energy
$L_{\rm kin} \approx 4-5\,L_{\rm edd}$ (blue). The dark red regions at high latitudes indicate that the energy flux is directed towards the funnel (outside black dashed line) instead of outwards through the bulk of the envelope closer to the midplane.

Panels (e) and (f) of Fig.~\ref{fig:fiducial_steady} demonstrate the evolution of the gas momentum.  Typically, the average net acceleration $\langle\partial_t(\rho\mathbf{v})\rangle_{t}/\langle\rho\rangle_{t}$ is found to be $\lesssim 10^{-2}$ of the gravitational acceleration
(panel (f), blue and red lines), 
indicating that the gas momentum has also reached a quasi-steady state at radii $r \lesssim 10\,R_c$. 
The imbalance in radiation, centrifugal and 
gravitational forces are $\lesssim 10\%$ of
the gravity (orange and green lines). We conclude that the envelope excluding the polar outflows is quasi-hydrostatic in this steady phase of evolution.

\begin{figure}
    \centering
    \includegraphics[width= 0.48\textwidth]{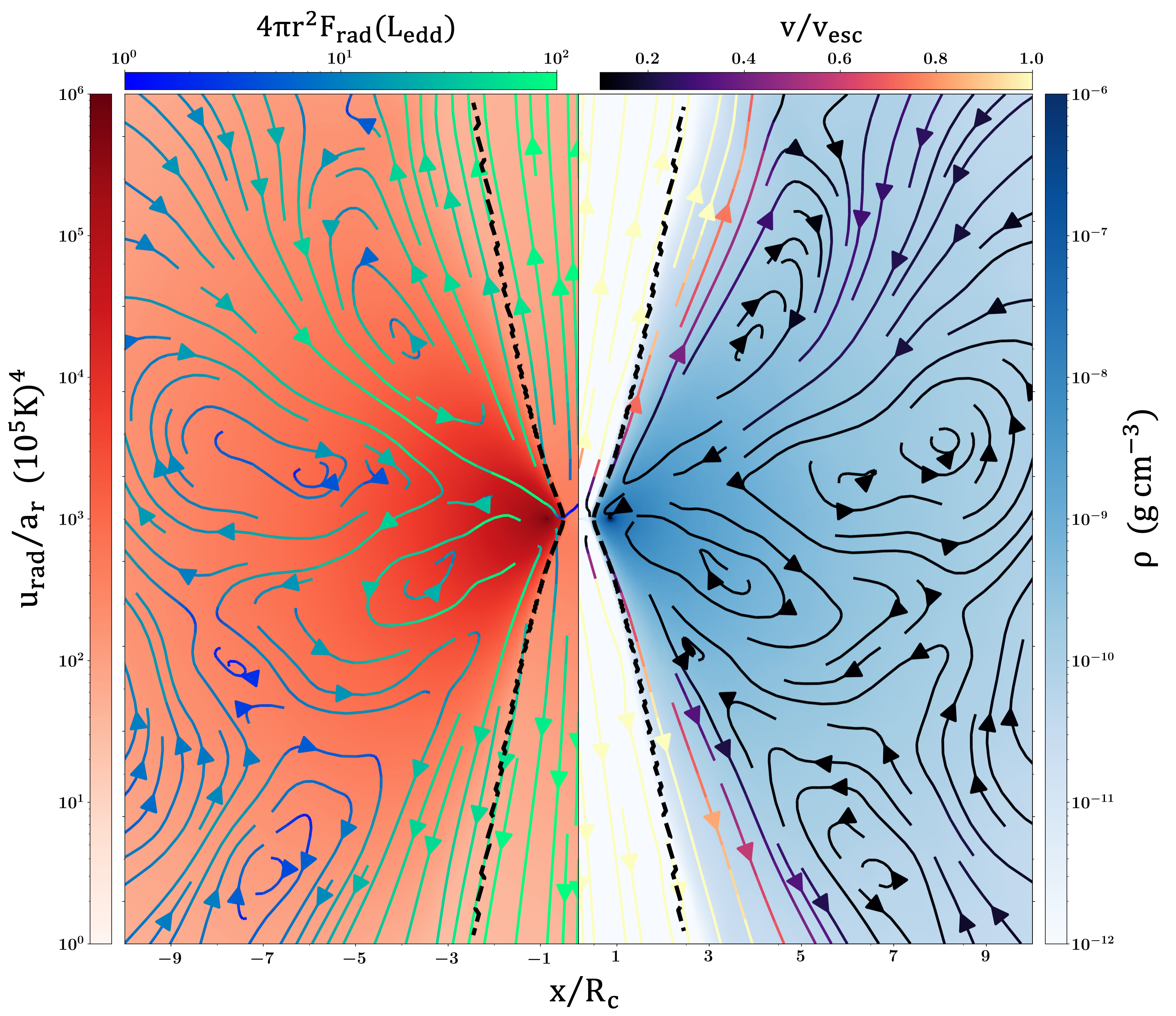}
    \caption{Quasi-steady-state distributions of radiation energy 
    and mass, averaged over the same time interval as in Fig.~\ref{fig:fiducial_steady}.
    \emph{Left panel}: Radiation energy density 
    $\langle u_{\rm rad} \rangle_t$
    normalized to $a_r\,T_0^4$ where
    $T_0 = 10^5\,K$, and $a_r$
    is radiation density constant. 
    Streamlines of lab frame radiation flux
    $\langle\mathbf{F}_{\rm rad}\rangle_t$
    are plotted on top, colored according to 
    their equivalent isotropic luminosity
    $4\pi\,r^2|\langle\mathbf{F}_{\rm rad}\rangle_t|$.
    \emph{Right panel:} Mass density $\rho$ 
    with streamlines of
    poloiodal velocity field $(v_r, v_\theta)$, weighted 
    by density $\langle\rho \mathbf{v}_{\rm pol}\rangle_t/\langle\rho\rangle_t$.
    The colorbar shows the ratio of poloidal speed 
    $|v_{\rm pol}|$
    to local escape speed
    $v_{\rm esc} \equiv \sqrt{2GM_\bullet/r}$.
    Black, dashed line marks the torus-funnel
    interface as in Fig.~\ref{fig:fiducial_early}. 
    A dense $\rho \sim 10^{-6}\,\rm g\, {\rm cm}^{-3}$,
    hot $T \gtrsim 10^6\,K$ ring of gas 
    is visible at $r \sim R_c$.
    Radiation is trapped in convective eddies within the envelope, and free to escape radially in the optically thin funnel.
    Gas flowing radially 
    inwards along intermediate latitudes either joins 
    the ring at $r= R_c$, or convective eddies close to the mid-plane,
    or gets unbound and ejected along the poles by the radiation force.
    (see also Fig.~\ref{fig:fiducial_steady}, panel (a)).
    }
    \label{fig:fiducial_snapshot}
\end{figure}

Fig.~\ref{fig:fiducial_snapshot} depicts the mass density (right panel) and radiation energy density (left panel) in the steady region $r \leq 10\,R_c$. 
The gas flow can be divided into flows in three distinct zones based on the streamlines shown on the right panel: (1) a convective mid-plane (vortices); (2) a hot ($T \gtrsim 10^6\,K)$ and dense 
($\rho \gtrsim 10^{-6}\,\rm g \rm\, cm^{-3}$) centrifugally-supported ring at $r \approx R_c$; and (3) a radial outflow along polar latitudes. Radiation is trapped in convective eddies close to the mid-plane, while it escapes radially outwards along the polar directions at (locally) super-Eddington rates, driving an outflow of gas accelerated to nearly the local escape speed, $v_{\rm esc} = \left(2GM_\bullet/r\right)^{1/2}$.

\begin{figure*}
    \centering
    \includegraphics[width= 0.9\textwidth]{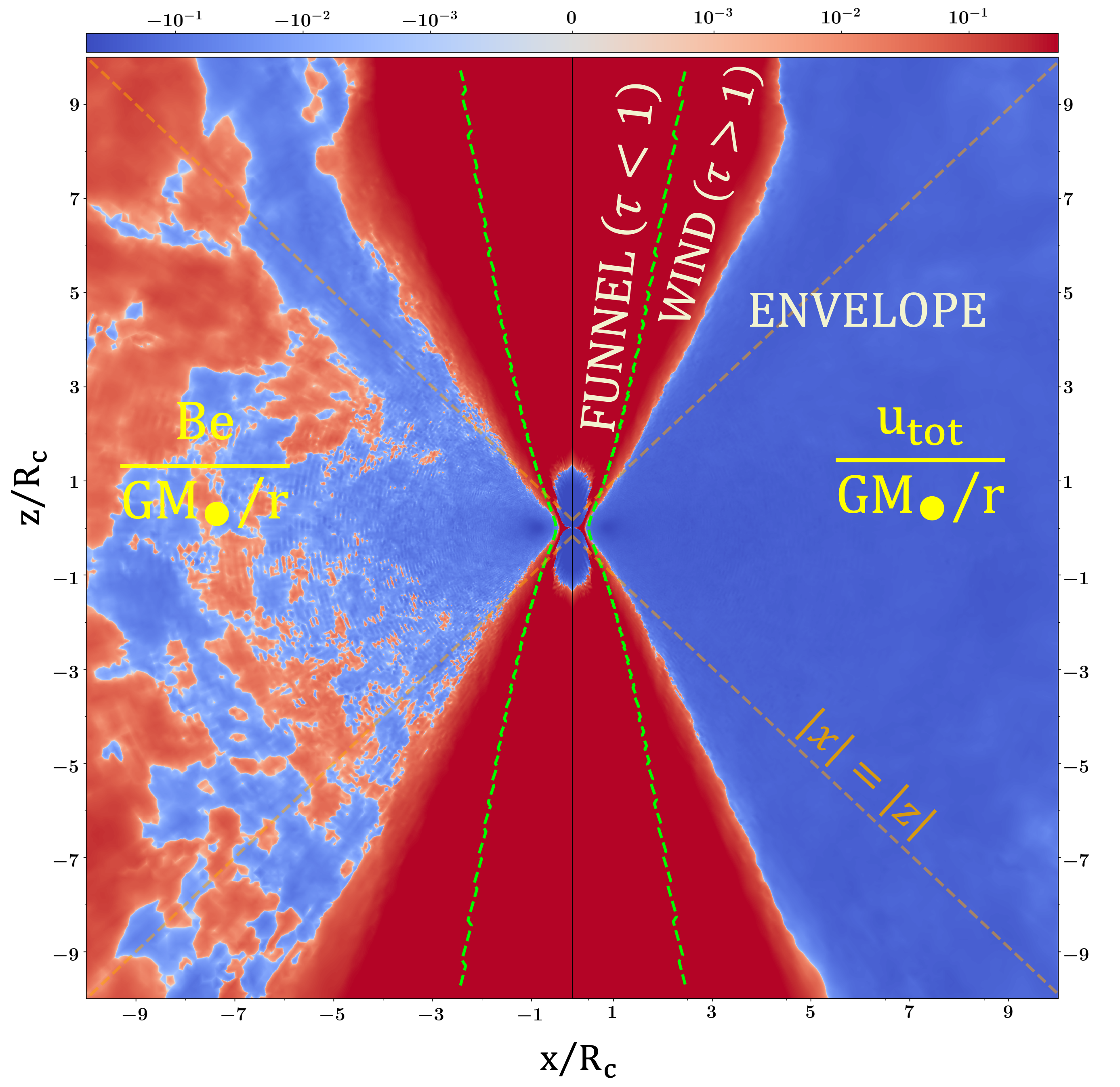}
    \caption{Averages of Bernoulli parameter $\langle\rho\,\rm{Be}\rangle_t/\langle\rho\rangle_t$ (left panel; Eq.~\eqref{eq:bernoulli_def}) and specific energy 
    $\langle\rho\,\rm{u_{\rm tot}}\rangle_t/\langle\rho\rangle_t$
    (right panel; Eq.~\eqref{eq:u_tot_def}) over  
    the same time interval as in Fig.~\ref{fig:fiducial_steady}, both scaled to the local gravitational potential energy  $GM_\bullet/r$. As illustrated most clearly on the right panel, the flow naturally divides into three distinct zones: a bound \emph{envelope} (blue region), an optically thick \emph{wind}, and an optically thin polar \emph{funnel} (both unbound, red regions). Green dashed lines mark the torus/funnel boundary of the initial equilibrium torus.
    Brown dashed lines show the $|x| = |z|$ lines, demonstrating that
    the envelope geometry is spherical $|z| \geq |x|$ beyond $r \approx 1.5\,R_c$. While the envelope is bound, in accordance with the virial theorem for a quasi-hydrostatic configuration, its Bernoulli parameter fluctuates around zero $|\rm{Be}| \lesssim 0.05\,GM_\bullet/r$ (see also Fig.~\ref{fig:wind_envelope_energy}).}
    \label{fig:energy_bernoulli}
\end{figure*}

Fig.~\ref{fig:energy_bernoulli} reveals a crucial aspect of the steady state system described above ($r \lesssim 10\,R_c$): The flow can be cleanly divided into separate bound ($\langle u_{\rm tot} \rangle_t < 0$; Eq.~\eqref{eq:u_tot_def}) 
and unbound ($\langle u_{\rm tot}\rangle_t > 0$)
components (blue and red regions in the right panel, respectively). In the bound portion of the envelope, the total specific energy scales roughly uniformly with the gravitational potential $\langle u_{\rm tot}\rangle_t \approx -0.2\,GM_\bullet/R_c$. 
By contrast, the Bernoulli parameter (Eq.~\eqref{eq:bernoulli_def}; left panel of Fig.~\ref{fig:energy_bernoulli}) does not exhibit a robust sign, instead fluctuating around a small value $|\rm Be| \lesssim 0.05\,GM_\bullet/r$.
Both the Bernoulli parameter and total specific energy are positive in the unbound component, taking nonuniform values $\gtrsim 0.5\,GM_\bullet/r$. 
The unbound component can be further divided into
an optically thick \emph{wind} ($\kappa\rho r \gg 1$) and an optically thin \emph{funnel} ($\kappa \rho r \ll 1$). In summary, the flow consists of a quasi-hydrostatic, bound envelope, an optically thick (unbound) wind, and an optically
thin funnel, as we further demonstrate below.

\begin{figure}
    \centering
    \includegraphics[width= 0.48\textwidth]{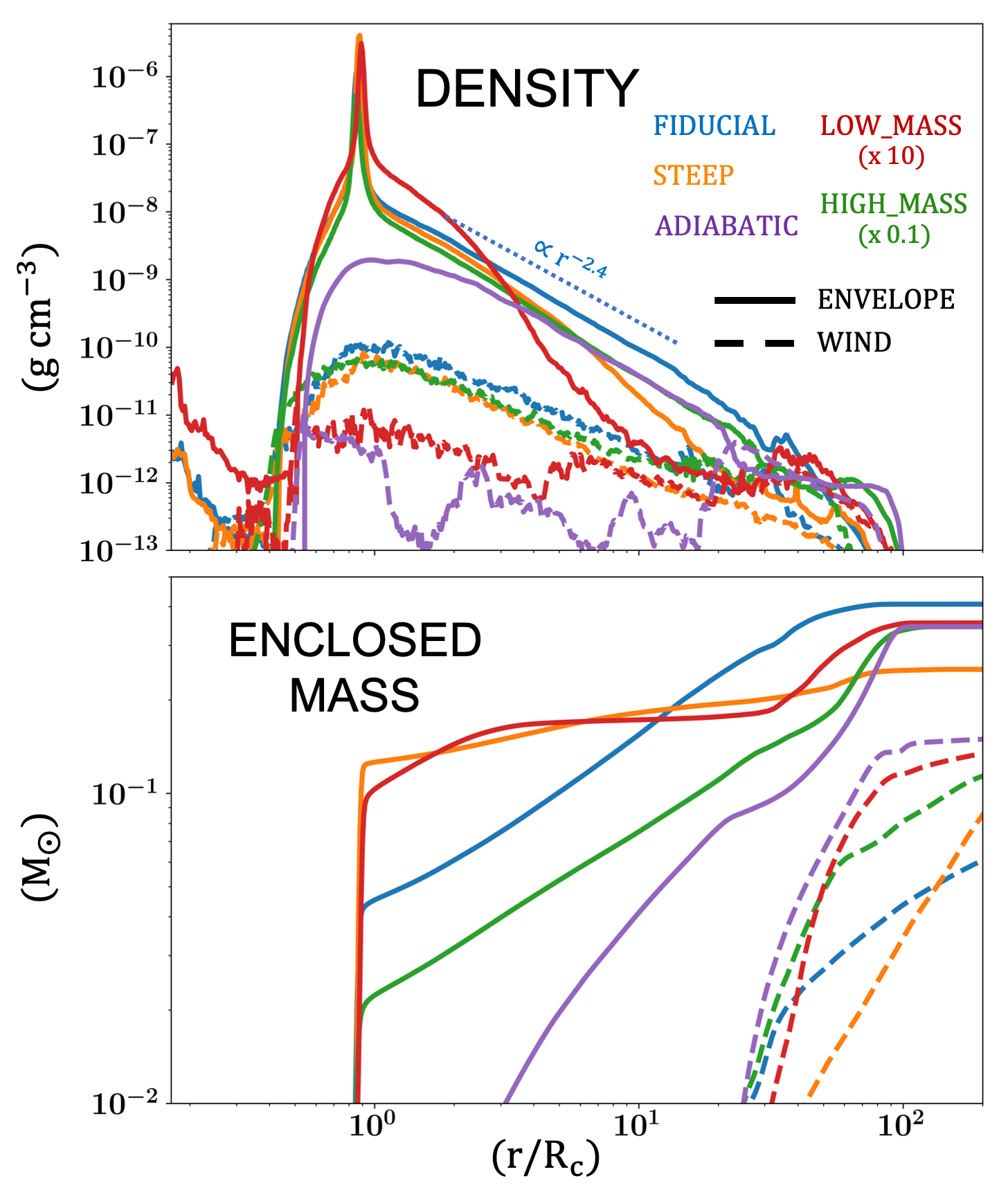}
    \caption{\emph{Top panel}: Steady state 
    average radial profiles of density
    $\langle\rho\rangle_{t, \theta}$ separated into contributions from the envelope (solid lines)
    and the wind (dashed lines), corresponding, respectively, to the blue and red regions in Fig.~\ref{fig:energy_bernoulli}.
    \emph{Bottom panel:} Enclosed mass $4\pi \int_0^rdr^\prime\,{r^\prime}^2\langle\rho\rangle_{t, \theta}$. 
    The profiles of models
    \texttt{HIGH$\_$MASS} and \texttt{LOW$\_$MASS}
    are scaled by 0.1 and 10, respectively (Table.~\ref{tab:sim_suite}). Time averages for 
    \texttt{FIDUCIAL}, \texttt{HIGH$\_$MASS} and
    \texttt{ADIABATIC} are taken in the same interval 
    as in Fig.~\ref{fig:fiducial_steady} where they 
    found to be in steady state. 
    \texttt{LOW$\_$MASS} is averaged over 
    $t\Omega_c \in [580, 840]$, where it finds a
    quasi-steady state similar to that achieved for model \texttt{FIDUCIAL} as shown in Fig.~\ref{fig:fiducial_steady}. All models evolve towards steeper density profiles. Radiative cooling allows formation of a dense ring or disk, at $r = R_c$, as seen as a spike in the density for those models which include radiative transport, but which is absent in \texttt{ADIABATIC}.
    }
    \label{fig:wind_envelope_density}
\end{figure}

Fig.~\ref{fig:wind_envelope_density} shows time- and angle-averaged radial profiles of the mass density $\langle\rho\rangle_{t, \theta}$ (top panel) and enclosed mass $4\pi \int_0^rdr^\prime\,{r^\prime}^2\langle\rho\rangle_{t, \theta}$ 
(bottom panel) throughout the envelope 
(blue regions in Fig.~\ref{fig:energy_bernoulli}), 
wind and funnel (red regions). The envelope density peaks sharply at $r = R_c$ at a value 
$\rho \sim 10^{-6}\,{\rm g}\,{\rm cm}^{-3}$. Across radii $2\,R_c \lesssim r \lesssim 20\,R_c$, the density profile is well fit by a power-law $\rho \propto r^{-2.4}$. The wind is roughly two orders of magnitude less dense than the envelope, but still optically thick to radiation, with $\tau \approx \kappa\rho r \gtrsim 100$.  More than $80\%$ ($0.4\,M_\odot$) of the envelope mass is bound (bottom panel), while $\approx 8\%$ ($0.04\,M_\odot$) is in a narrow ring, or disk, at $r = R_c$, and $\gtrsim 20\%$ is in the envelope $r \lesssim 10\,R_c$.

\begin{figure*}
    \centering
    \includegraphics[width= \textwidth]{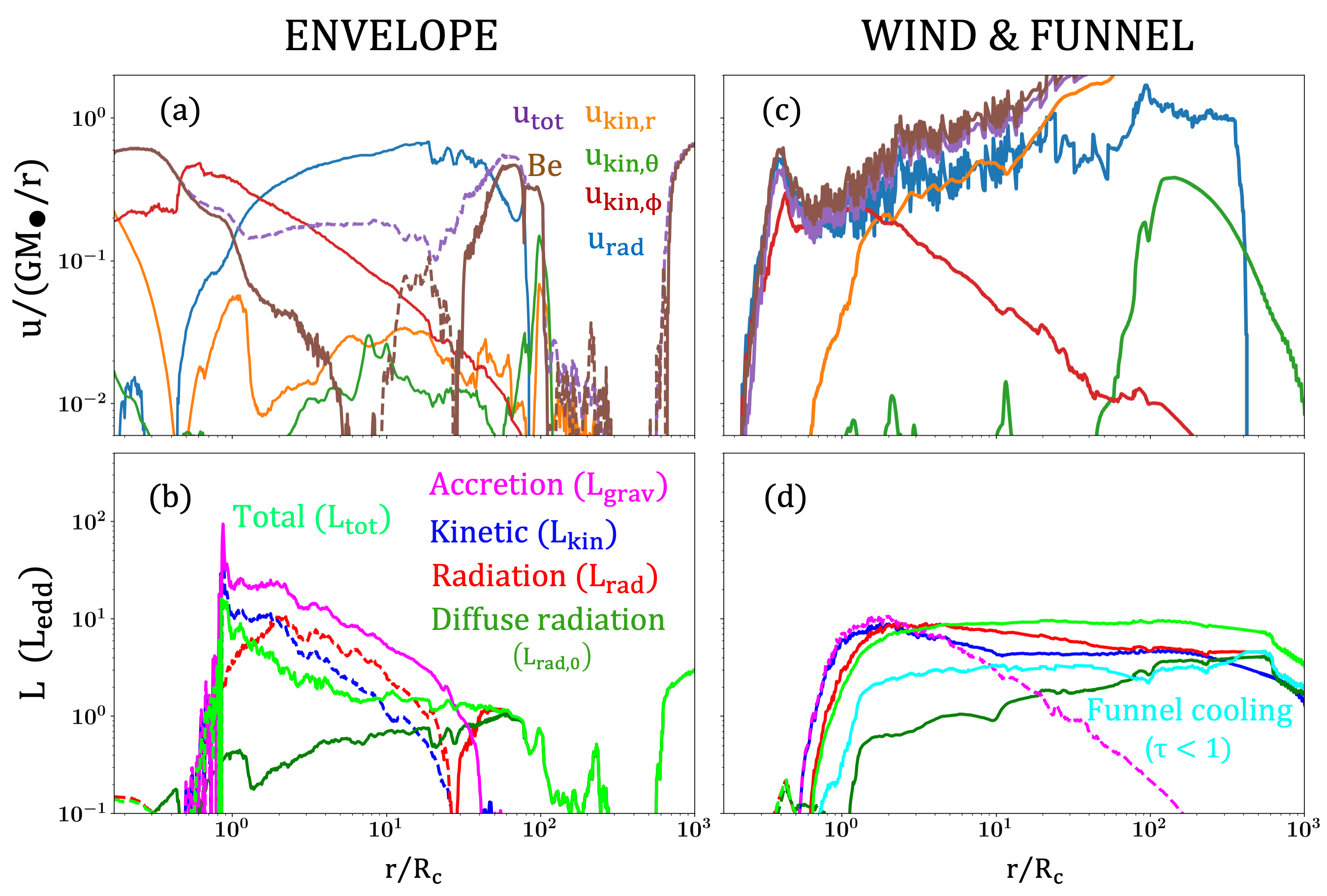}
    \caption{Radial profiles of specific energy density 
    (top panels) and flux (bottom panels)
    averaged over the same time interval as in 
    Fig.~\ref{fig:fiducial_steady}, separated into 
    \emph{envelope} (blue regions in the
    right panel of Fig.~\ref{fig:energy_bernoulli}),
    the \emph{wind} and the \emph{funnel} (red regions). 
    In panels (a), (c), values of specific energies are scaled 
    to gravitational potential $GM_\bullet/r$ and computed as
    $\langle\langle\rho\,u\rangle_t/\langle\rho\rangle_t\rangle_\theta$
    where $u$ stands for any component of the specific energy.
    In panels (b), (d), we plot the same quantities as in panel (d) of Fig.~\ref{fig:fiducial_steady}. In panel (d), we separately
    plot the radial flux of radiation leaving the funnel (teal line) 
    by restricting the shell average to regions $\kappa\rho r < 1$.
    The quasi-steady envelope ($r \lesssim 20\,R_c$) cools through the optically thin funnel at a rate
    $L \approx 3\,L_{\rm edd}$ at $r \approx 10\,R_c$, and the
    optically thick wind at a rate $L \approx 4-5\,L_{\rm edd}$.
    }
    \label{fig:wind_envelope_energy}
\end{figure*}

Fig.~\ref{fig:wind_envelope_energy} illustrates the 
distinct character of the energy distribution and transport rate through the envelope and the wind, respectively. Within the envelope (panel (a), $r \lesssim 10\,R_c$), 
rotational kinetic $u_{\rm kin, \phi} = v_\phi^2/2$
(red line) and radiation energy (red curve) dominate over the poloidal kinetic energies $u_{\rm kin, r} = v_r^2/2$, $u_{\rm kin, \theta}= v_\theta^2/2$
(orange and green lines).  Near the inner edge of the envelope ($r \sim R_c$), the specific rotational energy 
$u_{\rm kin, \phi}$ is close to its Keplerian value 
$\approx 0.4\,GM_\bullet/R_c$. The radiation energy increases with radius, such that the total energy density is proportional to gravitational potential $u_{\rm tot} \approx -0.2\,GM_\bullet/r$. The Bernoulli parameter takes values ${\rm Be} \lesssim 0.05\,GM_\bullet/R_{c}$ at radii $2 R_{\rm c}\lesssim r \lesssim 20R_{\rm c}$, akin to that expected according to the virial theorem for a radiation-dominated star. By contrast, the energy of the wind is dominated by radiation and radial kinetic energy. The growth in the wind kinetic energy from $\approx 0.2\,GM_\bullet/r$ at $r \approx 2\,R_c$ to $\gtrsim GM/r$ by $r \approx 30\,R_c$ is a consequence of the radiation force accelerating the gas as it expands outwards, converting thermal energy into bulk kinetic motion (panel (c)).

Panels (b) and (d) of Fig.~\ref{fig:wind_envelope_energy}
demonstrate how the steady, outwards energy 
flux $L_{\rm tot} \approx 10\,L_{\rm edd}$ found in Fig.~\ref{fig:fiducial_steady} is maintained by the
inflowing envelope, outflowing wind, and the funnel: 
Mass flows radially inwards through the envelope, generating accretion power at rates $L_{\rm grav} \approx 2 - 30\,L_{\rm edd}$ (magenta line), advecting 
radiation and kinetic energy (blue and red dashed lines)
towards $r \approx R_c$, such that a net 
energy flux $\approx 2\, - 10\,L_{\rm edd}$ (bright green line) is generated locally throughout the envelope. The wind and the funnel \emph{cool} 
the envelope by carrying kinetic/radiation energy away from the system. At envelope outer radius $r \approx 20\,R_c$ (boundary of the steady-zone), the radiation 
luminosity in the funnel and the wind are $\approx 3\,L_{\rm edd}$ (panel (d), teal line) and $4\,L_{\rm edd}$ (difference of red line from teal),
respectively. The remaining $\approx 4\,L_{\rm edd}$ is carried in the form of kinetic energy (blue line). Radiation luminosity in the fluid rest-frame (green line, panel (b)) grows close to Eddington value towards $r \approx 10\,R_c$, balancing a significant fraction of the gravitational force and centrifugal force across the envelope (see also Fig.~\ref{fig:fiducial_steady}).

\begin{figure}
    \centering
    \includegraphics[width= 0.48\textwidth]{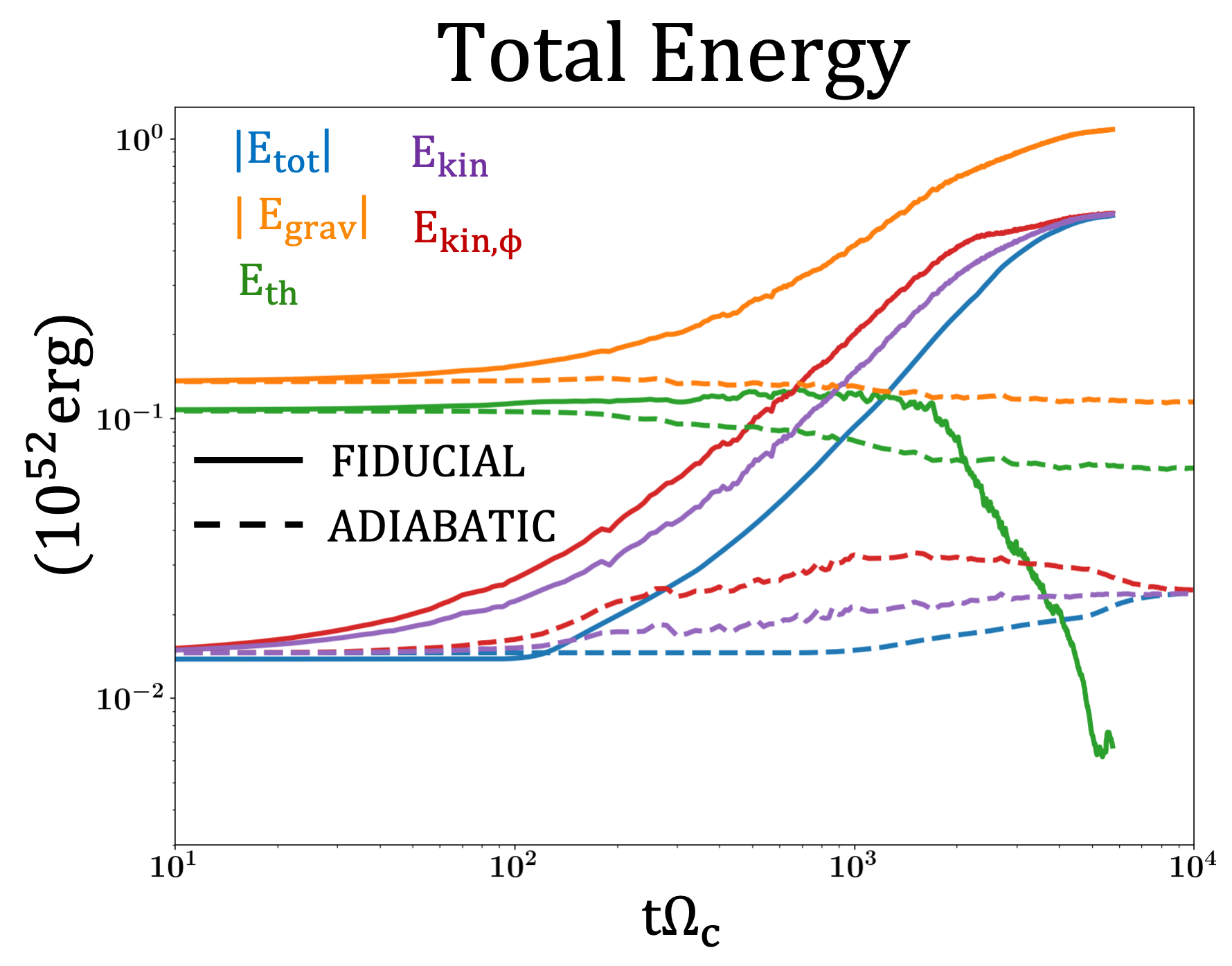}
    \caption{Evolution of total energy on the grid
    enclosed within $r = 600\,R_c$, in units of that required to circularize the initial envelope mass, $M_{\rm env}\epsilon_c \approx 10^{52}\,\rm erg$ (Eq.~\eqref{eq:epsilon_c}). For models \texttt{FIDUCIAL} (solid lines) and \texttt{ADIABATIC} (dashed lines), quantities shown include: total energy (absolute value; blue curve); total gravitational 
    potential energy (absolute value; orange); total radiation/thermal energy (green); kinetic and rotational energies (red, purple). In \texttt{ADIABATIC}, thermal energy (green dashed line) is used to expand the envelope (and unbind a fraction $\approx 45\%$, Fig.~\ref{fig:total_mass_evolution}), leading to a mass distribution marginally stable to convection.  In contrast, cooling enables the release of significant gravitational potential energy, which is eventually lost from the system in the form of radiation and outflow kinetic energy, at comparable rates (Fig.~\ref{fig:total_luminosity_comparison}).}
    \label{fig:total_energy_evolution}
\end{figure}
\begin{figure}
    \centering
    \includegraphics[width= 0.48\textwidth]{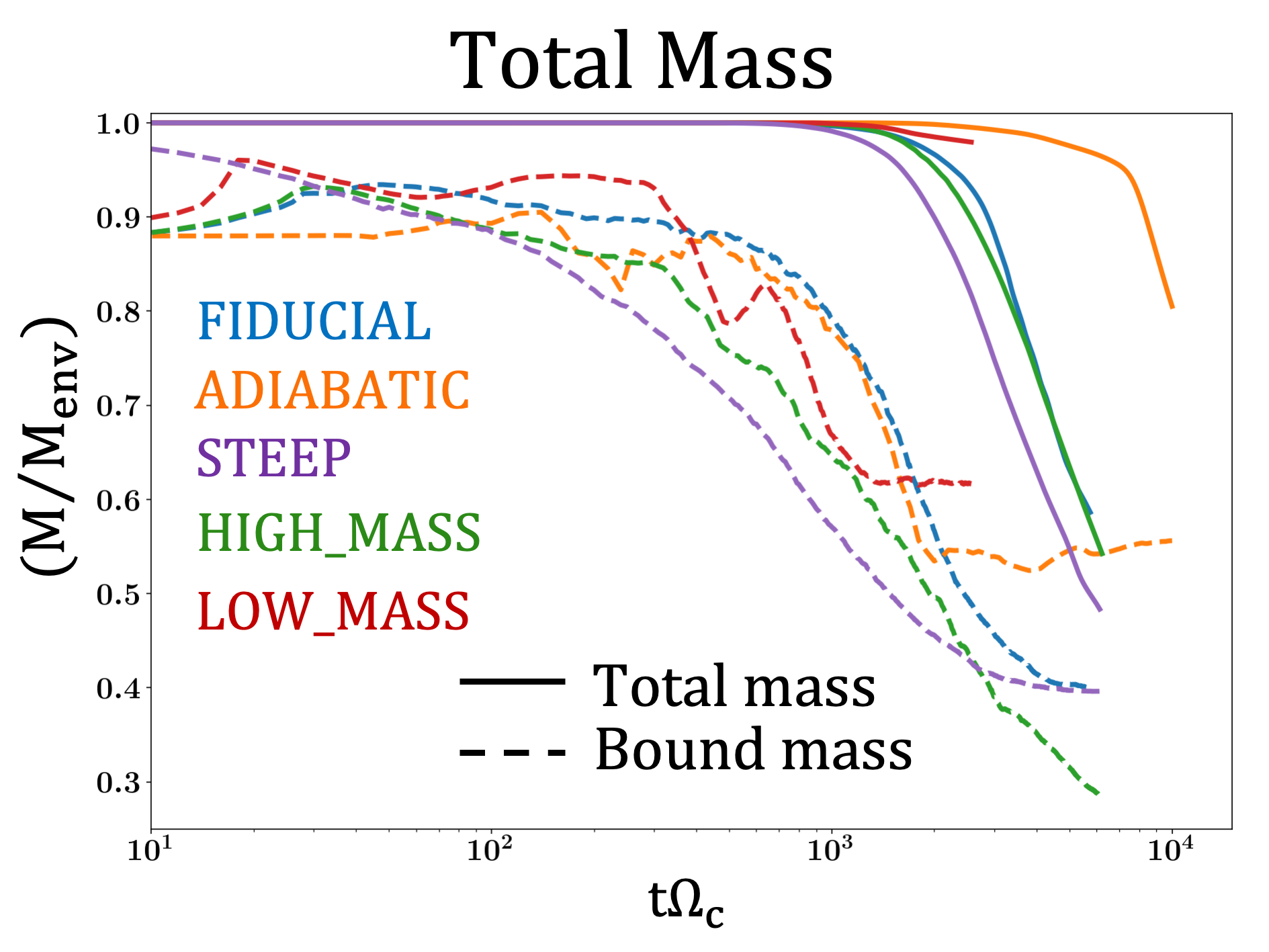}
    \caption{Total mass
    (solid lines), and total bound mass (Eq.~\eqref{eq:rho_bnd}) (dashed lines)
    enclosed within $r = 600\,R_c$, both scaled to initial mass $M_{\rm env}$. Cooling allows more mass to sink in the potential well of the SMBH, at locally super-Eddington rates (Sec.~\ref{sec:fiducial}), leading to more unbound mass compared to \texttt{ADIABATIC}.  The fraction of bound mass is greater for larger initial envelope mass and optical depth. Despite larger initial
    binding energy, a steeper density profile leads to 
    similar amount of unbound mass (purple lines).}
    \label{fig:total_mass_evolution}
\end{figure}

 Fig.~\ref{fig:total_energy_evolution} shows the time evolution of the total energy enclosed within $r = 600\,R_c$, as well as its breakdown into gravitational potential energy, radiation energy, kinetic energy, and 
rotational kinetic energy, all scaled to the energy $M_{\rm env}\epsilon_c \approx 10^{52}\,\rm erg$ required to form a thin disk at $R_{\rm c}$ 
from the initial envelope (Eq.~\eqref{eq:epsilon_c}). In Fig.~\ref{fig:total_mass_evolution}, we show the total mass (solid lines) and bound mass (dashed lines) in the same region.
Although our computational grid is larger ($r = 10^4\,R_c$, Table ~\ref{tab:fiducial_param_table}), we find that
gas leaving this region typically does not return,
and remains unbound at all times in all our models. Mass leaving
the computational domain is found to be negligible $\lesssim 10^{-4}\,M_{\rm env}$, due to (i) centrifugal force preventing accretion through the inner boundary absent 
explicit viscosity (ii) and outer boundary being at a large radius $r = 10^4\,R_c$. By the end of \texttt{FIDUCIAL}
$t\Omega_c \approx 6\,\times\,10^3$,
roughly $\approx 40\%$ of the initial envelope mass
($\approx 0.2\,M_\odot$) is bound 
(dashed blue line, Fig.~\ref{fig:total_energy_evolution}), 
of which the majority 
($\approx 0.18\,M_\odot$) resides in a narrow, dense, centrifugally-supported ring. As also evident from the density snapshots in Fig.~\ref{fig:fiducial_early},
there is negligible mass within $r \lesssim 0.5\,R_c$. As the envelope cools and contracts, total gravitational energy and rotational kinetic energy increase, consistent with angular momentum 
conservation. By the end of the simulation,
$E_{\rm grav} \approx -0.4\,M_{\rm env}\epsilon_c$
and $E_{\rm kin, \phi} \approx 0.2\,M_{\rm env}\epsilon_c$, representing the energy budget of the ring of mass $M \approx 0.4\,M_{\rm env}$.

\begin{figure}
    \centering
    \includegraphics[width= 0.48\textwidth]{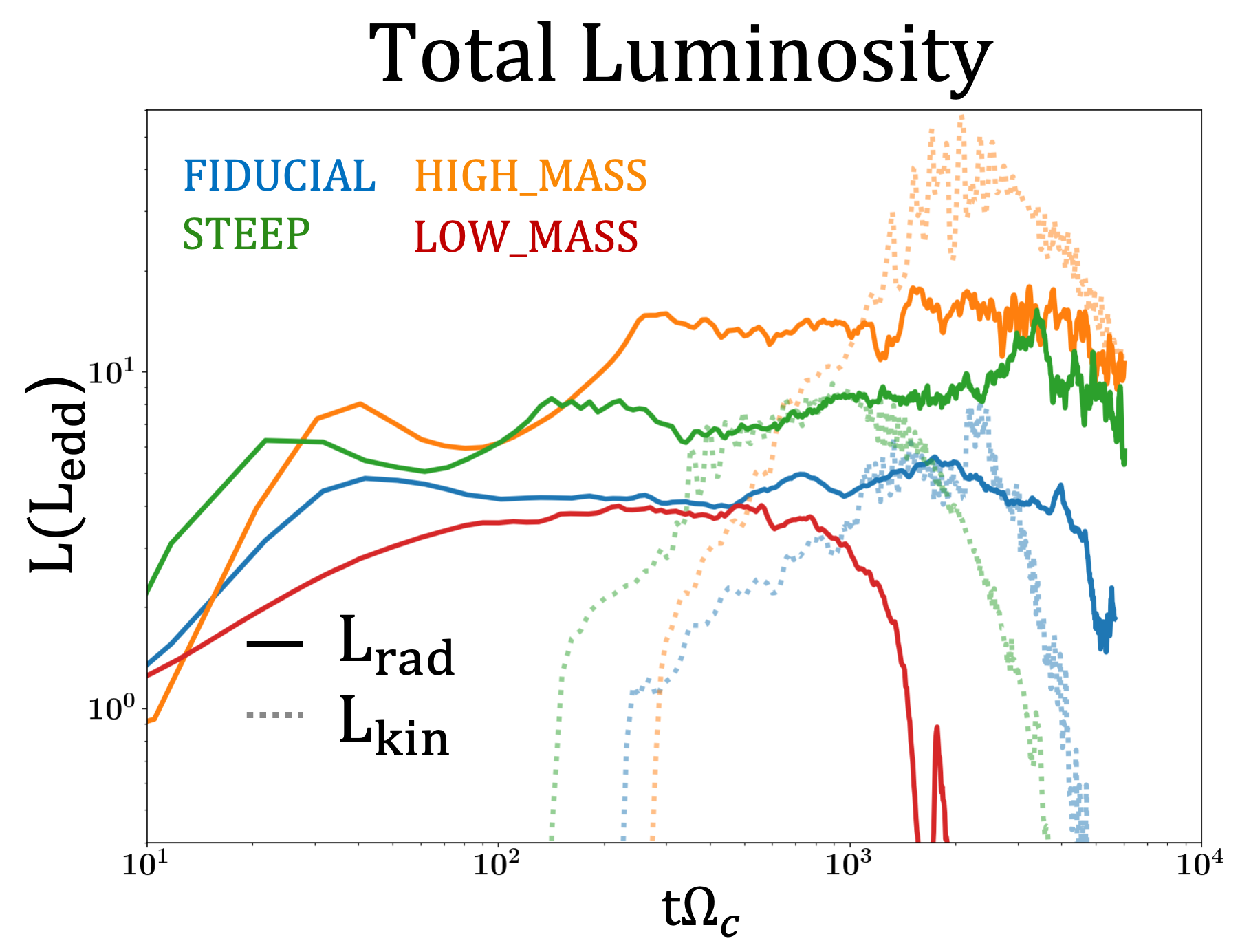}
    \caption{Time evolution of the net energy loss from the envelope/funnel structure for models \texttt{FIDUCIAL} (blue),
    \texttt{HIGH$\_$MASS} (orange),
    \texttt{LOW$\_$MASS} (red),
    \texttt{STEEP} (green), as 
    measured at radii
    $3\,\times\,10^{15}\,\rm cm$,
    $7\,\times\,10^{15}\,\rm cm$,
    $3\,\times\,10^{15}\,\rm cm$,
    $2\,\times\,10^{15}\,\rm cm$, respectively.
    These radii were chosen so as to enclose the photosphere at all times. The \emph{photon luminosity} (radiation flux) are shown with solid lines, while the \emph{kinetic luminosity} of outflows (Eq.~\eqref{eq:iso_fluxes}-\eqref{eq:shell_fluxes}) are shown with faint, dotted lines.  Compared to the \texttt{FIDUCIAL} model, the peak photon luminosity increases slowly with increasing envelope mass (\texttt{HIGH$\_$MASS}) compared to kinetic luminosity.  By contrast, the lower mass envelope in (\texttt{LOW$\_$MASS}) cools efficiently by photon diffusion, with less mass lost to outflows.
    }
    \label{fig:total_luminosity_comparison}
\end{figure}

Fig.~\ref{fig:total_luminosity_comparison} shows the energy flux as a function through a spherical surface of radius $r \approx 10^3\,R_c \approx 3\,\times\,10^{15} \rm cm$, so chosen because it is located beyond the photosphere at all times. The net radiation energy flux $L_{\rm rad}$ (\emph{photon luminosity}) and the flux of kinetic energy (\emph{kinetic luminosity}) are shown with solid and dotted (blue) lines, respectively (see Eq.~\eqref{eq:shell_fluxes} for definitions). The photon luminosity is approximately constant at $L_{\rm rad} \approx 4-6\,L_{\rm edd}$ across the time interval $t\Omega_c \in [100, 5\times\,10^3]$, while the kinetic luminosity peaks at $L_{\rm kin} \approx 8\,L_{\rm edd}$ by $t\Omega_c \approx 2\,\times\,10^3$. 
Following the approximate steady-state phase described above (Fig. ~\ref{fig:fiducial_steady}), the total net energy output of the envelope $L_{\rm tot} \approx 10\,L_{\rm edd}$ across the time interval $t\Omega_c \approx 10^3 - 3\,\times 10^3$, comprised of roughly equal parts kinetic energy radiative luminosity.
Eventually, the envelope begins to fall after about one fallback time $t\Omega_c \approx 5\,\times\,10^3$, with the photon luminosity dropping down to $L_{\rm rad} \approx L_{\rm edd}$.

\subsection{Comparison to \texttt{ADIABATIC} }
\label{sec:ADIABATIC}

\begin{figure}
    \centering
    \includegraphics[width= 0.48\textwidth]{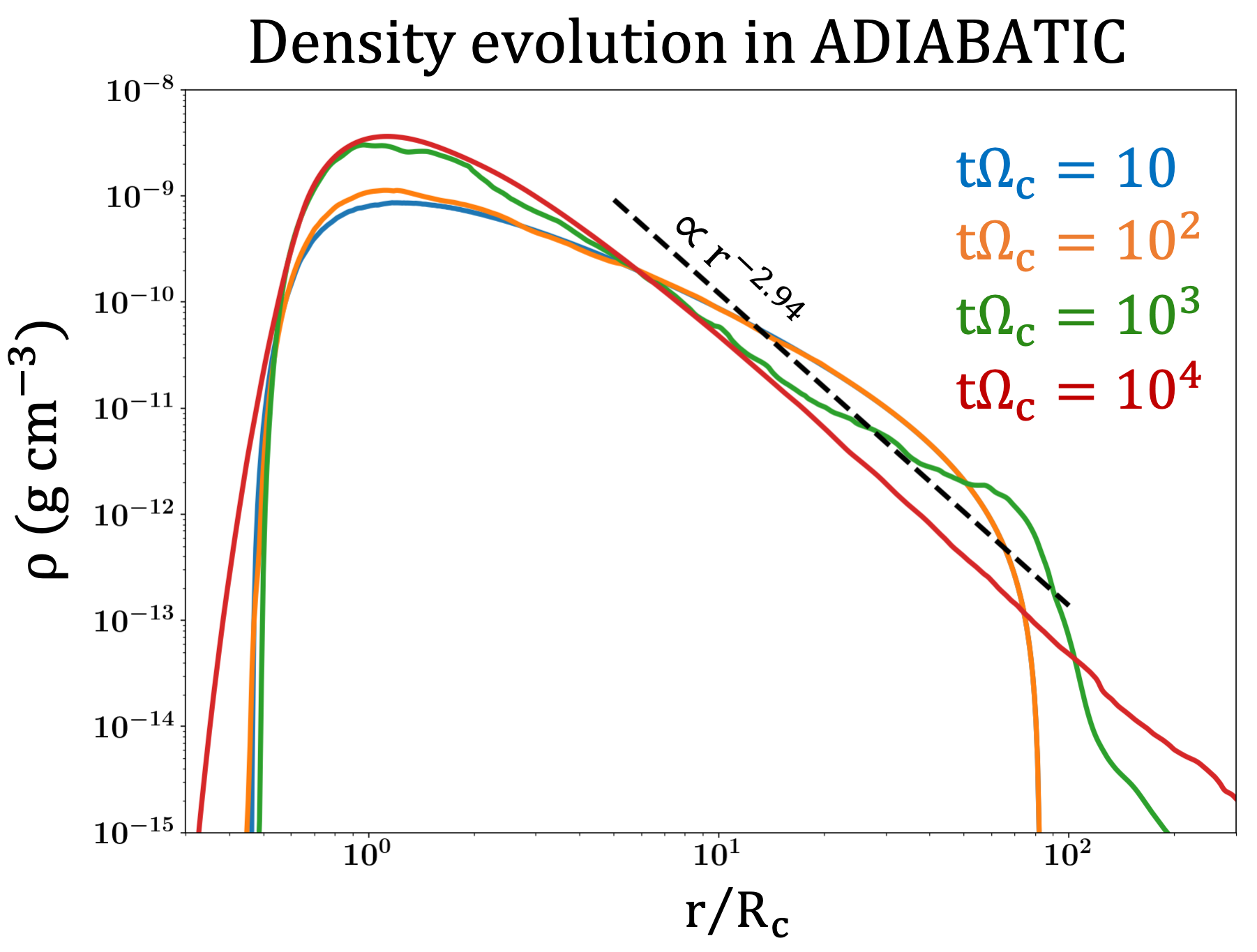}
    \caption{Snapshots of angle-averaged mass density in \texttt{ADIABATIC}.
    The initial density profile $r \propto r^{-1.6}$ (Eq.~\eqref{eq:init_rho_press}, Table~\ref{tab:sim_suite}) gradually evolves to the 
    distribution $\rho \propto r^{-3}$ marginally stable to convection (Sec.~\ref{sec:analytical_estimates}), loosing $\approx 45\%$ of
    the initial
    envelope mass to winds 
    (see Fig.~\ref{fig:total_mass_evolution}).}
    \label{fig:adiabatic_density}
\end{figure}

Given the important role that outflows appear to play in the cooling evolution of the envelope in the \texttt{FIDUCIAL} model, it is natural to ask how much of this evolution could be captured by a purely adiabatic evolution, i.e., neglecting the effects of radiative cooling. To address this, we ran \texttt{ADIABATIC} with otherwise identical 
initial conditions as in \texttt{FIDUCIAL} 
(see Tables ~\ref{tab:fiducial_param_table}, \ref{tab:sim_suite}), but instead assuming an adiabatic equation of state with $\gamma = 4/3$, appropriate
to radiation dominated conditions 
(Sec.~\ref{sec:analytical_estimates}), and with radiation transport artificially turned off.

\begin{figure*}
    \centering
    \includegraphics[width= \textwidth]{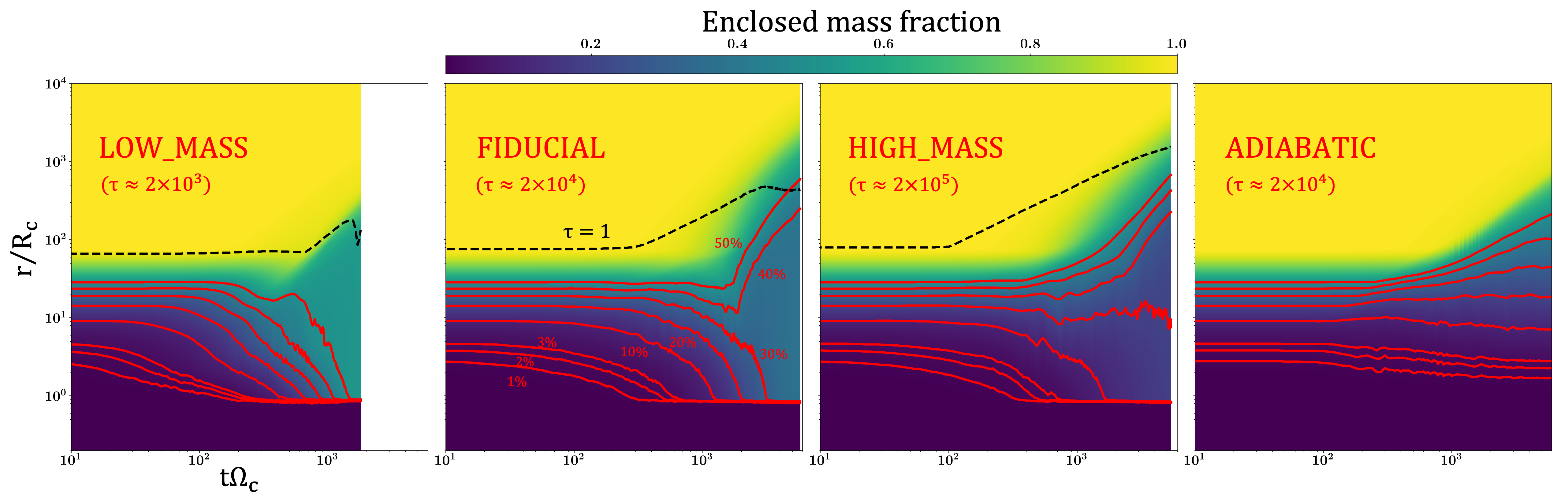}
    \caption{Radius-time diagrams of enclosed mass $4\pi\,\int_0^r\,dr^\prime{r^\prime}^2\,\langle\rho\rangle_\theta(t, r^\prime)$
    as a fraction of initial total mass 
    ($0.05\,M_\odot$,
    $0.5\,M_\odot$
    $5\,M_\odot$,
    $0.5\,M_\odot$ respectively, see Table.~\ref{tab:sim_suite})
    for models \texttt{LOW$\_$MASS},
    \texttt{FIDUCIAL},
    \texttt{HIGH$\_$MASS}.
    \texttt{ADIABATIC}. These models differ only
    in their density normalization (hence optical depth) at $t = 0$,
    other than radiation transport being turned off in \texttt{ADIABATIC}.
    Red curves mark the evolution of radii 
    that enclose chosen fractions ($\text{a few}\% - 50\%$) 
    of the total mass.
    Black dashed lines mark the average photosphere
    radii $\tau = 1$ (Eq.~\eqref{eq:tau_av}). All models
    evolve towards steeper density profiles by redistributing mass (see Fig.~\ref{fig:wind_envelope_density}). 
    \texttt{ADIABATIC} does so by moving more mass ($\gtrsim 80\%$) 
    to larger radii than inner radii. Cooling allows accretion of $\gtrsim 10\%$ from radii $r \approx 9\,R_c$ to $r \approx 2\,R_c$ by 
    $t\Omega_c \approx 10^3$. 
    Since support against gravity 
    at $r \gtrsim R_c$ is initially dominated by radiation pressure (Fig.~\ref{fig:initial_conditions}), 
    an \emph{effective (local) cooling timescale} can be 
    identified from these diagrams as the timescale over which a shell
    of given mass is displaced radially inwards 
    by an amount comparable to its initial value
    (e.g., half the initial radius).
    Cooling times of mass shells enclosing $\lesssim 30\%$ are 
    shorter in \texttt{LOW$\_$MASS} by a factor $\approx 5$ compared to \texttt{FIDUCIAL}, suggesting that \texttt{LOW$\_$MASS} is in 
    the diffusion-controlled regime of cooling.
    Radii $r \lesssim 10\,R_c$ 
    cool on similar timescales in \texttt{FIDUCIAL} and \texttt{HIGH$\_$MASS}, 
    indicating that these radii cool via convection in these models. The $\%10$
    mass shell at $r \approx 9\,R_c$, however, barely moves to
    $r = 7\,R_c$ in \texttt{ADIABATIC} by $t\Omega_c = 10^4$,
    whereas it contracts to $r = R_c$ in \texttt{HIGH$\_$MASS} by $t\Omega_c \approx 10^3$, demonstrating that cooling evolution 
    at these radii is qualitatively different from the adiabatic
    evolution (see also Sec.~\ref{sec:steep}, and Fig.~\ref{fig:steep_mass}).}
    \label{fig:mass_radius_time}
\end{figure*}

The evolution of \texttt{ADIABATIC} bears qualitative similarities 
to \texttt{FIDUCIAL}, but with crucial differences.  As in \texttt{FIDUCIAL}, the torus becomes unstable by $t\Omega_c \approx 10$, as expected from the initial entropy profile (Fig.~\ref{fig:initial_conditions})
being unstable to convection (Sec.~\ref{sec:analytical_estimates}).
Although this instability drives turbulence,
the irregularity seen in the spatial structure of the flow is less pronounced compared to \texttt{FIDUCIAL}. 
A similar envelope-wind system also forms, but accompanied by lower net mass inflow rates, giving rise to an envelope with a ``puffed up'' structure. In particular, the ring found to quickly 
form near the centrifugal barrier $r \approx R_c$ in \texttt{FIDUCIAL} is not seen in \texttt{ADIABATIC} (Fig.~\ref{fig:wind_envelope_density}). As Fig.~\ref{fig:adiabatic_density} illustrates, the envelope evolves towards 
a marginally stable, $\rho \propto r^{-3}$ density distribution, 
as expected (Sec.~\ref{sec:analytical_estimates}).

Fig.~\ref{fig:mass_radius_time} provides radius-time diagrams of enclosed mass in all of our simulations, including \texttt{FIDUCIAL} and
\texttt{ADIABATIC}. In \texttt{ADIABATIC}, the radius which encloses $\lesssim 20\%$ of the mass moves inwards, while the outer mass shells all move outwards to larger radii monotonically. The envelope evolution in \texttt{ADIABATIC} is attempting to alleviate its unstable entropy gradient (Sec.~\ref{sec:analytical_estimates}) 
by moving more mass outwards ($\gtrsim 80\%$) than inwards ($\lesssim 20\%$), by launching outflows.  In particular, the mass enclosed within $r = 2\,R_c$ remains $\lesssim 1.6\%$ at all times in \texttt{ADIABATIC}. By contrast, in \texttt{FIDUCIAL} radiative cooling allowed $\approx 8\%$ of the initial envelope mass to 
settle into a centrifugally supported ring at $r = R_c$ 
by $t\Omega_c \gtrsim 10^3$ (Fig.~\ref{fig:wind_envelope_density}), increasing to $\gtrsim 30\%$ by 
$t\Omega_c \approx 6\,\times\,10^3$. Crucially, in \texttt{FIDUCIAL} the inflow occurs significantly faster than the local diffusion timescale of the bulk envelope, especially close to the inner radii of the envelope ($r \lesssim 6-7\,R_c$) enclosing $\approx \text{a few}\,\%$ of the envelope mass initially, as discussed in Sec.~\ref{sec:fiducial}. The resulting highly super-Eddington accretion power (Fig.~\ref{fig:wind_envelope_energy}), acts to destabilize the flow in a more pronounced way than the initial unstable entropy gradient alone does. As a result, the evolution of \texttt{FIDUCIAL} and \texttt{ADIABATIC} differ qualitatively on timescales shorter than the estimated photon diffusion timescale in the bulk envelope, as we shall illustrate further when varying the initial entropy profile and envelope mass (Sec.~\ref{sec:steep} and Sec.~\ref{sec:envelope_mass}, respectively).

This distinct evolution of the envelope mass distribution is also reflected in the energetics of the flow, as revealed in Fig.~\ref{fig:total_energy_evolution}. 
In \texttt{ADIABATIC}, thermal energy (green dashed line) is used to expand the envelope to larger radii, increasing kinetic energy (purple dashed line) and driving outflows.  Because cooling-induced inflows do not occur, the center of mass moves to a larger radius during this process, causing the total gravitational potential energy to diminish in magnitude with time (orange dashed lines). This contrasts with \texttt{FIDUCIAL}, for which $\approx 10^{52}\,\rm erg$ of gravitational potential energy was released by the inward migration of mass driven by cooling to form a ring near $R_{\rm c}$. 
In \texttt{ADIABATIC}, $\approx 55\%$ of the envelope mass remains bound as the envelope expands (Fig.~\ref{fig:total_mass_evolution}, orange dashed line), while in \texttt{FIDUCIAL} this fraction is $\approx 40\%$ (blue dashed line).  Thus, in spite of the naive expectation that cooling should favor more bound mass, it ironically leads to greater mass ejection by enabling faster accretion and energy release through polar outflows and radiation.

\subsection{Role of initial envelope mass}
\label{sec:envelope_mass}
Having established the role played by radiative cooling 
in driving the envelope evolution on timescales $\lesssim 10^3\,\Omega_c^{-1}$, here we explore how much the effective cooling rate of the envelope 
is related to radiative diffusion by means of an experiment in which we vary the initial mass of the torus (Eq.~\eqref{eq:init_rho_press}) keeping all other parameters fixed at the \texttt{FIDUCIAL} model values.  A more (less) massive envelope possesses a higher (lower) initial optical depth, and hence a longer (shorter) characteristic diffusion timescale.  In particular, we should expect the envelope evolution to eventually converge to the \texttt{ADIABATIC} case as the envelope mass is increased and diffusion gets
more inefficient.  With these expectations, we increased/lowered the torus mass by a factor of 10 in models \texttt{HIGH$\_$MASS}/\texttt{LOW$\_$MASS}, respectively, leading to envelopes of $5\,M_\odot$/$0.05\,M_\odot$.  We caution that these models are not necessarily representative of an actual TDE of a more or less massive star, because we did not self-consistently change the other model parameters 
(e.g., $R_c$, Table.~\ref{tab:fiducial_param_table})
within the TDE theory laid out in Sec.~\ref{sec:analytical_estimates}. Rather, these models serve as a numerical experiment to better understand the dependence of the envelope evolution on the envelope optical depth.

Fig.~\ref{fig:mass_radius_time} illustrates that the evolution of the radial mass distribution indeed comes to more closely resemble \texttt{ADIABATIC} as the envelope mass is increased from $0.05\,M_\odot$ (\texttt{LOW$\_$MASS}) to $5\,M_\odot$ (\texttt{HIGH$\_$MASS}): more mass shells move 
radially outwards than inwards with increasing envelope mass. In \texttt{LOW$\_$MASS}, shells migrate inwards faster than for \texttt{FIDUCIAL} by factors of $4-5$, indicating that cooling of
\texttt{LOW$\_$MASS} is in the diffusion-controlled
regime. We find that mass shells at $r \lesssim 10\,R_c$ fall on similar timescales in \texttt{FIDUCIAL}
and \texttt{HIGH$\_$MASS}. This suggests that the cooling evolution at these radii is dominated by adiabatic convection instead of diffusion in \texttt{FIDUCIAL}
and \texttt{HIGH$\_$MASS}. Even in \texttt{HIGH$\_$MASS},
however, an accretion ring has already formed by $t\Omega_c \approx 10^3$, significantly earlier than the estimated diffusion time for this mass ($t\Omega_c \gtrsim 10^4$; Fig.~\ref{fig:initial_conditions}), enclosing
$\approx 4\%$ of the initial envelope mass ($0.2\,M_\odot$), as evident from the density profile (Fig.~\ref{fig:wind_envelope_density}; green solid lines). This suggests that the cooling of optically thin surface layers significantly
enhances convection due to initial unstable entropy gradient in \texttt{ADIABATIC}.

\begin{figure}
    \centering
    \includegraphics[width= 0.48\textwidth]{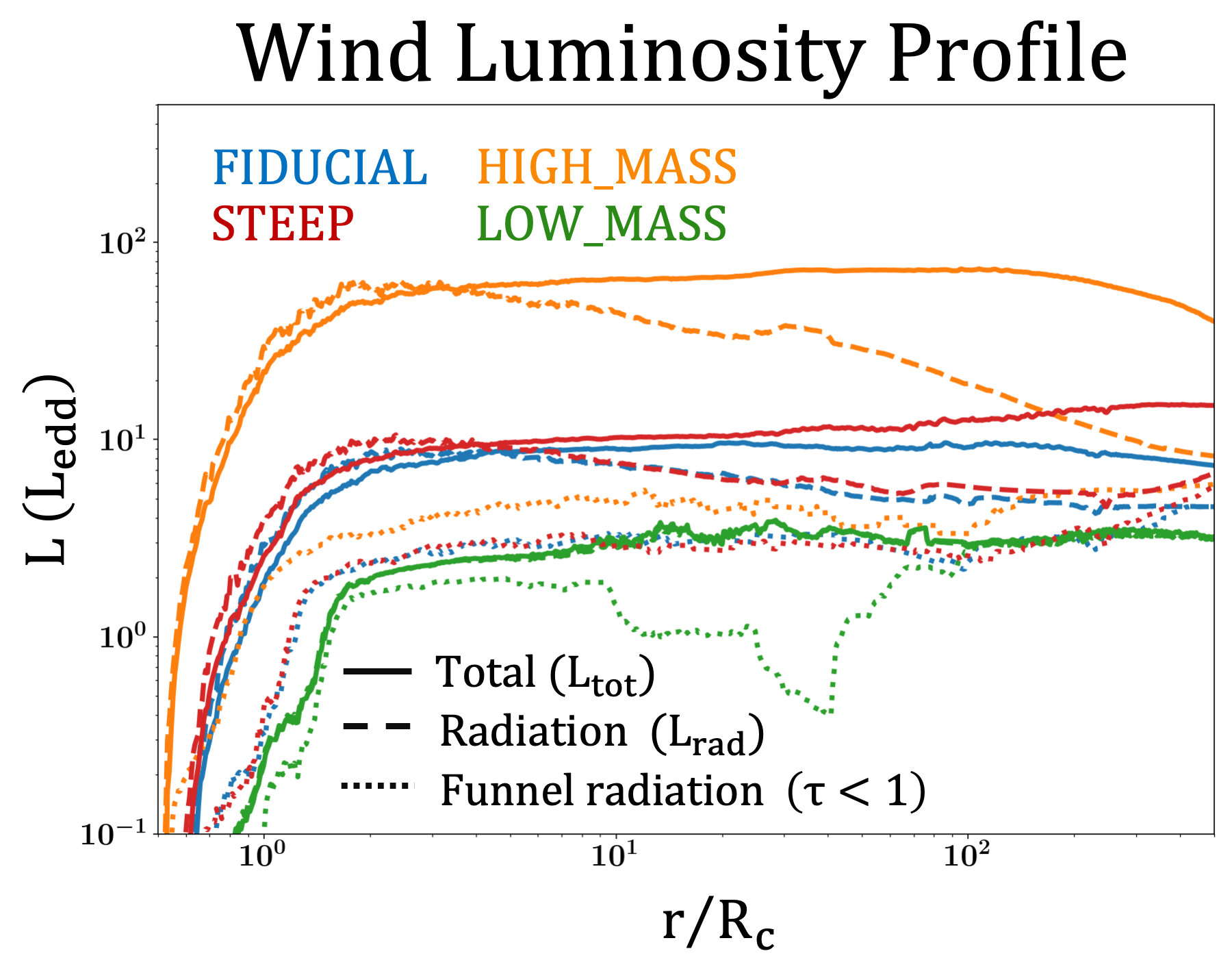}
    \caption{Comparison of steady-state wind profiles 
    (as in panel (d) of Fig.~\ref{fig:wind_envelope_energy}) for
    \texttt{FIDUCIAL} (blue),
    \texttt{HIGH$\_$MASS} (orange),
    \texttt{LOW$\_$MASS} (green)),
    \texttt{STEEP} (red),
    where time averages are taken in time intervals given in the
    caption of Fig.~\ref{fig:wind_envelope_density}.
    From total wind luminosity (solid lines),
    we infer that total energy output scales with envelope mass
    $\approx 4\,L_{\rm edd}$, 
    $\approx 10\,L_{\rm edd}$,
    $\approx 80\,L_{\rm edd}$. Total luminosity of the wind is set by
    accretion power at the envelope base 
    (Fig.~\ref{fig:wind_envelope_energy}), which mostly scales with
    the envelope mass, since effective cooling timescale of the envelope 
    only weakly depends on density at high enough optical depths (Sec.~\ref{sec:envelope_mass}, Fig.~\ref{fig:mass_radius_time}).
    }
    \label{fig:wind_cooling}
\end{figure}

Fig.~\ref{fig:wind_cooling} compares the radial wind profiles of \texttt{FIDUCIAL}, \texttt{LOW$\_$MASS}, and \texttt{HIGH$\_$MASS} in steady
state as in Panel (d) of Fig.~\ref{fig:wind_envelope_energy}. A good proxy of 
envelope accretion power is the total wind luminosity 
$L_{\rm tot}$ since it is found
to be roughly set by $L_{\rm grav}$ at $r = R_c$ (Fig.~\ref{fig:wind_envelope_energy}). 
We measure $L_{\rm tot} \approx 3-4\,L_{\rm edd}$ (green solid line), 
$L_{\rm tot} \approx 9-10\,L_{\rm edd}$ (blue solid line) and 
$L_{\rm tot} \approx 60-80\,L_{\rm edd}$ (orange solid line) 
for \texttt{LOW$\_$MASS},
\texttt{FIDUCIAL} and \texttt{HIGH$\_$MASS}. Since their cooling time is similar
(Fig.~\ref{fig:mass_radius_time}),
cooling rate roughly scales with mass as density is increased from \texttt{FIDUCIAL} to \texttt{HIGH$\_$MASS}: The wind luminosity of 10 times denser \texttt{HIGH$\_$MASS} is about $7-8$ times higher than the \texttt{FIDUCIAL}. The excess energy released by a more massive envelope goes, however, mainly to kinetic
energy of the outflows rather than radiation
in \texttt{HIGH$\_$MASS} (orange lines),
as can be noted from diminishing contribution of $L_{\rm rad}$ (dashed lines) in $L_{\rm tot}$ (solid lines) at radii $r > 10\,R_c$. Indeed, Fig.~\ref{fig:total_luminosity_comparison} shows that the photon luminosity achieved by \texttt{HIGH$\_$MASS} of $\approx 15\,L_{\rm edd}$ (orange solid line) is about three times as that of \texttt{FIDUCIAL} $4-5\,L_{\rm edd}$. By contrast, the kinetic luminosity (dotted lines) peaks at $\approx 60\,L_{\rm edd}$ towards the end of the simulation 
$t\Omega_c \approx 2\,\times\,10^3$, about an order of magnitude higher than the kinetic luminosity $\approx 5\,L_{\rm edd}$ found in \texttt{FIDUCIAL}. The fraction
of radiation flux leaving through the optically thin funnel (Fig.~\ref{fig:wind_cooling}, dotted lines)
is measured to be $L_{\rm funnel} \approx 1 - 2\,L_{\rm edd}$, $L_{\rm funnel} \approx 2 - 3\,L_{\rm edd}$,
$L_{\rm funnel} \approx 3 -  4\,L_{\rm edd}$
in \texttt{LOW$\_$MASS},
\texttt{FIDUCIAL} and \texttt{HIGH$\_$MASS}, respectively. The need for outflows is clear: Cooling through the optically-thin funnel is limited to the rate $\lesssim \rm few \,\, L_{\rm edd}$ as set by photon diffusion, in which case the excess energy must be carried away in kinetic form by winds.

By $t\Omega_c \approx 6\,\times\,10^3$, model \texttt{HIGH$\_$MASS} is found to have unbound a larger fraction of its mass $\approx 70\%$ compared to
\texttt{FIDUCIAL} $\approx 60\%$ 
(purple and blue lines in Fig.~\ref{fig:total_mass_evolution},
respectively). By contrast, \texttt{LOW$\_$MASS} loses $\approx 35\%$ of its mass to outflows, with the vast majority $\approx 65\%$ settling
into the ring at $r = R_c$ by $t\Omega_c \approx 3\times\,10^3$. The energy
generated by the quick almost-dynamical collapse in the \texttt{LOW$\_$MASS} model to form a dense ring 
at $r = R_c$ is almost completely released in radiation
(red solid lines, Fig.~\ref{fig:total_luminosity_comparison}) at rates 
$L_{\rm rad} \approx 3-4\,L_{\rm edd}$ on a timescale $\approx 10^3\,\Omega_c^{-1}$. 

More massive envelopes require more powerful winds to cool, considering that the radiated luminosity scales weakly with the envelope mass, and given the fixed potential well at the base of the envelope, this is achieved by a larger fraction of mass getting unbound. Perhaps surprisingly, the quantity of bound mass predicted by \texttt{ADIABATIC} becomes less accurate with increasing envelope mass (on timescales shorter than diffusion timescale), thereby further strengthening our conclusion in Sec.~\ref{sec:ADIABATIC} that cooling-induced convective transport in the envelope dominates the dynamics, as opposed to adiabatic convection driven purely by the initial entropy.

\subsection{Role of Initial Entropy Gradient}
\label{sec:steep}
\begin{figure}
    \centering
    \includegraphics[width= 0.48\textwidth]{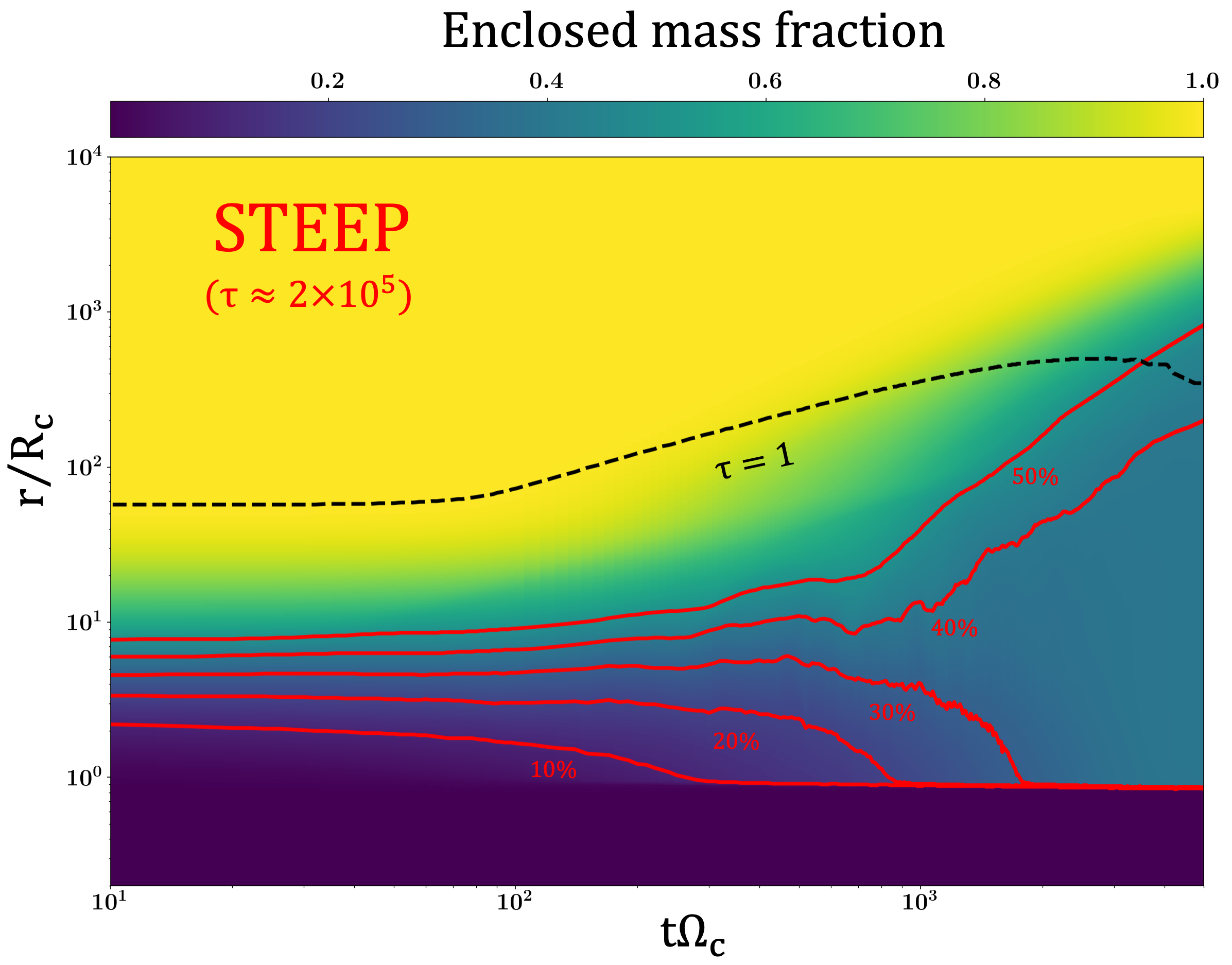}
    \caption{Radius-time diagram of enclosed mass
    in \texttt{STEEP} (as in Fig.~\ref{fig:mass_radius_time}). 
    Evolution of mass shells is qualitatively similar to \texttt{FIDUCIAL}, 
    despite the convective stability and larger binding energy compared to \texttt{FIDUCIAL}. Early evolution of \texttt{STEEP} is faster, 
    due to (i) the effective cooling timescale
    of the inner envelope $r \lesssim \, \text{a few}\,R_c$ depending only
    weakly on the density at high optical depths, as demonstrated
    in Sec.~\ref{sec:envelope_mass} (Fig.~\ref{fig:mass_radius_time}),
    and (ii) a larger fraction of the envelope mass ($\gtrsim 10\%$)
    initially residing at these radii compared to \texttt{FIDUCIAL} ($\lesssim \,\text{a few}\,\%$).}
    \label{fig:steep_mass}
\end{figure}

Our models with radiative transport reveal that turbulence drives convective energy transport towards the funnel region, leading to efficient cooling of the envelope through radiation and outflows.  It is therefore natural to ask what role the unstable initial entropy profile of the envelope played in this behavior, particularly given uncertainties associated in how the envelope forms.  Even in model \texttt{ADIABATIC}, in which radiative losses were turned off, redistribution of energy and mass associated with flattening the unstable entropy gradient alone led to the unbinding of $\approx 45\%$ of 
the envelope mass, as it evolved to a marginally stable state with $s \approx const$ (Sec.~\ref{sec:ADIABATIC}, Fig.~\ref{fig:total_mass_evolution}).  On the other hand, 
the steeper density profiles associated with stable entropy gradients (Sec.~\ref{sec:analytical_estimates}) result in the envelope being more tightly bound to the SMBH, thus potentially requiring more energy to unbind gas in outflows.  
Motivated thus, model \texttt{STEEP} explored an envelope with a steeper initial density profile (larger polytropic index $n = 3$ than the fiducial $n = 1.6$), close to marginal convective stability (Sec.~\ref{sec:analytical_estimates}). 
The envelope mass in \texttt{STEEP} ($0.5\,M_\odot$) and outer radius match that of \texttt{FIDUCIAL} (Table~\ref{tab:sim_suite}), but the binding energy is $\gtrsim 7$ times greater, and the diffusion timescale is much longer ($\gtrsim 10^4\,\Omega_c^{-1}$) because the mass is more concentrated towards small radii (panels(vi)-(vii) in Fig.~\ref{fig:initial_conditions}). These parameters are similar to those adopted by \citet{Curd&Narayan19}, except that their envelope contained lower
total mass ($M = 0.17\,M_\odot$).  As in models \texttt{LOW$\_$MASS}/\texttt{HIGH$\_$MASS}, this configuration does not correspond to the immediate aftermath of a physical TDE because the envelope binding energy is greater than that imparted by the disruption process.  Importantly, we confirmed that the initial torus of model \texttt{STEEP} remains essentially static when radiation transport is turned off, as expected from convective stability in the adiabatic limit. 

\begin{figure}
    \centering
    \includegraphics[width= 0.48\textwidth]{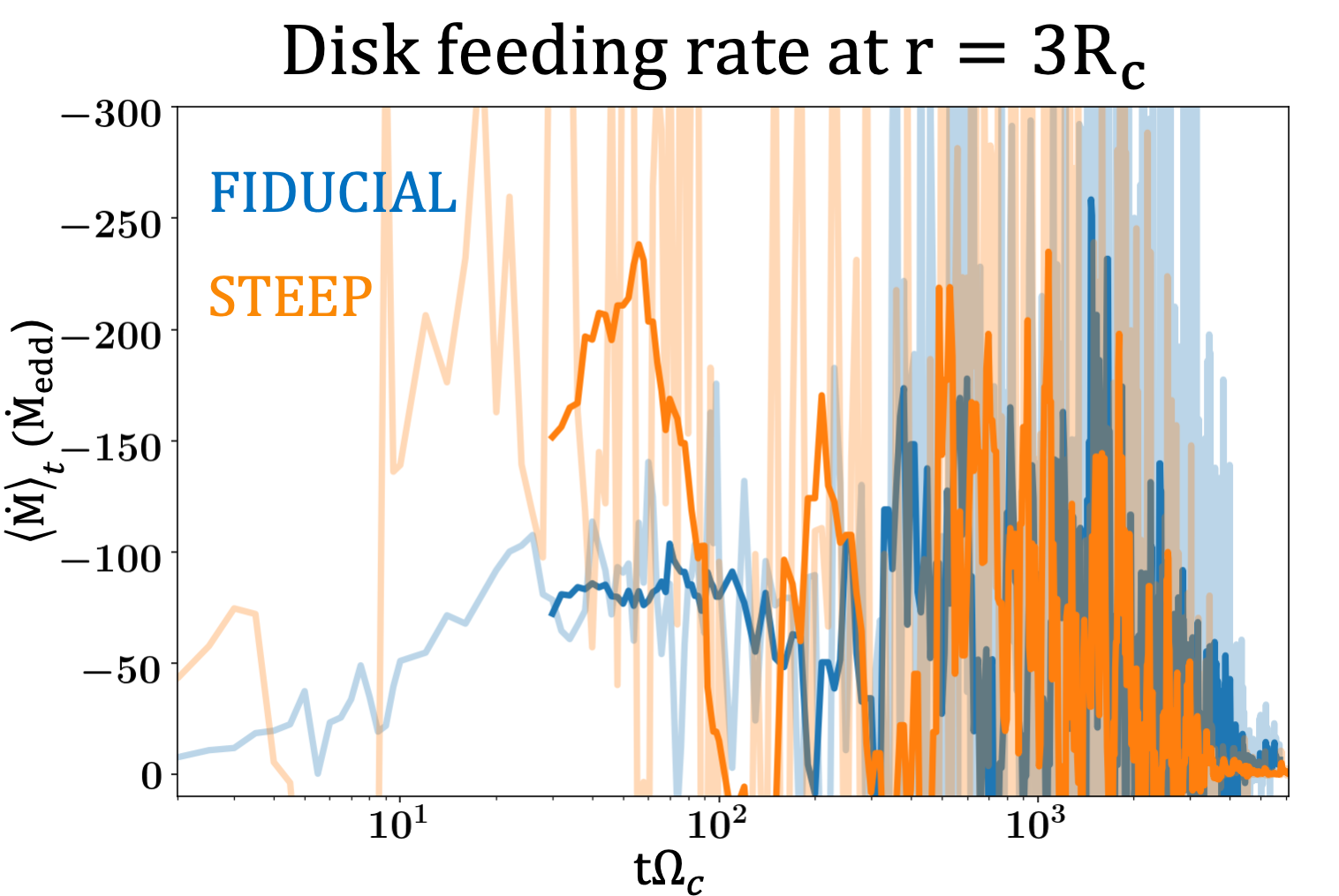}
    \caption{Angle-averaged net mass accretion rate 
    $\langle \dot{M} \rangle_\theta$ (faint colors)
    and its time average $\langle \dot{M}\rangle_{t,\theta}$
    measured at $r = 3\,R_c$ (Table~\ref{tab:fiducial_param_table})
    for \texttt{FIDUCIAL}, \texttt{STEEP}, in units
    of $\dot{M}_{\rm edd} \equiv 10\,L_{\rm edd}/c^2$. The envelope
    would feed the disk roughly at the
    given rates, if it were to form at $r \lesssim R_c$ with sufficiently
    weak (effective) viscosity (hence weak feedback onto the envelope). 
    In \texttt{FIDUCIAL}, peak mass accretion rate is delayed by cooling
    in contrast to \texttt{STEEP} where a significant fraction of 
    the envelope is readily available to be accreted
    at $r \lesssim 3\,R_c$.
    }
    \label{fig:disk_feeding_rate}
\end{figure}

The evolution of \texttt{STEEP} is qualitatively similar to \texttt{FIDUCIAL} despite its greater initial binding energy and marginally convectively-stable initial configuration.  A quasi-steady state is achieved similar to \texttt{FIDUCIAL} by around $t\Omega_c \gtrsim 500$, reaching a similar luminosity 
$L_{\rm tot}$ and mass inflow rates $\dot{M}$. However, for \texttt{STEEP} these properties quickly rise to high values $L_{\rm tot} \approx 40-50\,L_{\rm edd}$, $\dot{M} \approx -500-600\,\dot{M}_{\rm edd}$ even by early in the evolution $t\Omega_c \lesssim 500$, compared to \texttt{FIDUCIAL} where the evolution towards steady state is more gradual, with the luminosity rising slowly from sub-Eddington to $L_{\rm tot} \approx 10\,L_{\rm edd}$ and $|\dot{M}| \approx 50-100\,\dot{M}_{\rm edd}$.  This can be understood as a consequence of the flatter initial density profile of \texttt{FIDUCIAL}, which concentrates more of the envelope mass at larger radii (consistent with the weak binding energy of the TDE debris), leading to gradual feeding of the inner envelope from the outer envelope.  Since we have found that the envelope cooling timescale scales weakly with the envelope density as long as convection is efficient (Sec.~\ref{sec:envelope_mass}), the mass inflow rate will be (initially) higher if a larger fraction of the envelope mass starts closer to the black hole (see also Sec.~\ref{sec:tde_implications}, Fig.~\ref{fig:disk_feeding_rate}). 

Fig.~\ref{fig:steep_mass} shows that the mass shells of \texttt{STEEP} also evolve very similarly 
to \texttt{FIDUCIAL} (Fig.~\ref{fig:mass_radius_time}), revealing that the initial entropy gradient plays little if any role in the qualitative evolution of the envelope.  More important than the entropy gradient is the initial concentration of mass.  In \texttt{STEEP} a larger fraction ($10\%$) of the envelope mass is accreted from scales $r \approx \text{a few}\,R_c$ to $r = R_c$ compared to 
$\lesssim 1\%$ in \texttt{FIDUCIAL}, on a similar timescale ($\approx 200-300\,\Omega_c^{-1}$). The resulting higher accretion power is imprinted on the time evolution of total envelope luminosity measured outside the envelope throughout the simulation (Fig.~\ref{fig:total_luminosity_comparison}): \texttt{STEEP} has characteristic photon luminosities 
$L_{\rm rad} \approx 8 - 15\,L_{\rm edd}$ (green solid line) compared to $L_{\rm rad} \approx 5 - 6\,L_{\rm edd}$ measured in \texttt{FIDUCIAL},
lasting for a similar timescale $\approx 4-5\,\times\,10^3\,\Omega_c^{-1}$.
Kinetic luminosity (green dotted line) peaks at $L_{\rm kin} \approx 8-9\,L_{\rm edd}$ by $t\Omega_c \approx 800-10^3$. This is earlier than \texttt{FIDUCIAL} due to rapid early evolution in \texttt{STEEP}.

\begin{figure}
    \centering
    \includegraphics[width= 0.48\textwidth]{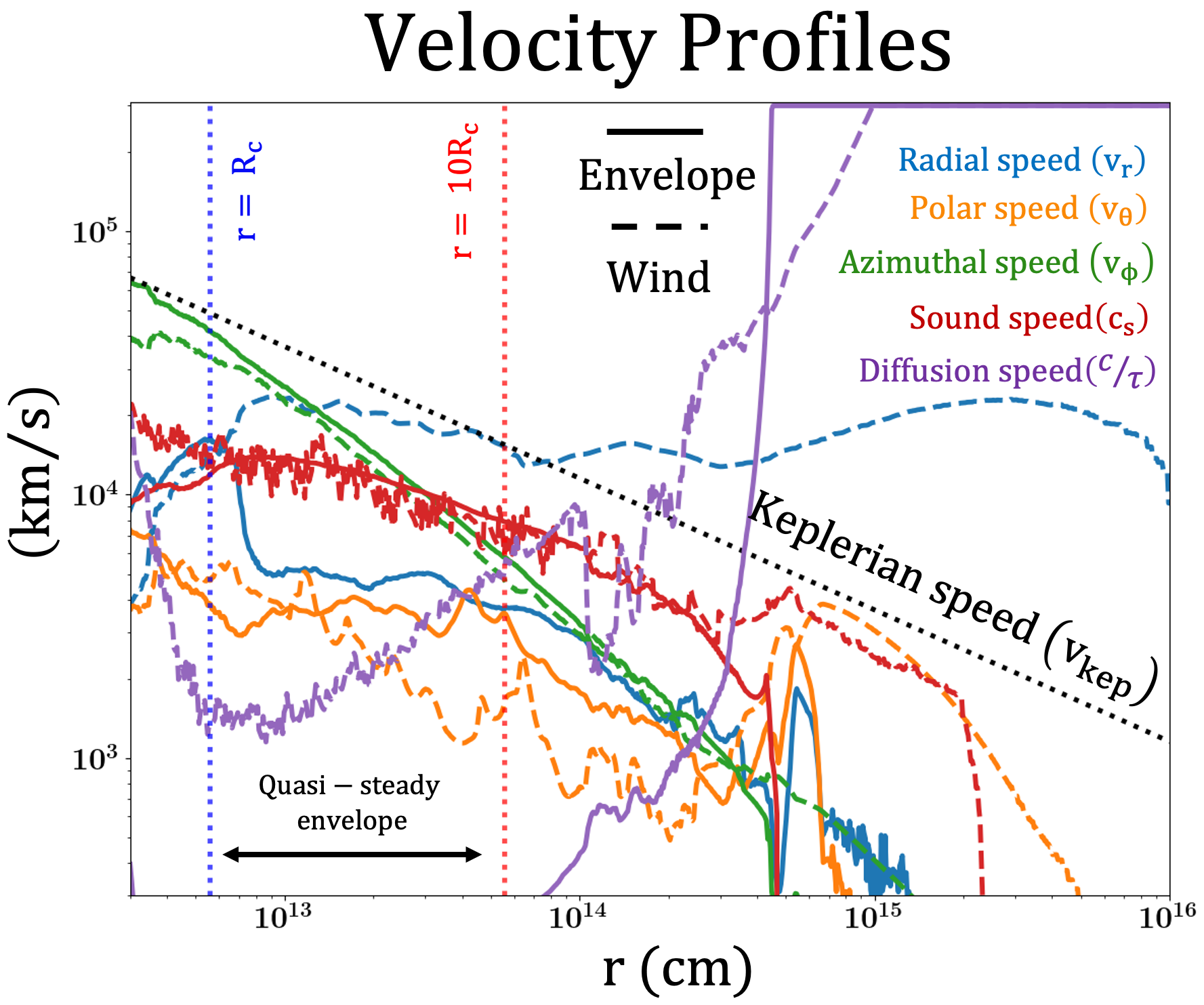}
    \caption{Velocity profiles in \texttt{FIDUCIAL}, averaged over the
    same time interval as in Fig.~\ref{fig:fiducial_steady}. Gas velocities are
    computed as $\left(\langle\langle \rho\,v^2\rangle_t/\langle \rho\rangle_t\rangle_\theta\right)^{1/2}$, where $v$ stands for the relevant component of the velocity. The (adiabatic) radiation sound speed and 
    diffusion speed are
    defined as $c_s^2 \equiv (4/9)\langle\langle \rho u_{\rm rad}\rangle_t/\langle\rho\rangle_t\rangle_\theta$ (Eq.~\eqref{eq:u_rad_def})
    and $v_{\rm diff} \equiv c/r\langle \kappa\rho \rangle_{t, \theta}$, respectively. In comparison, the
    local Keplerian speed $v_{\rm kep} \equiv \left(GM_\bullet/r\right)^{-1/2}$
    is shown (black dotted line).
    Dashed/solid lines are averages over envelope/wind zones as in Fig.~\ref{fig:wind_envelope_energy}. Blue and red vertical (dotted) lines
    mark the boundries of the quasi-hydrostatic envelope
    from $r \approx R_c \approx 6\,\times\,10^{12}\,\rm cm$
    to $r \approx 10\,R_c$ (larger radii are not in quasi-steady state at this
    time, see Fig.~\ref{fig:fiducial_steady}). If the angular momentum transport
    were efficient, an accretion disk would form at $r \lesssim R_c$, and be
    fed by the large scale envelope $r \approx 10\,R_c$. Much of the TDE phenomenology is consistent with the spherical accretion of a large scale 
    $r \gtrsim 10\,R_c$ envelope and its winds powered by cooling induced accretion (Fig.~\ref{fig:cartoon}): (i) A quasi-spherical photosphere 
    at $r \sim 10^{14} - 10^{15}\,\rm cm$ (roughly where $c/\tau \sim c$, solid purple line), (ii) (thermal) velocity fluctuations of order the Keplerian speed $c_s \approx 10^4\,\rm km\,s^{-1}$ (red solid line), (iii) a radial,
    optically thick outflow  (blue dashed lines) carrying its photosphere at a radius $r \gtrsim 10^{15}\,\rm cm$ (purple, dashed line) at speeds $v_r \lesssim 0.1\,c$.}
    \label{fig:velocities}
\end{figure}

Similar to \texttt{FIDUCIAL}, \texttt{STEEP} unbinds 
$\approx 60\%$ of its initial mass by $t\Omega_c \approx 6\,\times\,10^3$, contrary to the naive expectation that a more tightly bound envelope would be more challenging to unbind (purple lines in Fig.~\ref{fig:total_mass_evolution}). Fig.~\ref{fig:wind_envelope_density} illustrates that the distribution of the bound mass is more sharply peaked near the circularization radius, with the ring at radius $r = R_c$
enclosing $\approx 0.15\,M_\odot$ ($\approx 30\%$) compared to 
$\approx 0.04\,M_\odot$ ($\approx 8\%$). For reference, the enclosed
mass within $r = 2\,R_c$ in \texttt{STEEP} initially is 
$\approx 0.05\,M_\odot$ (Fig.~\ref{fig:initial_conditions}).

\section{Discussion}
\label{sec:discussion}

\begin{figure*}
    \centering
    \includegraphics[width= \textwidth]{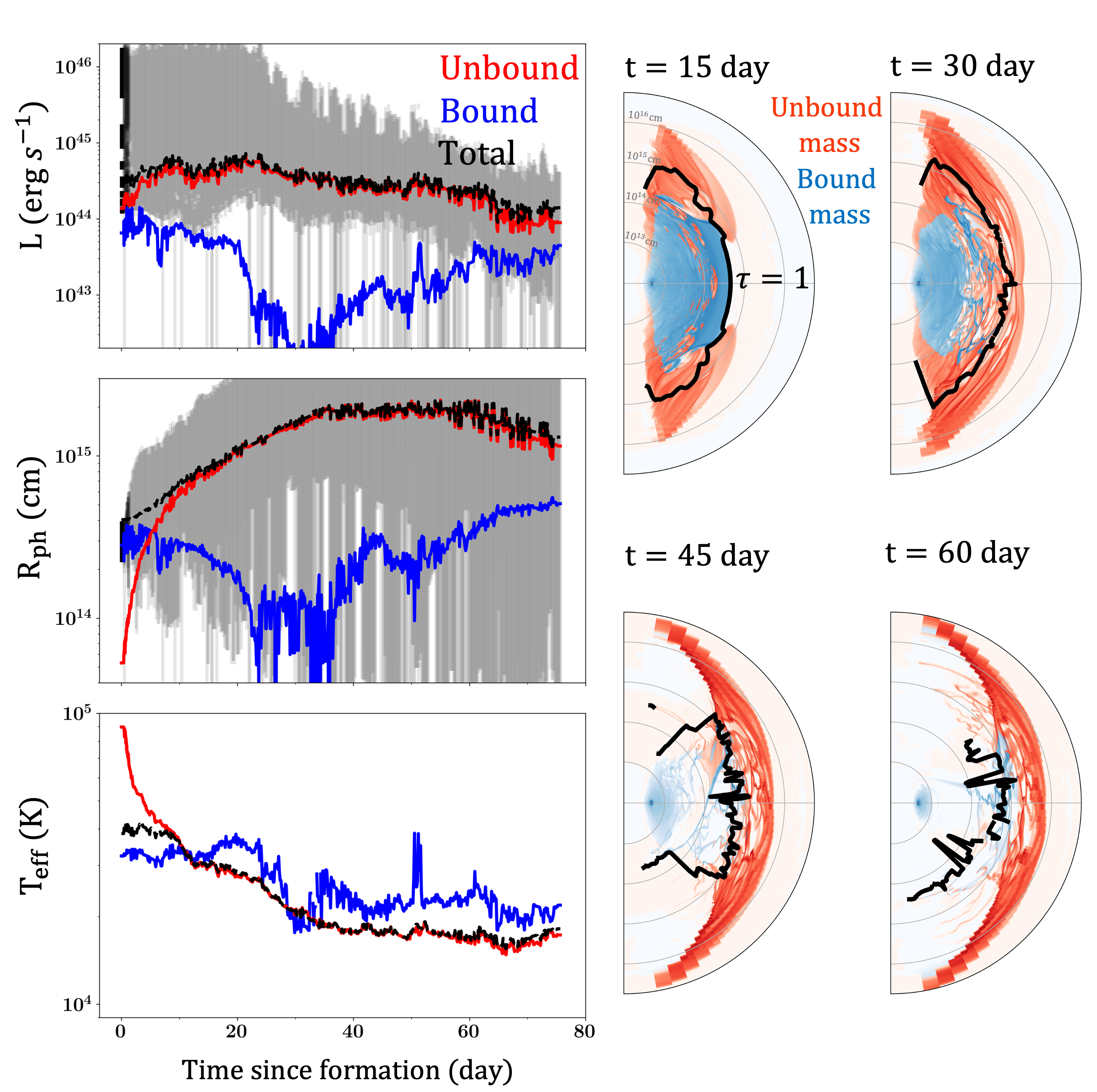}
    \caption{\emph{Left panel:} Evolution of light curve $L(t)$, 
    photosphere radius $R_{\rm ph}(t)$ and effective
    temperature $T_{\rm eff}(t)$ for \texttt{FIDUCIAL}. 
    Photosphere surface $r_{\rm ph}(t,\theta)$ is defined by 
    $1 = \int_\infty^{r_{\rm ph}(t,\theta)}\,dr^\prime\,\kappa\rho$. 
    Luminosity and effective photosphere radius are defined as $L(t) \equiv \sum_i\,F_{r,i}\,dA_i$, and $R^2_{\rm ph}(t) \equiv \sum_i\,dA_i/4\pi$,
    where $F_r$ is the radial component of the radiation flux, $dA$ 
    is the area element perpendicular to radial direction, and the
    index $i$ runs through the cells on the photosphere surface. Effective temperature is computed as $\sigma\,T_{\rm eff}^4(t) \equiv L(t)/4\pi R_{\rm ph}^2(t)$, where $\sigma$ is the
    Stefan-Boltzman constant. Breakdown of the sums into bound (blue)
    and unbound (red) cells are shown separately as in Fig.~\ref{fig:wind_envelope_energy}.
    Black shaded lines show the values of the (angle-dependent) photosphere radius $r_{\rm ph}(t, \theta)$ and isotropic luminosity $4\pi r^2_{\rm ph}(t, \theta)\,F_r(t, r_{\rm ph}(t, \theta), \theta)$ at varying latitudes. \emph{Right panel:} Snapshots of the mass
    distribution and photosphere surface, where density is colored in
    log-scale. If an area element on
    the photosphere surface (black curves) is bound (unbound),
    it contributes to blue (red) line on the left panel.
    }
    \label{fig:light_curve}
\end{figure*}
\begin{figure}
    \centering
    \includegraphics[width= 0.48\textwidth]{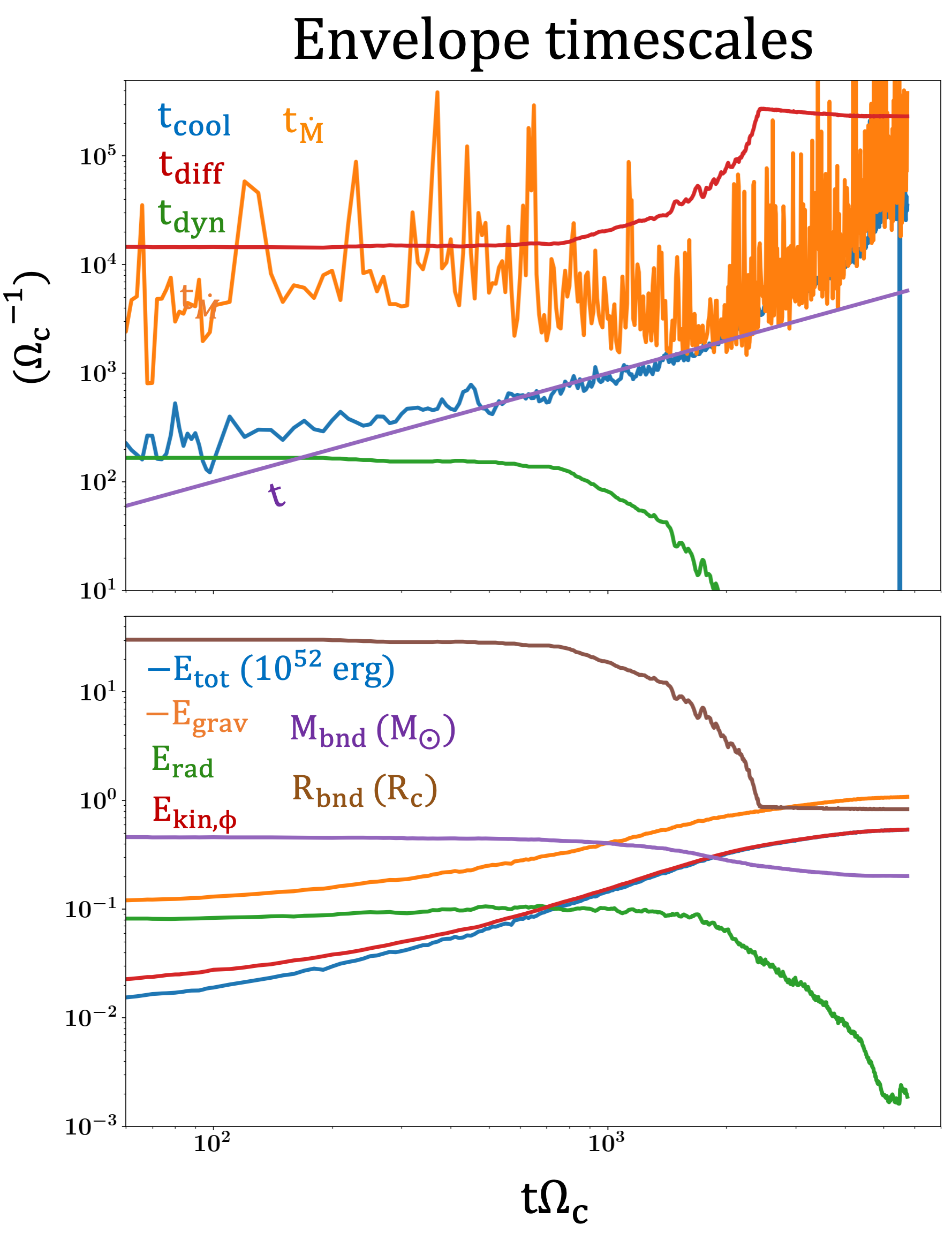}
    \caption{Time evolution of characteristic timescales within the bound envelope for model \texttt{FIDUCIAL}.  \emph{Top panel:} Cooling time and mass-loss time are defined as $t_{\rm cool}(t) \equiv |d\log u_{\rm tot, bnd}/dt|^{-1}$, $t_{\rm \dot{M}}(t) \equiv |d\log M_{\rm bnd}/dt|^{-1}$, where the bound envelope is defined according to Eq.~\ref{eq:rho_bnd}.  The diffusion and dynamical times of the envelope are defined according to
    $t_{\rm diff} \equiv 3\kappa\,M_{\rm bnd}/4\pi R_{\rm bnd}^2(t)c$, and $t_{\rm dyn} \equiv \left(GM_\bullet/R_{\rm bnd}(t)\right)^{-1/2}$, where $R_{\rm bnd}(t)$ is an effective envelope radius within which half the instantaneous bound mass $M_{\rm bnd}(t)$ is enclosed.  \emph{Bottom panel:} Time evolution of the energy, mass and radius of the bound envelope.
    }
    \label{fig:bound_timescales}
\end{figure}

\subsection{Implications for TDE Observations}
\label{sec:tde_implications}

Fig.~\ref{fig:velocities} shows radial profiles of the flow speed,
sound speed and photon diffusion speed of model \texttt{FIDUCIAL}, averaged over the quasi-steady state ($\approx 7- 15\,\rm days$ after the envelope formation, Fig.~\ref{fig:fiducial_steady}) and spanning a wide range of scales $10^{13}\,\rm cm \lesssim r \lesssim 10^{16}\,\rm cm$ surrounding the black hole.  The large photosphere radii of the wind and quasi-hydrostatic envelope at $r \sim 10^{14} - 10^{15}\,\rm cm$ is broadly consistent with those inferred by optical/near-UV observations of TDEs (e.g., \citealt{Gezari21}), while the mildly sub-Keplerian thermal speed $c_s \sim \,10^4\,\rm km/s$ agrees with spectroscopic line widths \citep{Charalampopoulos+21}.

Fig.~\ref{fig:disk_feeding_rate} shows the time-dependent
mass accretion rate measured just outside the circularization radius $r = 3\,R_c$. If the envelope we have modeled has sufficient time to form, this indicates that the inner accretion disk would be supplied mass at a rate  $|\dot{M}| \approx 50\,\dot{M}_{\rm edd}$ over the first few days of evolution $t \lesssim 200\,\Omega_c^{-1} \approx 2.6\,\rm day$. However, the peak of the accretion rate is delayed until $t \approx 10^3\,\Omega_c^{-1} \approx 13\,\rm day$, when the majority of the envelope mass$-$initially located at large radius$-$has time to settle down to $r \lesssim 3\,R_c$.  By comparison, for model \texttt{STEEP}, the accretion rate around $r = 3\,R_c$ peaks much faster, over the first $\approx 2 \,\rm days$ of the evolution, because the initial mass distribution is more tightly concentrated near the inner radius of the envelope.  We therefore confirm that for a TDE envelope endowed with the correct binding energy in the initial state, feeding of the inner disk can indeed be significantly delayed by cooling \citep{Metzger22}, although the initial accretion rate is still super-Eddington even if very little mass starts at radii $r \lesssim 3\,R_c$.

Fig.~\ref{fig:light_curve} shows the time-evolution of the emergent photon luminosity (light curve), photosphere radius, effective temperature, as well as snapshots of the photosphere surface. The photosphere is initially an almost spherical surface at radius $r \approx \, 5\,\times\,10^{14}\,\rm cm$. However, as the envelope evolves, gas at lower latitudes progressively joins an unbound wind starting from the poles, which carries the photosphere to $r \approx 10^{15}\,\rm cm$. The quasi-hydrostatic portion of the bound envelope described in Sec.~\ref{sec:fiducial} does not carry the photosphere, as it extends to an outer radius of $\lesssim 10^{14}\,\rm cm$ and lasts between days $13-30$.  The photosphere is instead a more dynamical structure which consists primarily of unbound material and encases the quasi-hydrostatic portion of the envelope, extending to $\sim 10$ times its outer radius.  After $\approx 45\,\rm days$,
the wind becomes optically thin in polar directions, reducing the
emitting area and causing the effective photosphere radius to recede.  By $t\approx 60\, \rm days$, the accretion ring at $r = R_c$ is surrounded by a spherical shell of unbound wind that is now optically thin at most latitudes, except along the mid-plane. Inner layers of the wind become bound as they cool, and fall back onto the ring.

The overall light curve evolution suggests that passive cooling of an Eddington envelope of fixed initial mass alone likely cannot account for the TDE emission in optical/UV bands:  
Instead of powering the optical/UV light curve for many months as required by observations (e.g., \citealt{Hammerstein+23}), the envelope runs out of ``fuel'' by $t \approx 75\,\rm day$ after forming, with $\approx 0.2\,M_\odot$ being accreted onto the accretion ring at $r = R_c$ and the remaining $\approx 0.3\,M_\odot$ ejected in unbound winds (see the snapshot in Fig.~\ref{fig:light_curve}). The bolometric light curve is approximately constant for $\approx 60\,\rm days$ at 
$L \approx 2-5\,\times\,10^{44}\,\rm erg\,\rm s^{-1}$, before dropping to near the Eddington luminosity $L \approx 10^{44}\,\rm erg\,\rm s^{-1}$ between days $60-70$. As polar winds are launched, the effective photosphere radius expands from its initial value $\approx 5\,\times\,10^{14}\,\rm cm$ to $\approx 10^{15}\,\rm cm$ as the wind expands to larger radii, before receding between days $60-70$ as the wind expands and dilutes.  The effective temperature 
declines over time from $T_{\rm eff} \approx 4\,\times\,10^4\,K$ to 
$T_{\rm eff} \approx 2\,\times\,10^4\,K$ as a consequence of increasing photosphere radius. This behavior contrasts with the typically shorter peak durations of TDE optical light curves $\lesssim 30\,\rm day$, and the late-time gradual decline of the luminosity after peak (e.g., \citealt{Gezari21}).

As compared to predictions \citep{Metzger22}, the inconsistency of our predicted light curve behavior with TDE observations can be understood as a consequence of the unexpectedly fast cooling rate of the envelope, as demonstrated in Sec.~\ref{sec:fiducial}. Fig.~\ref{fig:bound_timescales} shows the evolution of characteristic timescales over which the energy and 
mass of the bound envelope change, as well as an estimate of the dynamical and diffusion timescales of the bound envelope. Following a brief transient phase $\approx 400\,\Omega_c^{-1}$, the envelope cools in a self-similar manner (top panel, blue line), as it contracts from $\approx 30\,R_c$ to $\approx 10\,R_c$ (bottom panel, brown line). This cooling is gradual and sub-dynamical (i.e., quasi-hydrostatic; $t_{\rm cool} \gtrsim t_{\rm dyn}$), until becoming dominated by mass-loss (i.e., when $t_{\dot{M}} \approx t_{\rm cool}$).

In Sec.~\ref{sec:analytical_estimates}, we argued that a quasi-hydrostatic envelope could form following the tidal disruption of the star, assuming no mass/energy is lost from the system as orbital energy is dissipated in shocks (Fig.~\ref{fig:TDEtree}).
We find, however, that such an envelope cools efficiently within one outer dynamical time $\approx 13\,\rm day \ll t_{\rm fb} \approx 57\,\rm day$, i.e., much faster than were the luminosity Eddington-limited.  This implies that cooling may play a key role already within a fallback time, over which the assumed envelope would form.  We conclude that, if an envelope similar to what we explored in this work forms, it will possess a lower mass, potentially larger binding energy, and that its gradual feeding by the fallback streams should play an important role in the dynamics. 
Stated another way, although the gravitational potential energy liberated as matter contracts quasi-spherically from large radii to $R_{\rm c}$ is still a promising source of power for the luminosities of TDEs, the rate-limiting step to feeding the black hole and powering the light-curve may still be driven by fall-back.   

In principle, it is possible to incorporate the effects of fallback 
feeding in the envelope evolution in our setup, by injecting mass
into the envelope at the fallback rate (Eq.~\ref{eq:mdot_fb}). 
We aim to explore the role of such fallback feedback, as well as additional energy liberated by the inner accretion disk, in future work.

\subsection{On the nature of envelope cooling:
The funnel, lateral energy transport and initial conditions}
\label{sec:cooling_nature_discussion}
\begin{figure}
    \centering
    \includegraphics[width= 0.48\textwidth]{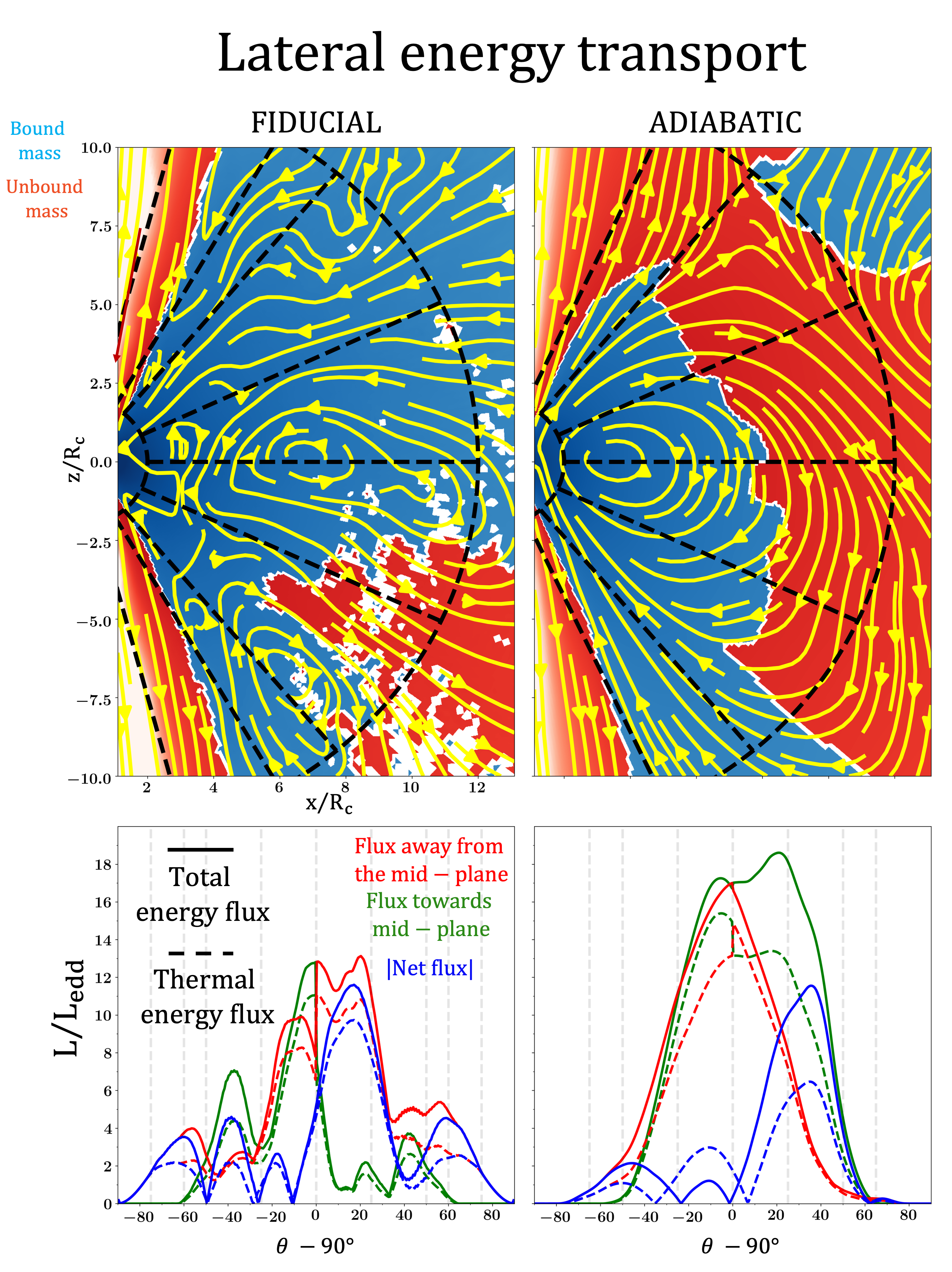}
    \caption{\emph{Top panel}: Streamlines of 
    time-averaged radiation
    energy flux (left panel), and enthalpy flux (right panel) in \texttt{FIDUCIAL}, and \texttt{ADIABATIC},
    respectively. Bound and unbound mass density
    is colored in blue and red, respectively, 
    as in Fig.~\ref{fig:energy_bernoulli}. Time average
    is taken in the time interval $100 < t\Omega_c < 500$, prior to the steady state described
    for \texttt{FIDUCIAL} in Sec.~\ref{sec:fiducial} (Fig.\ref{fig:fiducial_steady}), and much earlier than the estimated
    diffusion time in the bulk envelope $\approx 10^{3}\,\Omega_c^{-1}$ (Fig.~\ref{fig:initial_conditions}). 
    Radii $r = 2\,R_c$
    and $r = 12\,R_c$  as well as chosen $\theta = \rm constant$ lines are marked in dashed black lines.
    \emph{Bottom panel:} Dashed curves 
    show polar radiation
    flux $\langle F_{\rm rad, \theta} \rangle_t$ (left panel) and enthalpy flux $\langle 4\,v_\theta\,P\rangle_t$ (right panel) integrated
    along $\theta = \rm constant$ lines, between radii 
    $2\,R_c \leq r \leq 12\,R_c$. Solid 
    lines show total energy flux including kinetic energy. Selected $\theta = \rm constant$ lines in top panels are marked with (shaded) vertical dashed lines in  bottom panels. Contribution to net flux
    is broken down into contributions directed 
    away from the mid-plane (red) and towards 
    the mid-plane (green). The magnitude of the net lateral flux is shown in blue.  Net flux (blue lines) is similar close to mid-plane $|\theta - 90^\circ| \lesssim 40^\circ$ in \texttt{ADIABATIC} and \texttt{FIDUCIAL}, as expected given the
    diffusion time is longer in those layers. At low density surface layers $|\theta - 90^\circ| \gtrsim 60^\circ$ facing the funnel, a net flux of thermal energy (red dashed lines) 
    $L \approx 4\,L_{\rm edd}$ and total energy (blue lines) $L \approx 8\,L_{\rm edd}$ leaves the envelope and allows material to sink 
    in \texttt{FIDUCIAL}. In \texttt{ADIABATIC}, envelope loses energy at
    a rate $L \approx 2\,L_{\rm edd}$ at $\theta \approx 50^\circ$, and gains
    at roughly the same rate from lower half-plane $\theta \approx 140^\circ$. This explains higher net inflow rates at early times in \texttt{FIDUCIAL}, and
    its absence in \texttt{ADIABATIC}.}
    \label{fig:fid_lateral_energy_transport}
\end{figure}
In Sec.~\ref{sec:fiducial}, we found that 
the steady-state TDE envelope
has an envelope-wind structure where energy and mass fall radially inward through the envelope in a quasi-hydrostatic manner, and are ejected radially outwards through 
the wind. In order for gas to fall inwards, radiation
support needs to be removed in the first place, which we find to 
eventually 
emerge in the polar wind and the funnel (Fig.~\ref{fig:wind_envelope_energy}). 
This suggests an overall net lateral
transfer of energy from the mid-plane, where
the bulk of the envelope resides, into the polar direction, where the wind takes over.  

In Fig.~\ref{fig:fid_lateral_energy_transport},
we demonstrate that this is indeed the case
in the time interval $t\Omega_c \leq 500$,
i.e. the phase leading up to the envelope
steady state interior to $r \lesssim 10\,R_c$ described in Fig.~\ref{fig:fiducial_steady}.
In the top (left) panel, we see that the flow near the mid-plane in \texttt{FIDUCIAL } is dominated by large eddies of size $r \sim 5\,R_c$ and angular extent $\approx 50^\circ$, 
similar to \texttt{ADIABATIC}. 
About $\approx 60^\circ$ away from the mid-plane, energy transport is towards the wind and the funnel, at a net rate $\approx 7-8\,L_{\rm edd}$ (blue solid lines in bottom-left panel), of which $\approx 4\,L_{\rm edd}$
is radiation energy flux. This is in agreement 
with our finding in Fig.~\ref{fig:total_luminosity_comparison} 
that the net energy output of the system
is roughly equipartitioned in the forms 
of kinetic and radiation energy (Fig.~\ref{fig:total_luminosity_comparison})
in the steady-state. The envelope $r \lesssim 12\,R_c$
in \texttt{ADIABATIC}, in comparison, loses total energy in 
the upper half plane $\theta \approx 55^\circ$ at a rate 
$L \approx 2\,L_{\rm edd}$ and gains in the lower half-plane $\theta \approx 140^\circ$ at roughly the same rate, hence the net transport is in the
radial direction, on average. This
explains the higher inflow rates observed in the early evolution of \texttt{FIDUCIAL} at $t\Omega_c \ll 10^3$ (e.g. Fig.~\ref{fig:fiducial_early}): 
Lateral flux of thermal energy is enhanced by the optically thin 
surface layers facing the funnel, which facilitates the inflow. 
These findings support our observation in 
Sec.~\ref{sec:ADIABATIC} that the cooling is the main driver of 
the turbulence even at relatively early times, 
compared to the initially present, convectively unstable entropy gradient. As the envelope continues to evolve, we find that turbulence in \texttt{ADIABATIC} gets weaker, whereas it is sustained by cooling on timescales longer than a bulk diffusion time in \texttt{FIDUCIAL}, as expected. 

Our results indicate that the existence of a low-density funnel, although it is relatively narrow $\Delta\theta \approx 15^\circ$
(Fig.~\ref{fig:fid_lateral_energy_transport}), 
plays a central role in the dynamics of the envelope: It provides 
a surface layer through which the low density layers of the envelope
$15^\circ \lesssim \theta \lesssim 30^\circ$ quickly cool, driving 
and maintaining turbulence. Radiation bursting into the funnel 
unbinds these layers, which launches a relatively steady wind that cools the envelope at super-Eddington rates. Our simulations are initialized with an already existing funnel, due to the assumption of uniform angular momentum distribution. In reality, it is possible for angular momentum to be concentrated on the mid-plane, and some low-angular momentum, optically thick gas to fill the funnel, which,
in principle, could change the evolution qualitatively.
We find, however, that  such non-uniform angular momentum distributions are difficult to reconcile with the energy, mass, angular momentum budget available from tidal disruption process.
We therefore expect a TDE envelope to indeed have a funnel, and the qualitative evolution to be similar to our simulations, if it were
to form.

\section{Conclusions}
\label{sec:conclusion}

We have carried out axisymmetric, radiation–hydrodynamic simulations to investigate how the shock-heated debris from a tidal disruption event cools and contracts to form a compact, rotationally supported accretion structure. Motivated by the expectation that stream-stream or stream-disk dissipation is rapid, but subsequent cooling may proceed more slowly than the fallback of debris, we initialized our calculations with quasi-hydrostatic, rotating envelopes whose mass, energy, and angular momentum reflect those imparted by a realistic stellar disruption. Our study quantifies the effective cooling timescale of such envelopes, the balance between radiative and advective energy transport, and the amount of mass ultimately available to assemble an accretion disk at the circularization radius.

The envelopes become turbulent within only a few orbital periods at $R_{\rm c}$, and within roughly a week of evolution settle into a quasi-steady configuration across radii $r \lesssim 10R_{\rm c}$. This state is characterized by radially constant mass inflow rates $|\dot M| \sim 70\, \dot M_{\rm Edd}$ and luminosities $L_{\rm tot} \sim 10 L_{\rm Edd}$, supported by a broad, bound, quasi-spherical envelope feeding an unbound, optically thick wind. The outflow carries comparable kinetic and radiative luminosities and is powered by the gravitational energy released as matter cools and sinks quasi-spherically toward the centrifugal barrier. Viscous angular-momentum transport being absent, material accumulates into a dense ring at $r \approx R_{\rm c}$, which contains the bulk of the bound mass by the end of the simulations.

Comparison with adiabatic control runs demonstrates that radiative losses are essential for this evolution. Without radiation transport, the envelope expands and drives weaker outflows as it relaxes toward a steeper density and entropy profile that is marginally stable against convective overturn, i.e., one in which the entropy gradient is just sufficient to suppress buoyant instability. In this adiabatic case no ring forms at $R_{\rm c}$, and the binding energy released in the radiative models is instead retained as thermal support. Radiative cooling therefore not only enables mass to settle inward but also strongly enhances mass loss: for our fiducial model, roughly $60\%$ of the envelope becomes unbound within a fallback time, despite starting from a less bound initial configuration than the corresponding adiabatic model.

Although the envelopes are extremely optically thick, the effective cooling rate is not set by photon diffusion through the bulk interior. Instead, once turbulence develops, energy is redistributed laterally toward low-density polar regions, where it escapes efficiently through a combination of radiation and mass-loaded winds. As a result, the cooling timescale of the inner envelope depends only weakly on the overall density normalization and therefore on optical depth. Varying the envelope mass by an order of magnitude confirms this behavior: lighter envelopes cool predominantly through photon diffusion, while more massive ones cool chiefly through stronger winds, yet the characteristic inflow times at $r \lesssim 10 R_{\rm c}$ remain similar. The presence of a low-density funnel is thus crucial, as it provides a surface channel through which super-Eddington energy fluxes emerge and maintain turbulence throughout the envelope.

These results bear directly on the phenomenology of TDEs. The large radii of the photosphere ($\sim 10^{14}$--$10^{15},{\rm cm}$), the moderate effective temperatures ($\sim 2$--$4\times10^{4},{\rm K}$), and the characteristic gas velocities ($\sim 10^{4}\,{\rm km\,s^{-1}}$) produced by our models align well with observational inferences from optical--UV TDEs. Moreover, the inflow rates supplied to the circularization region are initially super-Eddington but peak only after a delay, supporting scenarios in which the build-up of the inner accretion flow, and hence the emergence of X-ray or radio activity, lags behind the optical/UV emission. However, a passively cooling envelope with a fixed initial mass cannot sustain a months-long luminous phase: in our simulations, the envelope exhausts its available binding energy within roughly a fallback time, and the luminosity then declines too rapidly and not in a fashion resembling observed light curves. The discrepancy arises because cooling proceeds far more efficiently than anticipated in analytic models that envisioned Eddington-limited Kelvin-Helmholtz-like contraction \citep{Metzger22}.

Taken together, these findings suggest that while the quasi-spherical envelope envisioned in prior work remains a promising ingredient in understanding TDE emission, its long-term evolution cannot be captured by a passively cooling configuration alone. Continuous feeding by fallback streams---or energy injection from the growing accretion disk---may be required both to sustain the envelope and to regulate its cooling over many fallback times. Such feedback may in turn influence how mass is partitioned between the eventual disk and unbound outflows, and how the system transitions from optical/UV-dominated emission to later X-ray and radio activity. Incorporating these effects, together with angular-momentum transport and inner-disk feedback, will be crucial for developing a fully self-consistent picture of TDE envelope formation, evolution, and observable signatures.

\acknowledgements

S.T. thanks Xiaoshan Huang and Chris Fragile for insightful discussions.  ST and BDM acknowledge support from NASA (grant 80NSSC24K0934), the National Science Foundation (grant AST-2406637), and the Simons Foundation (727700).  The Flatiron Institute is supported by the Simons Foundation. 

\appendix

\section{Details of the numerical setup}
\label{app:numerical_details}

\texttt{Athena++} \citep{Stone+20} solves the equations of radiation hydrodynamics:
\begin{align}
    &\partial_t\,\rho\,
    + \nabla\left(\rho\mathbf{v}\right) = 0
    \nonumber\\
    &\partial_t\,(\rho\,\mathbf{v})
    + \nabla\left(\rho\mathbf{v}\mathbf{v}
    + P\right) = -\mathbf{S_{\rm rad, M}} -\rho\cdot\nabla\Phi
    \nonumber\\
    &\partial_t\,E
    + \nabla\left(\mathbf{v}(E + P_g)\right) = 
    -{S_{\rm rad, E}} - \rho\mathbf{v}\cdot\nabla\Phi
    \nonumber\\
    &\partial_t\,I + c\mathbf{n}\cdot\nabla\,I = c\,S_I
    \nonumber\\
    S_I &= \Gamma^{-3}\Biggl[
    \rho\,(\kappa_s + \kappa_a)\,
    (J_0 - I_0)  + \rho(\kappa_a + \kappa_{\delta{P}})
    \left(\frac{a_rT^4}{4\pi} - J_0\right)\Biggr]
    \nonumber\\
    S_{\rm rad, E} &= 4\pi\,c\int\,d\Omega\,S_I;\,\,\,\,
    \mathbf{S}_{\rm rad, M} = 4\pi\,c\int\,d\Omega\,\mathbf{n}S_I
    \label{eq:RHD_equations}
\end{align}
Here, $\rho$, $T_{\rm g}$, $\mathbf{v}$, $P_g = k_B\rho\,T_g/\mu\,m_p$, and $E = \rho v^{2}/2 + P_g/(\gamma_{\rm ad} -1)$ are the density, temperature, velocity, pressure, and energy of the gas ($\gamma_{\rm ad} = 5/3$); $I$ is the specific intensity of the radiation field, which equals $I_0$ in the fluid rest frame and the factors $\Gamma = \gamma(1 - \mathbf{n}\cdot\mathbf{v}/c)$, $\gamma = (1 - v^2/c^2)^{-1/2}$ encode relativistic corrections to frame transformations between the fluid rest-frame and lab-frame.  We adopt a Newtonian gravitational potential for the SMBH of mass $M_{\bullet}$ (Eq.~\ref{eq:Phi_BH}). $\kappa_s$, $\kappa_a$ denote Rosseland mean of scattering and absorption opacities whereas $\kappa_{\delta P}$ denotes the Planck mean of the absorption
opacity. $J_0 = \int\,d\Omega\,I_0$ is the mean intensity in fluid rest-frame. We considered absorption opacity only
by setting $\kappa_s = 0, \kappa_{\delta P} = 0$,
and $\kappa_{\rm a}$ is defined in Eq.~\eqref{eq:opacity}. 

We neglect angular momentum transport due to viscosity in this work by not prescribing a viscous tensor; the effects of viscous accretion, and associated energy release from the inner accretion disk on the system evolution, will be explored in future work (Tuna et al., in prep).

We set floor and ceiling values on the temperature and density according to:
\begin{align}
    \rho_{\rm floor} = 
    10^{-11}\,\rho\,(r= R_c, \theta = \pi/2); \,\,\,
    T_{\rm gas, floor} = 10\,K; \,\,\,
    T_{\rm gas, ceiling} &= 10^{12}\,K \label{eq:floor_values}
\end{align}
Our simulation domain covers the radial interval $[6\,R_{\rm g}, 4\,\times\,10^5\,R_{\rm g}]$, extending from the approximate scale of the SMBH horizon to an outer radius sufficiently large to contain the photosphere radius at all times.  We employ a logarithmic grid of 128 points in the radial direction, and 36 grid points in the polar direction. 
We use 3 levels of static mesh refinement at radii $6\,R_{\rm g} < r < 400\,R_c \approx 1.5\,\times\,10^4\,R_g$ and 1 level of 
refinement at radii $400\,R_{\rm g} < r < 1200\,R_{\rm g}$. The radiation intensity field is discretized in the space of directions by  adapting a coordinate system $(\zeta, \psi)$, where
$\zeta, \psi$ label the angles the direction vector $\vu{n}$ makes with the radial $\vu{r}$ and polar $\vu{\theta}$ unit vectors in spherical coordinates, respectively \citep{Davis&Gammie20}. 
We use 20 rays in the $\zeta$ direction and $4$ rays in $\psi$ direction, for a total of 80 ray directions.

When the intensity field is initialized isotropically as in Eq.~\ref{eq:intensity_init}, the net radiation force is zero (Eq.~\ref{eq:RHD_equations}), hence the early evolution is out of hydrostatic equilibrium for a brief relaxation period. To asses the consequences of this, we ran separate simulations where the initial density and temperature profile are fixed, but the intensity field is evolved until the radiation flux streamlines find an equilibrium. We found that following the transition
to turbulence within a dynamical time at $r = R_c$, the evolution is essentially the same with or without initial relaxation. 

\section{Initial conditions}
\label{app:initial_conditions}

Rotating, axisymmetric tori in mechanical equilibrium are described by \citep{Papaloizou&Pringle84}:
\begin{align}
    \frac{\nabla\,P}{\rho} = - \nabla\phi + 
    \frac{l^2(\varpi)}{r^3}\vu{\varpi},
    \label{eq:torus_force_balance}
\end{align}
where $(\varpi, \theta, z)$ are cylindrical polar coordinates,
and $l(\varpi)$ is the specific angular momentum of the fluid (see also the relativistic Polish doughnut; \citealt{Kozlowski+78}). This equation can be written in terms of an
effective potential:
\begin{align}
    \frac{\nabla\,P}{\rho} = -\nabla\Psi;\,\,\,\,\,\,\Psi(\varpi, z) \equiv -\frac{GM_\bullet}{(\varpi^2 + z^2)^{1/2}} - \int\,d\varpi\,\frac{l^2}{\varpi^3}
\end{align}
For a polytropic relationship between pressure and density of the form $P\,\propto\,\rho^{1 + 1/n}$, the solutions are given by:
\begin{align}
    \rho &= \rho_0\left(\frac{Be_0 - \Psi}
    {Be_0 - \Psi_0}\right)^n
\end{align}
characterized by three parameters $\{Be_0, \Psi_0, R_0\}$, where $R_0$ is a reference
radius, $\rho_0, \Psi_0$ are values of density and the effective potential at $r= R_0$, and $Be_0$ is the Bernoulli parameter
of the polytropic gas, which takes a constant value for this family of solutions :
\begin{align}
    Be_0 = (1 + n)\frac{P}{\rho} - 
    \frac{GM_\bullet}{(\omega^2 + z^2)^{1/2}}
    - \int\,d\varpi\frac{l^2}{\varpi^3}.
\end{align}
We consider angular momentum distributions $l = l_0 (\varpi/R_c)^p$, for which we obtain
\begin{align}
    \Psi(r, \theta) = -\frac{GM_\bullet}{r} - 
    \frac{l_0^2}{(2p - 2)R_c^2}\left(\frac{r\sin\theta}{R_c}\right)^{2p - 2}
\label{eq:lprofile}
\end{align}
where we switched to spherical polar coordinates $(r, \theta, \phi)$. In all of our simulations, we take $p = 0,$ corresponding to uniform specific angular momentum.
This family of initial conditions is widely used as initial conditions in numerical simulations of AGN \citep{Kato+04, Jiang+17} and TDE \citep{Curd&Narayan19} disks. In our setup,
 the parameters $\rm Be_0$ and $n$ set by specific binding energy
 requirement, whereas $\rho_0$ is fixed by the total envelope mass (Sec.~\ref{sec:analytical_estimates}). A uniform specific
 angular momentum automatically satisfies the angular momentum
 constraint.

\begin{figure*}
    \centering
    \includegraphics[width= \textwidth]{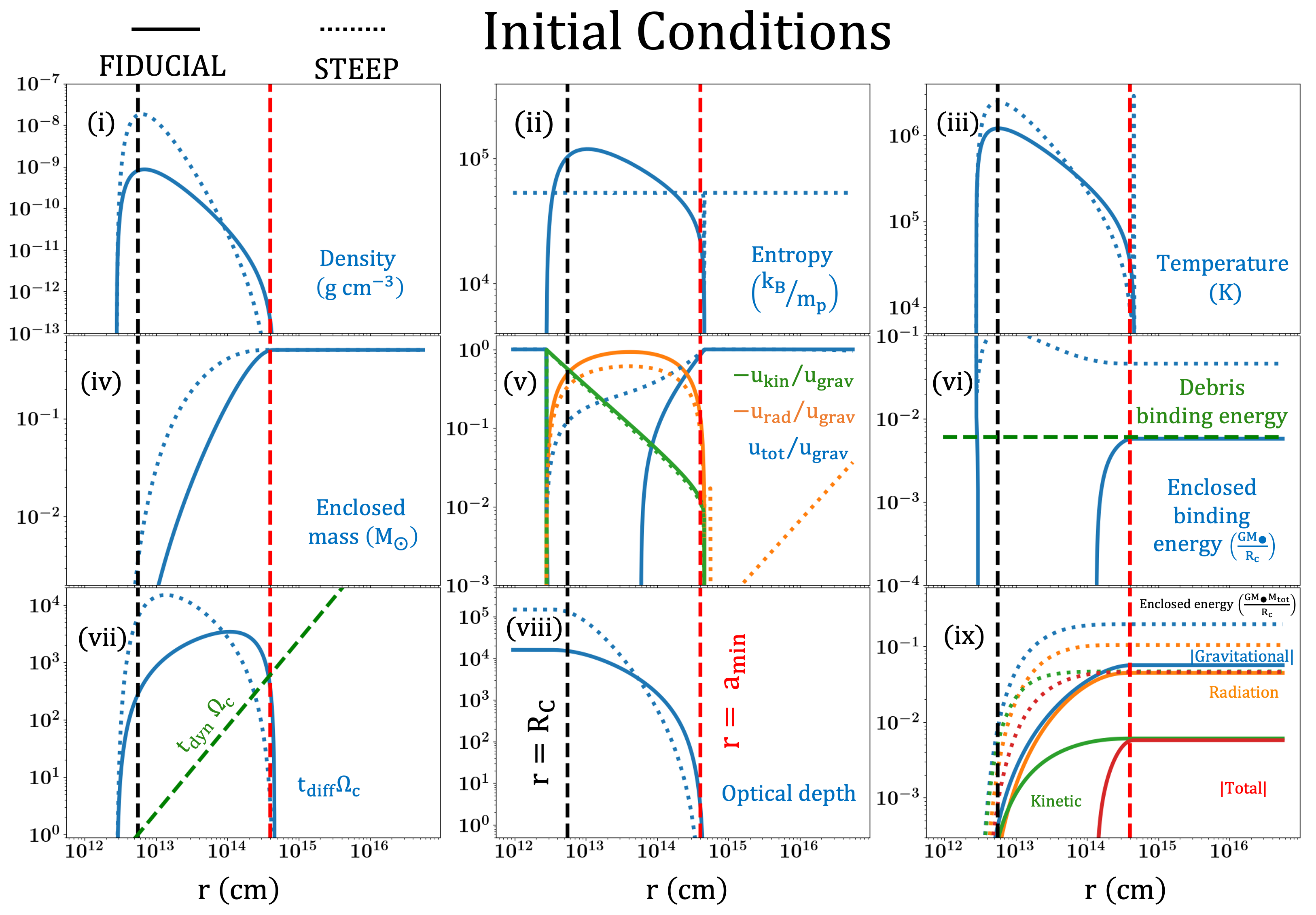}
    \caption{Angle-averaged initial conditions
    of the envelope
    at $t=0$ for models \texttt{FIDUCIAL} (solid lines)
    and \texttt{STEEP} (dotted lines).
    (i) Density profile 
    $\langle\rho\rangle_\theta$,
    (ii) (radiation) entropy per baryon 
    $\langle s_{\rm rad}\rangle$, with $s_{\rm rad} \equiv 4P_{\rm rad}/P_{\rm gas}$ (gas entropy
    is negligible), 
    (iii) gas temperature $T_g$ (initialized in 
    equilibrium with radiation), 
    (iv) enclosed mass $4\pi\,\int_0^r\,dr\,\,r^2\,\langle\rho\rangle_\theta$,
    (v) breakdown of specific energy to kinetic (due to rotation) and radiation as a fraction of average
    gravitational potential $GM_\bullet\langle\rho\rangle_\theta/r$,
    (vi) total energy per unit mass of the envelope 
    enclosed within radius $r$ in comparison to specific
    energy of stellar debris following tidal disruption
    (Eq.~\eqref{eq:epsilon_t}),
    (vii) local diffusion time $\kappa_{\rm es}\langle\rho\rangle_\theta r^2/c$ in units of
    dynamical time at $r = R_c$ ($\Omega_c^{-1}$),
    in comparison to local dynamical time 
    $t_{\rm dyn} \equiv (GM_\bullet/r^3)^{-1/2}$ 
    (green dashed line),
    (viii) optical depth to electron scattering
    integrated along radial
    direction 
    $\tau(r) = \int_r^\infty\,dr^\prime\,\kappa_{\rm es}\langle\rho\rangle_\theta$,
    (ix) enclosed energies in units of $\epsilon_c$ 
    (green dashed, Eq.~\eqref{eq:epsilon_c}). Black dashed vertical
    lines mark the circularization radius
    $r= R_c$ (Eq.\eqref{eq:Rc}), 
    red dashed vertical lines mark the
    minimum semi-major axis of bound debris streams
    $a_{0}$ (Eq.~\eqref{eq:a_min},
    also the characteristic outer radius of the envelope due to binding energy constraints).
    The envelope
    is in force balance (Eq.~\ref{eq:torus_force_balance} above),
    where gravity is balanced by radiation pressure at
    scales $r \sim a_{0}$, and rotation 
    at scales $r \sim R_c$. The scale
    separation $a_{0} \gg R_c$ is required
    by constraints of total energy/angular momentum 
    available from tidal disruption (Sec.~\ref{sec:analytical_estimates}). 
    Specific angular momentum distribution 
    is taken to be 
    uniform (panel (vi)). Although 
    \texttt{FIDUCIAL} is unstable 
    and \texttt{STEEP} is marginally stable 
    to convection
    in the adiabatic limit according to Hoiland criterion (as can be noted from entropy profiles,
    panel (ii)), both models evolve similarly. (Sec.~\ref{sec:fiducial}, Sec.~\ref{sec:steep}).
    }
    \label{fig:initial_conditions}
\end{figure*}

\bibliographystyle{aasjournal}
\bibliography{refs2}

@article{Lodato+09,
    author = {Lodato, G. and King, A. R. and Pringle, J. E.},
    title = {Stellar disruption by a supermassive black hole: is the light curve really proportional to t−5/3?},
    journal = {Monthly Notices of the Royal Astronomical Society},
    volume = {392},
    number = {1},
    pages = {332-340},
    year = {2008},
    month = {12},
    issn = {0035-8711},
    doi = {10.1111/j.1365-2966.2008.14049.x},
    url = {https://doi.org/10.1111/j.1365-2966.2008.14049.x},
    eprint = {https://academic.oup.com/mnras/article-pdf/392/1/332/3715415/mnras0392-0332.pdf},
}

@ARTICLE{Evans&Kochanek89,
       author = {{Evans}, Charles R. and {Kochanek}, Christopher S.},
        title = "{The Tidal Disruption of a Star by a Massive Black Hole}",
      journal = {\apjl},
         year = 1989,
        month = nov,
       volume = {346},
        pages = {L13},
          doi = {10.1086/185567},
       adsurl = {https://ui.adsabs.harvard.edu/abs/1989ApJ...346L..13E}
}

@ARTICLE{Rees88,
       author = {{Rees}, Martin J.},
        title = "{Tidal disruption of stars by black holes of {}10$^{6}$-{}10$^{8}$ solar masses in nearby galaxies}",
      journal = {\nat},
         year = 1988,
        month = jun,
       volume = {333},
       number = {6173},
        pages = {523-528},
          doi = {10.1038/333523a0},
       adsurl = {https://ui.adsabs.harvard.edu/abs/1988Natur.333..523R}
}

@ARTICLE{Lacy+82,
       author = {{Lacy}, J.~H. and {Townes}, C.~H. and {Hollenbach}, D.~J.},
        title = "{The nature of the central parsec of the Galaxy}",
      journal = {\apj},
         year = 1982,
        month = nov,
       volume = {262},
        pages = {120-134},
          doi = {10.1086/160402},
       adsurl = {https://ui.adsabs.harvard.edu/abs/1982ApJ...262..120L}
}

@ARTICLE{Andalman+22,
       author = {{Andalman}, Zachary L. and {Liska}, Matthew T.~P. and {Tchekhovskoy}, Alexander and {Coughlin}, Eric R. and {Stone}, Nicholas},
        title = "{Tidal disruption discs formed and fed by stream-stream and stream-disc interactions in global GRHD simulations}",
      journal = {\mnras},
         year = 2022,
        month = feb,
       volume = {510},
       number = {2},
        pages = {1627-1648},
          doi = {10.1093/mnras/stab3444},
archivePrefix = {arXiv},
       eprint = {2008.04922},
 primaryClass = {astro-ph.HE},
       adsurl = {https://ui.adsabs.harvard.edu/abs/2022MNRAS.510.1627A}
}

@ARTICLE{Steinberg&Stone24,
       author = {{Steinberg}, Elad and {Stone}, Nicholas C.},
        title = "{Stream-disk shocks as the origins of peak light in tidal disruption events}",
      journal = {\nat},
         year = 2024,
        month = jan,
       volume = {625},
       number = {7995},
        pages = {463-467},
          doi = {10.1038/s41586-023-06875-y},
archivePrefix = {arXiv},
       eprint = {2206.10641},
 primaryClass = {astro-ph.HE},
       adsurl = {https://ui.adsabs.harvard.edu/abs/2024Natur.625..463S}
}

@ARTICLE{Guillochon+14,
       author = {{Guillochon}, James and {Manukian}, Haik and {Ramirez-Ruiz}, Enrico},
        title = "{PS1-10jh: The Disruption of a Main-sequence Star of Near-solar Composition}",
      journal = {\apj},
         year = 2014,
        month = mar,
       volume = {783},
       number = {1},
          eid = {23},
        pages = {23},
          doi = {10.1088/0004-637X/783/1/23},
archivePrefix = {arXiv},
       eprint = {1304.6397},
 primaryClass = {astro-ph.HE},
       adsurl = {https://ui.adsabs.harvard.edu/abs/2014ApJ...783...23G}
}

@ARTICLE{Coughlin&Begelman14,
       author = {{Coughlin}, Eric R. and {Begelman}, Mitchell C.},
        title = "{Hyperaccretion during Tidal Disruption Events: Weakly Bound Debris Envelopes and Jets}",
      journal = {\apj},
         year = 2014,
        month = feb,
       volume = {781},
       number = {2},
          eid = {82},
        pages = {82},
          doi = {10.1088/0004-637X/781/2/82},
archivePrefix = {arXiv},
       eprint = {1312.5314},
 primaryClass = {astro-ph.HE},
       adsurl = {https://ui.adsabs.harvard.edu/abs/2014ApJ...781...82C}
}

@ARTICLE{Frank&Rees76,
       author = {{Frank}, J. and {Rees}, M.~J.},
        title = "{Effects of massive black holes on dense stellar systems.}",
      journal = {\mnras},
         year = 1976,
        month = sep,
       volume = {176},
        pages = {633-647},
          doi = {10.1093/mnras/176.3.633},
       adsurl = {https://ui.adsabs.harvard.edu/abs/1976MNRAS.176..633F}
}

@ARTICLE{Carter&Luminet83,
       author = {{Carter}, B. and {Luminet}, J. -P.},
        title = "{Tidal compression of a star by a large black hole. I Mechanical evolution and nuclear energy release by proton capture}",
      journal = {\aap},
         year = 1983,
        month = may,
       volume = {121},
       number = {1},
        pages = {97-113},
       adsurl = {https://ui.adsabs.harvard.edu/abs/1983A&A...121...97C}
}

@ARTICLE{Roth+16,
       author = {{Roth}, Nathaniel and {Kasen}, Daniel and {Guillochon}, James and {Ramirez-Ruiz}, Enrico},
        title = "{The X-Ray through Optical Fluxes and Line Strengths of Tidal Disruption Events}",
      journal = {\apj},
         year = 2016,
        month = aug,
       volume = {827},
       number = {1},
          eid = {3},
        pages = {3},
          doi = {10.3847/0004-637X/827/1/3},
archivePrefix = {arXiv},
       eprint = {1510.08454},
 primaryClass = {astro-ph.HE},
       adsurl = {https://ui.adsabs.harvard.edu/abs/2016ApJ...827....3R}
}

@ARTICLE{Arcavi+14,
       author = {{Arcavi}, Iair and {Gal-Yam}, Avishay and {Sullivan}, Mark and {Pan}, Yen-Chen and {Cenko}, S. Bradley and {Horesh}, Assaf and {Ofek}, Eran O. and {De Cia}, Annalisa and {Yan}, Lin and {Yang}, Chen-Wei and {Howell}, D.~A. and {Tal}, David and {Kulkarni}, Shrinivas R. and {Tendulkar}, Shriharsh P. and {Tang}, Sumin and {Xu}, Dong and {Sternberg}, Assaf and {Cohen}, Judith G. and {Bloom}, Joshua S. and {Nugent}, Peter E. and {Kasliwal}, Mansi M. and {Perley}, Daniel A. and {Quimby}, Robert M. and {Miller}, Adam A. and {Theissen}, Christopher A. and {Laher}, Russ R.},
        title = "{A Continuum of H- to He-rich Tidal Disruption Candidates With a Preference for E+A Galaxies}",
      journal = {\apj},
         year = 2014,
        month = sep,
       volume = {793},
       number = {1},
          eid = {38},
        pages = {38},
          doi = {10.1088/0004-637X/793/1/38},
archivePrefix = {arXiv},
       eprint = {1405.1415},
 primaryClass = {astro-ph.HE},
       adsurl = {https://ui.adsabs.harvard.edu/abs/2014ApJ...793...38A}
}

@ARTICLE{Stone+13,
       author = {{Stone}, Nicholas and {Sari}, Re'em and {Loeb}, Abraham},
        title = "{Consequences of strong compression in tidal disruption events}",
      journal = {\mnras},
         year = 2013,
        month = nov,
       volume = {435},
       number = {3},
        pages = {1809-1824},
          doi = {10.1093/mnras/stt1270},
archivePrefix = {arXiv},
       eprint = {1210.3374},
 primaryClass = {astro-ph.HE},
       adsurl = {https://ui.adsabs.harvard.edu/abs/2013MNRAS.435.1809S}
}

@ARTICLE{Ulmer99,
       author = {{Ulmer}, Andrew},
        title = "{Flares from the Tidal Disruption of Stars by Massive Black Holes}",
      journal = {\apj},
         year = 1999,
        month = mar,
       volume = {514},
       number = {1},
        pages = {180-187},
          doi = {10.1086/306909},
       adsurl = {https://ui.adsabs.harvard.edu/abs/1999ApJ...514..180U}
}

@ARTICLE{Kochanek94,
       author = {{Kochanek}, Christopher S.},
        title = "{The Aftermath of Tidal Disruption: The Dynamics of Thin Gas Streams}",
      journal = {\apj},
         year = 1994,
        month = feb,
       volume = {422},
        pages = {508},
          doi = {10.1086/173745},
       adsurl = {https://ui.adsabs.harvard.edu/abs/1994ApJ...422..508K}
}

@ARTICLE{Guillochon&RR13,
       author = {{Guillochon}, James and {Ramirez-Ruiz}, Enrico},
        title = "{Hydrodynamical Simulations to Determine the Feeding Rate of Black Holes by the Tidal Disruption of Stars: The Importance of the Impact Parameter and Stellar Structure}",
      journal = {\apj},
         year = 2013,
        month = apr,
       volume = {767},
       number = {1},
          eid = {25},
        pages = {25},
          doi = {10.1088/0004-637X/767/1/25},
archivePrefix = {arXiv},
       eprint = {1206.2350},
 primaryClass = {astro-ph.HE},
       adsurl = {https://ui.adsabs.harvard.edu/abs/2013ApJ...767...25G}
}

@ARTICLE{Strubbe&Quataert09,
       author = {{Strubbe}, Linda E. and {Quataert}, Eliot},
        title = "{Optical flares from the tidal disruption of stars by massive black holes}",
      journal = {\mnras},
         year = 2009,
        month = dec,
       volume = {400},
       number = {4},
        pages = {2070-2084},
          doi = {10.1111/j.1365-2966.2009.15599.x},
archivePrefix = {arXiv},
       eprint = {0905.3735},
 primaryClass = {astro-ph.CO},
       adsurl = {https://ui.adsabs.harvard.edu/abs/2009MNRAS.400.2070S}
}

@ARTICLE{Strubbe&Quataert11,
       author = {{Strubbe}, Linda E. and {Quataert}, Eliot},
        title = "{Spectroscopic signatures of the tidal disruption of stars by massive black holes}",
      journal = {\mnras},
         year = 2011,
        month = jul,
       volume = {415},
       number = {1},
        pages = {168-180},
          doi = {10.1111/j.1365-2966.2011.18686.x},
archivePrefix = {arXiv},
       eprint = {1008.4131},
 primaryClass = {astro-ph.CO},
       adsurl = {https://ui.adsabs.harvard.edu/abs/2011MNRAS.415..168S}
}

@ARTICLE{RR&Rosswog09,
       author = {{Ramirez-Ruiz}, Enrico and {Rosswog}, Stephan},
        title = "{The Star Ingesting Luminosity of Intermediate-Mass Black Holes in Globular Clusters}",
      journal = {\apjl},
         year = 2009,
        month = jun,
       volume = {697},
       number = {2},
        pages = {L77-L80},
          doi = {10.1088/0004-637X/697/2/L77},
archivePrefix = {arXiv},
       eprint = {0808.3847},
 primaryClass = {astro-ph},
       adsurl = {https://ui.adsabs.harvard.edu/abs/2009ApJ...697L..77R}
}

@ARTICLE{Ayal+00,
       author = {{Ayal}, Shai and {Livio}, Mario and {Piran}, Tsvi},
        title = "{Tidal Disruption of a Solar-Type Star by a Supermassive Black Hole}",
      journal = {\apj},
         year = 2000,
        month = dec,
       volume = {545},
       number = {2},
        pages = {772-780},
          doi = {10.1086/317835},
archivePrefix = {arXiv},
       eprint = {astro-ph/0002499},
 primaryClass = {astro-ph},
       adsurl = {https://ui.adsabs.harvard.edu/abs/2000ApJ...545..772A}
}

@ARTICLE{Komossa&Bade99,
       author = {{Komossa}, Stefanie and {Bade}, Norbert},
        title = "{The giant X-ray outbursts in NGC 5905 and IC 3599:() hfill Follow-up observations and outburst scenarios}",
      journal = {\aap},
         year = 1999,
        month = mar,
       volume = {343},
        pages = {775-787},
          doi = {10.48550/arXiv.astro-ph/9901141},
archivePrefix = {arXiv},
       eprint = {astro-ph/9901141},
 primaryClass = {astro-ph},
       adsurl = {https://ui.adsabs.harvard.edu/abs/1999A&A...343..775K}
}

@ARTICLE{Rosswog+09,
       author = {{Rosswog}, S. and {Ramirez-Ruiz}, E. and {Hix}, W.~R.},
        title = "{Tidal Disruption and Ignition of White Dwarfs by Moderately Massive Black Holes}",
      journal = {\apj},
         year = 2009,
        month = apr,
       volume = {695},
       number = {1},
        pages = {404-419},
          doi = {10.1088/0004-637X/695/1/404},
archivePrefix = {arXiv},
       eprint = {0808.2143},
 primaryClass = {astro-ph},
       adsurl = {https://ui.adsabs.harvard.edu/abs/2009ApJ...695..404R}
}

@ARTICLE{Bonnerot&Lu20,
       author = {{Bonnerot}, Cl{\'e}ment and {Lu}, Wenbin},
        title = "{Simulating disc formation in tidal disruption events}",
      journal = {\mnras},
         year = 2020,
        month = jun,
       volume = {495},
       number = {1},
        pages = {1374-1391},
          doi = {10.1093/mnras/staa1246},
archivePrefix = {arXiv},
       eprint = {1906.05865},
 primaryClass = {astro-ph.HE},
       adsurl = {https://ui.adsabs.harvard.edu/abs/2020MNRAS.495.1374B}
}

@ARTICLE{Cannizzo+90,
       author = {{Cannizzo}, John K. and {Lee}, Hyung Mok and {Goodman}, Jeremy},
        title = "{The Disk Accretion of a Tidally Disrupted Star onto a Massive Black Hole}",
      journal = {\apj},
         year = 1990,
        month = mar,
       volume = {351},
        pages = {38},
          doi = {10.1086/168442},
       adsurl = {https://ui.adsabs.harvard.edu/abs/1990ApJ...351...38C}
}

@ARTICLE{Piran+15,
       author = {{Piran}, Tsvi and {Svirski}, Gilad and {Krolik}, Julian and {Cheng}, Roseanne M. and {Shiokawa}, Hotaka},
        title = "{‧Disk Formation Versus Disk Accretion{\textemdash}What Powers Tidal Disruption Events?}",
      journal = {\apj},
         year = 2015,
        month = jun,
       volume = {806},
       number = {2},
          eid = {164},
        pages = {164},
          doi = {10.1088/0004-637X/806/2/164},
archivePrefix = {arXiv},
       eprint = {1502.05792},
 primaryClass = {astro-ph.HE},
       adsurl = {https://ui.adsabs.harvard.edu/abs/2015ApJ...806..164P}
}

@ARTICLE{Hayasaki+16,
       author = {{Hayasaki}, Kimitake and {Stone}, Nicholas and {Loeb}, Abraham},
        title = "{Circularization of tidally disrupted stars around spinning supermassive black holes}",
      journal = {\mnras},
         year = 2016,
        month = oct,
       volume = {461},
       number = {4},
        pages = {3760-3780},
          doi = {10.1093/mnras/stw1387},
archivePrefix = {arXiv},
       eprint = {1501.05207},
 primaryClass = {astro-ph.HE},
       adsurl = {https://ui.adsabs.harvard.edu/abs/2016MNRAS.461.3760H}
}

@ARTICLE{Hayasaki+13,
       author = {{Hayasaki}, Kimitake and {Stone}, Nicholas and {Loeb}, Abraham},
        title = "{Finite, intense accretion bursts from tidal disruption of stars on bound orbits}",
      journal = {\mnras},
         year = 2013,
        month = sep,
       volume = {434},
       number = {2},
        pages = {909-924},
          doi = {10.1093/mnras/stt871},
archivePrefix = {arXiv},
       eprint = {1210.1333},
 primaryClass = {astro-ph.HE},
       adsurl = {https://ui.adsabs.harvard.edu/abs/2013MNRAS.434..909H}
}

@ARTICLE{Guillochon&RR15,
       author = {{Guillochon}, James and {Ramirez-Ruiz}, Enrico},
        title = "{A Dark Year for Tidal Disruption Events}",
      journal = {\apj},
         year = 2015,
        month = aug,
       volume = {809},
       number = {2},
          eid = {166},
        pages = {166},
          doi = {10.1088/0004-637X/809/2/166},
archivePrefix = {arXiv},
       eprint = {1501.05306},
 primaryClass = {astro-ph.HE},
       adsurl = {https://ui.adsabs.harvard.edu/abs/2015ApJ...809..166G}
}

@ARTICLE{Giannios&Metzger11,
       author = {{Giannios}, Dimitrios and {Metzger}, Brian D.},
        title = "{Radio transients from stellar tidal disruption by massive black holes}",
      journal = {\mnras},
         year = 2011,
        month = sep,
       volume = {416},
       number = {3},
        pages = {2102-2107},
          doi = {10.1111/j.1365-2966.2011.19188.x},
archivePrefix = {arXiv},
       eprint = {1102.1429},
 primaryClass = {astro-ph.HE},
       adsurl = {https://ui.adsabs.harvard.edu/abs/2011MNRAS.416.2102G}
}

@ARTICLE{Blandford&Begelman04,
       author = {{Blandford}, Roger D. and {Begelman}, Mitchell C.},
        title = "{Two-dimensional adiabatic flows on to a black hole - I. Fluid accretion}",
      journal = {\mnras},
         year = 2004,
        month = mar,
       volume = {349},
       number = {1},
        pages = {68-86},
          doi = {10.1111/j.1365-2966.2004.07425.x},
archivePrefix = {arXiv},
       eprint = {astro-ph/0306184},
 primaryClass = {astro-ph},
       adsurl = {https://ui.adsabs.harvard.edu/abs/2004MNRAS.349...68B}
}

@ARTICLE{Dai+18,
       author = {{Dai}, Lixin and {McKinney}, Jonathan C. and {Roth}, Nathaniel and {Ramirez-Ruiz}, Enrico and {Miller}, M. Coleman},
        title = "{A Unified Model for Tidal Disruption Events}",
      journal = {\apjl},
         year = 2018,
        month = jun,
       volume = {859},
       number = {2},
          eid = {L20},
        pages = {L20},
          doi = {10.3847/2041-8213/aab429},
archivePrefix = {arXiv},
       eprint = {1803.03265},
 primaryClass = {astro-ph.HE},
       adsurl = {https://ui.adsabs.harvard.edu/abs/2018ApJ...859L..20D}
}

@ARTICLE{Hu+24,
       author = {{Hu}, Fangyi (Fitz) and {Price}, Daniel J. and {Mandel}, Ilya},
        title = "{Optical Appearance of Eccentric Tidal Disruption Events}",
      journal = {\apjl},
         year = 2024,
        month = mar,
       volume = {963},
       number = {1},
          eid = {L27},
        pages = {L27},
          doi = {10.3847/2041-8213/ad29ec},
archivePrefix = {arXiv},
       eprint = {2312.03210},
 primaryClass = {astro-ph.HE},
       adsurl = {https://ui.adsabs.harvard.edu/abs/2024ApJ...963L..27H}
}

@ARTICLE{Shiokawa+15,
       author = {{Shiokawa}, Hotaka and {Krolik}, Julian H. and {Cheng}, Roseanne M. and {Piran}, Tsvi and {Noble}, Scott C.},
        title = "{General Relativistic Hydrodynamic Simulation of Accretion Flow from a Stellar Tidal Disruption}",
      journal = {\apj},
         year = 2015,
        month = may,
       volume = {804},
       number = {2},
          eid = {85},
        pages = {85},
          doi = {10.1088/0004-637X/804/2/85},
archivePrefix = {arXiv},
       eprint = {1501.04365},
 primaryClass = {astro-ph.HE},
       adsurl = {https://ui.adsabs.harvard.edu/abs/2015ApJ...804...85S}
}

@ARTICLE{Metzger22,
       author = {{Metzger}, Brian D.},
        title = "{Cooling Envelope Model for Tidal Disruption Events}",
      journal = {\apjl},
         year = 2022,
        month = sep,
       volume = {937},
       number = {1},
          eid = {L12},
        pages = {L12},
          doi = {10.3847/2041-8213/ac90ba},
archivePrefix = {arXiv},
       eprint = {2207.07136},
 primaryClass = {astro-ph.HE},
       adsurl = {https://ui.adsabs.harvard.edu/abs/2022ApJ...937L..12M}
}

@ARTICLE{Gezari21,
       author = {{Gezari}, Suvi},
        title = "{Tidal Disruption Events}",
      journal = {\araa},
         year = 2021,
        month = sep,
       volume = {59},
        pages = {21-58},
          doi = {10.1146/annurev-astro-111720-030029},
archivePrefix = {arXiv},
       eprint = {2104.14580},
 primaryClass = {astro-ph.HE},
       adsurl = {https://ui.adsabs.harvard.edu/abs/2021ARA&A..59...21G}
}

@ARTICLE{Bonnerot&Stone21,
       author = {{Bonnerot}, C. and {Stone}, N.~C.},
        title = "{Formation of an Accretion Flow}",
      journal = {\ssr},
         year = 2021,
        month = feb,
       volume = {217},
       number = {1},
          eid = {16},
        pages = {16},
          doi = {10.1007/s11214-020-00789-1},
archivePrefix = {arXiv},
       eprint = {2008.11731},
 primaryClass = {astro-ph.HE},
       adsurl = {https://ui.adsabs.harvard.edu/abs/2021SSRv..217...16B}
}

@ARTICLE{Sarin&Metzger24,
       author = {{Sarin}, Nikhil and {Metzger}, Brian D.},
        title = "{Tidal Disruption Events through the Lens of the Cooling Envelope Model}",
      journal = {\apjl},
         year = 2024,
        month = jan,
       volume = {961},
       number = {1},
          eid = {L19},
        pages = {L19},
          doi = {10.3847/2041-8213/ad16d8},
archivePrefix = {arXiv},
       eprint = {2307.15121},
 primaryClass = {astro-ph.HE},
       adsurl = {https://ui.adsabs.harvard.edu/abs/2024ApJ...961L..19S}
}

@ARTICLE{Calderon+24,
       author = {{Calder{\'o}n}, Diego and {Pejcha}, Ond{\v{r}}ej and {Metzger}, Brian D. and {Duffell}, Paul C.},
        title = "{The effect of relativistic precession on light curves of tidal disruption events}",
      journal = {\mnras},
         year = 2024,
        month = feb,
       volume = {528},
       number = {2},
        pages = {2568-2587},
          doi = {10.1093/mnras/stae194},
archivePrefix = {arXiv},
       eprint = {2309.10040},
 primaryClass = {astro-ph.HE},
       adsurl = {https://ui.adsabs.harvard.edu/abs/2024MNRAS.528.2568C}
}

@ARTICLE{Teboul&Metzger23,
       author = {{Teboul}, Odelia and {Metzger}, Brian D.},
        title = "{A Unified Theory of Jetted Tidal Disruption Events: From Promptly Escaping Relativistic to Delayed Transrelativistic Jets}",
      journal = {\apjl},
         year = 2023,
        month = nov,
       volume = {957},
       number = {1},
          eid = {L9},
        pages = {L9},
          doi = {10.3847/2041-8213/ad0037},
archivePrefix = {arXiv},
       eprint = {2308.05161},
 primaryClass = {astro-ph.HE},
       adsurl = {https://ui.adsabs.harvard.edu/abs/2023ApJ...957L...9T}
}

@ARTICLE{Stone&Metzger16,
       author = {{Stone}, Nicholas C. and {Metzger}, Brian D.},
        title = "{Rates of stellar tidal disruption as probes of the supermassive black hole mass function}",
      journal = {\mnras},
         year = 2016,
        month = jan,
       volume = {455},
       number = {1},
        pages = {859-883},
          doi = {10.1093/mnras/stv2281},
archivePrefix = {arXiv},
       eprint = {1410.7772},
 primaryClass = {astro-ph.HE},
       adsurl = {https://ui.adsabs.harvard.edu/abs/2016MNRAS.455..859S}
}

@ARTICLE{Kesden12,
       author = {{Kesden}, Michael},
        title = "{Black-hole spin dependence in the light curves of tidal disruption events}",
      journal = {\prd},
         year = 2012,
        month = sep,
       volume = {86},
       number = {6},
          eid = {064026},
        pages = {064026},
          doi = {10.1103/PhysRevD.86.064026},
archivePrefix = {arXiv},
       eprint = {1207.6401},
 primaryClass = {astro-ph.CO},
       adsurl = {https://ui.adsabs.harvard.edu/abs/2012PhRvD..86f4026K}
}

@ARTICLE{Cheng+14,
       author = {{Cheng}, Roseanne M. and {Bogdanovi{\'c}}, Tamara},
        title = "{Tidal disruption of a star in the Schwarzschild spacetime: Relativistic effects in the return rate of debris}",
      journal = {\prd},
         year = 2014,
        month = sep,
       volume = {90},
       number = {6},
          eid = {064020},
        pages = {064020},
          doi = {10.1103/PhysRevD.90.064020},
archivePrefix = {arXiv},
       eprint = {1407.3266},
 primaryClass = {gr-qc},
       adsurl = {https://ui.adsabs.harvard.edu/abs/2014PhRvD..90f4020C}
}

@ARTICLE{Zhang+25,
       author = {{Zhang}, Lizhong and {Stone}, James M. and {Mullen}, Patrick D. and {Davis}, Shane W. and {Jiang}, Yan-Fei and {White}, Christopher J.},
        title = "{Radiation GRMHD Models of Accretion onto Stellar-Mass Black Holes: I. Survey of Eddington Ratios}",
      journal = {arXiv e-prints},
         year = 2025,
        month = jun,
          eid = {arXiv:2506.02289},
        pages = {arXiv:2506.02289},
          doi = {10.48550/arXiv.2506.02289},
archivePrefix = {arXiv},
       eprint = {2506.02289},
 primaryClass = {astro-ph.HE},
       adsurl = {https://ui.adsabs.harvard.edu/abs/2025arXiv250602289Z}
}

@ARTICLE{Huang+23,
       author = {{Huang}, Xiaoshan and {Davis}, Shane W. and {Jiang}, Yan-fei},
        title = "{A Bright First Day for Tidal Disruption Events}",
      journal = {\apj},
         year = 2023,
        month = aug,
       volume = {953},
       number = {1},
          eid = {117},
        pages = {117},
          doi = {10.3847/1538-4357/ace0be},
archivePrefix = {arXiv},
       eprint = {2303.17443},
 primaryClass = {astro-ph.HE},
       adsurl = {https://ui.adsabs.harvard.edu/abs/2023ApJ...953..117H}
}

@ARTICLE{Meza+25,
       author = {{Meza}, Maria Renee and {Huang}, Xiaoshan and {Davis}, Shane W. and {Jiang}, Yan-Fei},
        title = "{Radiation-magnetohydrodynamic Simulations of Accretion Flow Formation After a Tidal Disruption Event}",
      journal = {arXiv e-prints},
         year = 2025,
        month = may,
          eid = {arXiv:2506.00109},
        pages = {arXiv:2506.00109},
          doi = {10.48550/arXiv.2506.00109},
archivePrefix = {arXiv},
       eprint = {2506.00109},
 primaryClass = {astro-ph.HE},
       adsurl = {https://ui.adsabs.harvard.edu/abs/2025arXiv250600109M}
}

@ARTICLE{Mummery&VanVelzen24,
       author = {{Mummery}, Andrew and {van Velzen}, Sjoert},
        title = "{The optical, UV-plateau and X-ray tidal disruption event luminosity functions reproduced from first principles}",
      journal = {arXiv e-prints},
         year = 2024,
        month = oct,
          eid = {arXiv:2410.17087},
        pages = {arXiv:2410.17087},
          doi = {10.48550/arXiv.2410.17087},
archivePrefix = {arXiv},
       eprint = {2410.17087},
 primaryClass = {astro-ph.HE},
       adsurl = {https://ui.adsabs.harvard.edu/abs/2024arXiv241017087M}
}

@ARTICLE{Metzger&Stone16,
       author = {{Metzger}, Brian D. and {Stone}, Nicholas C.},
        title = "{A bright year for tidal disruptions}",
      journal = {\mnras},
         year = 2016,
        month = sep,
       volume = {461},
       number = {1},
        pages = {948-966},
          doi = {10.1093/mnras/stw1394},
archivePrefix = {arXiv},
       eprint = {1506.03453},
 primaryClass = {astro-ph.HE},
       adsurl = {https://ui.adsabs.harvard.edu/abs/2016MNRAS.461..948M}
}

@ARTICLE{Quataert+16,
       author = {{Quataert}, Eliot and {Fern{\'a}ndez}, Rodrigo and {Kasen}, Daniel and {Klion}, Hannah and {Paxton}, Bill},
        title = "{Super-Eddington stellar winds driven by near-surface energy deposition}",
      journal = {\mnras},
         year = 2016,
        month = may,
       volume = {458},
       number = {2},
        pages = {1214-1233},
          doi = {10.1093/mnras/stw365},
archivePrefix = {arXiv},
       eprint = {1509.06370},
 primaryClass = {astro-ph.SR},
       adsurl = {https://ui.adsabs.harvard.edu/abs/2016MNRAS.458.1214Q}
}

@ARTICLE{Loeb&Ulmer97,
       author = {{Loeb}, Abraham and {Ulmer}, Andrew},
        title = "{Optical Appearance of the Debris of a Star Disrupted by a Massive Black Hole}",
      journal = {\apj},
         year = 1997,
        month = nov,
       volume = {489},
       number = {2},
        pages = {573-578},
          doi = {10.1086/304814},
archivePrefix = {arXiv},
       eprint = {astro-ph/9703079},
 primaryClass = {astro-ph},
       adsurl = {https://ui.adsabs.harvard.edu/abs/1997ApJ...489..573L}
}

@article{Curd&Narayan19,
    author = "Curd, Brandon and Narayan, Ramesh",
    title = "{GRRMHD simulations of tidal disruption event accretion discs around supermassive black holes: jet formation, spectra, and detectability}",
    eprint = "1811.06971",
    archivePrefix = "arXiv",
    primaryClass = "astro-ph.HE",
    doi = "10.1093/mnras/sty3134",
    journal = "Mon. Not. Roy. Astron. Soc.",
    volume = "483",
    number = "1",
    pages = "565--592",
    year = "2019"
}

@ARTICLE{Thomsen+22,
       author = {{Thomsen}, Lars L. and {Kwan}, Tom M. and {Dai}, Lixin and {Wu}, Samantha C. and {Roth}, Nathaniel and {Ramirez-Ruiz}, Enrico},
        title = "{Dynamical Unification of Tidal Disruption Events}",
      journal = {\apjl},
         year = 2022,
        month = oct,
       volume = {937},
       number = {2},
          eid = {L28},
        pages = {L28},
          doi = {10.3847/2041-8213/ac911f},
archivePrefix = {arXiv},
       eprint = {2206.02804},
 primaryClass = {astro-ph.HE},
       adsurl = {https://ui.adsabs.harvard.edu/abs/2022ApJ...937L..28T}
}

@ARTICLE{Sadowski+16,
       author = {{S{{a}}dowski}, Aleksander and {Tejeda}, Emilio and {Gafton}, Emanuel and {Rosswog}, Stephan and {Abarca}, David},
        title = "{Magnetohydrodynamical simulations of a deep tidal disruption in general relativity}",
      journal = {\mnras},
         year = 2016,
        month = jun,
       volume = {458},
       number = {4},
        pages = {4250-4268},
          doi = {10.1093/mnras/stw589},
archivePrefix = {arXiv},
       eprint = {1512.04865},
 primaryClass = {astro-ph.HE},
       adsurl = {https://ui.adsabs.harvard.edu/abs/2016MNRAS.458.4250S}
}

@ARTICLE{Curd21,
       author = {{Curd}, Brandon},
        title = "{Global simulations of tidal disruption event disc formation via stream injection in GRRMHD}",
      journal = {\mnras},
         year = 2021,
        month = nov,
       volume = {507},
       number = {3},
        pages = {3207-3227},
          doi = {10.1093/mnras/stab2172},
archivePrefix = {arXiv},
       eprint = {2105.09904},
 primaryClass = {astro-ph.HE},
       adsurl = {https://ui.adsabs.harvard.edu/abs/2021MNRAS.507.3207C}
}

@article{Hammerstein+23,
    author = "Hammerstein, Erica and others",
    title = "{The Final Season Reimagined: 30 Tidal Disruption Events from the ZTF-I Survey}",
    eprint = "2203.01461",
    archivePrefix = "arXiv",
    primaryClass = "astro-ph.HE",
    doi = "10.3847/1538-4357/aca283",
    journal = "Astrophys. J.",
    volume = "942",
    number = "1",
    pages = "9",
    year = "2023"
}

@ARTICLE{Mainetti+17,
       author = {{Mainetti}, Deborah and {Lupi}, Alessandro and {Campana}, Sergio and {Colpi}, Monica and {Coughlin}, Eric R. and {Guillochon}, James and {Ramirez-Ruiz}, Enrico},
        title = "{The fine line between total and partial tidal disruption events}",
      journal = {\aap},
         year = 2017,
        month = apr,
       volume = {600},
          eid = {A124},
        pages = {A124},
          doi = {10.1051/0004-6361/201630092},
archivePrefix = {arXiv},
       eprint = {1702.07730},
 primaryClass = {astro-ph.HE},
       adsurl = {https://ui.adsabs.harvard.edu/abs/2017A&A...600A.124M}
}

@ARTICLE{Lu&Bonnerot20,
       author = {{Lu}, Wenbin and {Bonnerot}, Cl{\'e}ment},
        title = "{Self-intersection of the fallback stream in tidal disruption events}",
      journal = {\mnras},
         year = 2020,
        month = feb,
       volume = {492},
       number = {1},
        pages = {686-707},
          doi = {10.1093/mnras/stz3405},
archivePrefix = {arXiv},
       eprint = {1904.12018},
 primaryClass = {astro-ph.HE},
       adsurl = {https://ui.adsabs.harvard.edu/abs/2020MNRAS.492..686L}
}

@ARTICLE{Bonnerot+16,
       author = {{Bonnerot}, Cl{\'e}ment and {Rossi}, Elena M. and {Lodato}, Giuseppe and {Price}, Daniel J.},
        title = "{Disc formation from tidal disruptions of stars on eccentric orbits by Schwarzschild black holes}",
      journal = {\mnras},
         year = 2016,
        month = jan,
       volume = {455},
       number = {2},
        pages = {2253-2266},
          doi = {10.1093/mnras/stv2411},
archivePrefix = {arXiv},
       eprint = {1501.04635},
 primaryClass = {astro-ph.HE},
       adsurl = {https://ui.adsabs.harvard.edu/abs/2016MNRAS.455.2253B}
}

@ARTICLE{Bonnerot+21,
       author = {{Bonnerot}, Cl{\'e}ment and {Lu}, Wenbin and {Hopkins}, Philip F.},
        title = "{First light from tidal disruption events}",
      journal = {\mnras},
         year = 2021,
        month = jul,
       volume = {504},
       number = {4},
        pages = {4885-4905},
          doi = {10.1093/mnras/stab398},
archivePrefix = {arXiv},
       eprint = {2012.12271},
 primaryClass = {astro-ph.HE},
       adsurl = {https://ui.adsabs.harvard.edu/abs/2021MNRAS.504.4885B}
}

@article{Liptai+19,
    author = "Liptai, David and Price, Daniel J. and Mandel, Ilya and Lodato, Giuseppe",
    title = "{Disc formation from tidal disruption of stars on eccentric orbits by Kerr black holes using GRSPH}",
    eprint = "1910.10154",
    archivePrefix = "arXiv",
    primaryClass = "astro-ph.HE",
    month = "10",
    year = "2019"
}

@ARTICLE{Price+24,
       author = {{Price}, Daniel J. and {Liptai}, David and {Mandel}, Ilya and {Shepherd}, Joanna and {Lodato}, Giuseppe and {Levin}, Yuri},
        title = "{Eddington Envelopes: The Fate of Stars on Parabolic Orbits Tidally Disrupted by Supermassive Black Holes}",
      journal = {\apjl},
         year = 2024,
        month = aug,
       volume = {971},
       number = {2},
          eid = {L46},
        pages = {L46},
          doi = {10.3847/2041-8213/ad6862},
archivePrefix = {arXiv},
       eprint = {2404.09381},
 primaryClass = {astro-ph.HE},
       adsurl = {https://ui.adsabs.harvard.edu/abs/2024ApJ...971L..46P}
}

@ARTICLE{Bonnerot&Lu22,
       author = {{Bonnerot}, Cl{\'e}ment and {Lu}, Wenbin},
        title = "{The nozzle shock in tidal disruption events}",
      journal = {\mnras},
         year = 2022,
        month = apr,
       volume = {511},
       number = {2},
        pages = {2147-2169},
          doi = {10.1093/mnras/stac146},
archivePrefix = {arXiv},
       eprint = {2106.01376},
 primaryClass = {astro-ph.HE},
       adsurl = {https://ui.adsabs.harvard.edu/abs/2022MNRAS.511.2147B}
}

@article{Kato+04,
    author = "Kato, Yoshiaki and Mineshige, Shin and Shibata, Kazunari",
    title = "{Magnetohydrodynamical accretion flows: Formation of magnetic tower jet and subsequent quasi-steady state}",
    eprint = "astro-ph/0307306",
    archivePrefix = "arXiv",
    reportNumber = "ASTRO-PH-YKATO-29686",
    doi = "10.1086/381234",
    journal = "Astrophys. J.",
    volume = "605",
    pages = "307--320",
    year = "2004"
}

@ARTICLE{Jiang+17,
       author = {{Jiang}, Yan-Fei and {Stone}, James M. and {Davis}, Shane W.},
        title = "{Super-Eddington Accretion Disks around Supermassive Black Holes}",
      journal = {\apj},
         year = 2019,
        month = aug,
       volume = {880},
       number = {2},
          eid = {67},
        pages = {67},
          doi = {10.3847/1538-4357/ab29ff},
archivePrefix = {arXiv},
       eprint = {1709.02845},
 primaryClass = {astro-ph.HE},
       adsurl = {https://ui.adsabs.harvard.edu/abs/2019ApJ...880...67J}
}

@ARTICLE{Ryu+20a,
       author = {{Ryu}, Taeho and {Krolik}, Julian and {Piran}, Tsvi and {Noble}, Scott C.},
        title = "{Tidal Disruptions of Main-sequence Stars. I. Observable Quantities and Their Dependence on Stellar and Black Hole Mass}",
      journal = {\apj},
         year = 2020,
        month = dec,
       volume = {904},
       number = {2},
          eid = {98},
        pages = {98},
          doi = {10.3847/1538-4357/abb3cf},
archivePrefix = {arXiv},
       eprint = {2001.03501},
 primaryClass = {astro-ph.HE},
       adsurl = {https://ui.adsabs.harvard.edu/abs/2020ApJ...904...98R}
}

@ARTICLE{Ryu+20b,
       author = {{Ryu}, Taeho and {Krolik}, Julian and {Piran}, Tsvi and {Noble}, Scott C.},
        title = "{Tidal Disruptions of Main-sequence Stars. II. Simulation Methodology and Stellar Mass Dependence of the Character of Full Tidal Disruptions}",
      journal = {\apj},
         year = 2020,
        month = dec,
       volume = {904},
       number = {2},
          eid = {99},
        pages = {99},
          doi = {10.3847/1538-4357/abb3cd},
archivePrefix = {arXiv},
       eprint = {2001.03502},
 primaryClass = {astro-ph.HE},
       adsurl = {https://ui.adsabs.harvard.edu/abs/2020ApJ...904...99R}
}

@ARTICLE{Hills75,
       author = {{Hills}, J.~G.},
        title = "{Possible power source of Seyfert galaxies and QSOs}",
      journal = {\nat},
         year = 1975,
        month = mar,
       volume = {254},
       number = {5498},
        pages = {295-298},
          doi = {10.1038/254295a0},
       adsurl = {https://ui.adsabs.harvard.edu/abs/1975Natur.254..295H}
}

@ARTICLE{Lidskii&Ozernoi79,
       author = {{Lidskii}, V.~V. and {Ozernoi}, L.~M.},
        title = "{Tidal triggering of stellar flares by a massive black hole}",
      journal = {Soviet Astronomy Letters},
         year = 1979,
        month = jan,
       volume = {5},
        pages = {16-19},
       adsurl = {https://ui.adsabs.harvard.edu/abs/1979SvAL....5...16L}
}

@INPROCEEDINGS{Phinney89,
       author = {{Phinney}, E.~S.},
        title = "{Manifestations of a Massive Black Hole in the Galactic Center}",
    booktitle = {The Center of the Galaxy},
         year = 1989,
       editor = {{Morris}, Mark},
       series = {IAU Symposium},
       volume = {136},
        month = jan,
        pages = {543},
       adsurl = {https://ui.adsabs.harvard.edu/abs/1989IAUS..136..543P}
}

@ARTICLE{Hammerstein+23a,
       author = {{Hammerstein}, Erica and {van Velzen}, Sjoert and {Gezari}, Suvi and {Cenko}, S. Bradley and {Yao}, Yuhan and {Ward}, Charlotte and {Frederick}, Sara and {Villanueva}, Natalia and {Somalwar}, Jean J. and {Graham}, Matthew J. and {Kulkarni}, Shrinivas R. and {Stern}, Daniel and {Andreoni}, Igor and {Bellm}, Eric C. and {Dekany}, Richard and {Dhawan}, Suhail and {Drake}, Andrew J. and {Fremling}, Christoffer and {Gatkine}, Pradip and {Groom}, Steven L. and {Ho}, Anna Y.~Q. and {Kasliwal}, Mansi M. and {Karambelkar}, Viraj and {Kool}, Erik C. and {Masci}, Frank J. and {Medford}, Michael S. and {Perley}, Daniel A. and {Purdum}, Josiah and {van Roestel}, Jan and {Sharma}, Yashvi and {Sollerman}, Jesper and {Taggart}, Kirsty and {Yan}, Lin},
        title = "{The Final Season Reimagined: 30 Tidal Disruption Events from the ZTF-I Survey}",
      journal = {\apj},
         year = 2023,
        month = jan,
       volume = {942},
       number = {1},
          eid = {9},
        pages = {9},
          doi = {10.3847/1538-4357/aca283},
archivePrefix = {arXiv},
       eprint = {2203.01461},
 primaryClass = {astro-ph.HE},
       adsurl = {https://ui.adsabs.harvard.edu/abs/2023ApJ...942....9H}
}

@ARTICLE{Gezari+17,
       author = {{Gezari}, S. and {Cenko}, S.~B. and {Arcavi}, I.},
        title = "{X-Ray Brightening and UV Fading of Tidal Disruption Event ASASSN-15oi}",
      journal = {\apjl},
         year = 2017,
        month = dec,
       volume = {851},
       number = {2},
          eid = {L47},
        pages = {L47},
          doi = {10.3847/2041-8213/aaa0c2},
archivePrefix = {arXiv},
       eprint = {1712.03968},
 primaryClass = {astro-ph.HE},
       adsurl = {https://ui.adsabs.harvard.edu/abs/2017ApJ...851L..47G}
}

@ARTICLE{vanVelzen+19,
       author = {{van Velzen}, Sjoert and {Stone}, Nicholas C. and {Metzger}, Brian D. and {Gezari}, Suvi and {Brown}, Thomas M. and {Fruchter}, Andrew S.},
        title = "{Late-time UV Observations of Tidal Disruption Flares Reveal Unobscured, Compact Accretion Disks}",
      journal = {\apj},
         year = 2019,
        month = jun,
       volume = {878},
       number = {2},
          eid = {82},
        pages = {82},
          doi = {10.3847/1538-4357/ab1844},
archivePrefix = {arXiv},
       eprint = {1809.00003},
 primaryClass = {astro-ph.HE},
       adsurl = {https://ui.adsabs.harvard.edu/abs/2019ApJ...878...82V}
}

@ARTICLE{Mummery+24,
       author = {{Mummery}, Andrew and {van Velzen}, Sjoert and {Nathan}, Edward and {Ingram}, Adam and {Hammerstein}, Erica and {Fraser-Taliente}, Ludovic and {Balbus}, Steven},
        title = "{Fundamental scaling relationships revealed in the optical light curves of tidal disruption events}",
      journal = {\mnras},
         year = 2024,
        month = jan,
       volume = {527},
       number = {2},
        pages = {2452-2489},
          doi = {10.1093/mnras/stad3001},
archivePrefix = {arXiv},
       eprint = {2308.08255},
 primaryClass = {astro-ph.HE},
       adsurl = {https://ui.adsabs.harvard.edu/abs/2024MNRAS.527.2452M}
}

@ARTICLE{Shen&Matzner14,
       author = {{Shen}, Rong-Feng and {Matzner}, Christopher D.},
        title = "{Evolution of Accretion Disks in Tidal Disruption Events}",
      journal = {\apj},
         year = 2014,
        month = apr,
       volume = {784},
       number = {2},
          eid = {87},
        pages = {87},
          doi = {10.1088/0004-637X/784/2/87},
archivePrefix = {arXiv},
       eprint = {1305.5570},
 primaryClass = {astro-ph.HE},
       adsurl = {https://ui.adsabs.harvard.edu/abs/2014ApJ...784...87S}
}

@ARTICLE{Nolthenius&Katz82,
       author = {{Nolthenius}, R.~A. and {Katz}, J.~I.},
        title = "{The passage of a star by a massive black hole}",
      journal = {\apj},
         year = 1982,
        month = dec,
       volume = {263},
        pages = {377-385},
          doi = {10.1086/160511},
       adsurl = {https://ui.adsabs.harvard.edu/abs/1982ApJ...263..377N}
}

@ARTICLE{Miller15,
       author = {{Miller}, M. Coleman},
        title = "{Disk Winds as an Explanation for Slowly Evolving Temperatures in Tidal Disruption Events}",
      journal = {\apj},
         year = 2015,
        month = may,
       volume = {805},
       number = {1},
          eid = {83},
        pages = {83},
          doi = {10.1088/0004-637X/805/1/83},
archivePrefix = {arXiv},
       eprint = {1502.03284},
 primaryClass = {astro-ph.GA},
       adsurl = {https://ui.adsabs.harvard.edu/abs/2015ApJ...805...83M}
}

@ARTICLE{Bade+96,
       author = {{Bade}, N. and {Komossa}, S. and {Dahlem}, M.},
        title = "{Detection of an extremely soft X-ray outburst in the HII-like nucleus of NGC 5905.}",
      journal = {\aap},
         year = 1996,
        month = may,
       volume = {309},
        pages = {L35-L38},
       adsurl = {https://ui.adsabs.harvard.edu/abs/1996A&A...309L..35B}
}

@ARTICLE{Sazonov+21,
       author = {{Sazonov}, S. and {Gilfanov}, M. and {Medvedev}, P. and {Yao}, Y. and {Khorunzhev}, G. and {Semena}, A. and {Sunyaev}, R. and {Burenin}, R. and {Lyapin}, A. and {Meshcheryakov}, A. and {Uskov}, G. and {Zaznobin}, I. and {Postnov}, K.~A. and {Dodin}, A.~V. and {Belinski}, A.~A. and {Cherepashchuk}, A.~M. and {Eselevich}, M. and {Dodonov}, S.~N. and {Grokhovskaya}, A.~A. and {Kotov}, S.~S. and {Bikmaev}, I.~F. and {Zhuchkov}, R. Ya and {Gumerov}, R.~I. and {van Velzen}, S. and {Kulkarni}, S.},
        title = "{First tidal disruption events discovered by SRG/eROSITA: X-ray/optical properties and X-ray luminosity function at z < 0.6}",
      journal = {\mnras},
         year = 2021,
        month = dec,
       volume = {508},
       number = {3},
        pages = {3820-3847},
          doi = {10.1093/mnras/stab2843},
archivePrefix = {arXiv},
       eprint = {2108.02449},
 primaryClass = {astro-ph.HE},
       adsurl = {https://ui.adsabs.harvard.edu/abs/2021MNRAS.508.3820S}
}

@ARTICLE{vanVelzen+11,
       author = {{van Velzen}, Sjoert and {Farrar}, Glennys R. and {Gezari}, Suvi and {Morrell}, Nidia and {Zaritsky}, Dennis and {{\"O}stman}, Linda and {Smith}, Mathew and {Gelfand}, Joseph and {Drake}, Andrew J.},
        title = "{Optical Discovery of Probable Stellar Tidal Disruption Flares}",
      journal = {\apj},
         year = 2011,
        month = nov,
       volume = {741},
       number = {2},
          eid = {73},
        pages = {73},
          doi = {10.1088/0004-637X/741/2/73},
archivePrefix = {arXiv},
       eprint = {1009.1627},
 primaryClass = {astro-ph.CO},
       adsurl = {https://ui.adsabs.harvard.edu/abs/2011ApJ...741...73V}
}

@ARTICLE{Jonker+20,
       author = {{Jonker}, P.~G. and {Stone}, N.~C. and {Generozov}, A. and {van Velzen}, S. and {Metzger}, B.},
        title = "{Implications from Late-time X-Ray Detections of Optically Selected Tidal Disruption Events: State Changes, Unification, and Detection Rates}",
      journal = {\apj},
         year = 2020,
        month = feb,
       volume = {889},
       number = {2},
          eid = {166},
        pages = {166},
          doi = {10.3847/1538-4357/ab659c},
archivePrefix = {arXiv},
       eprint = {1906.12236},
 primaryClass = {astro-ph.HE},
       adsurl = {https://ui.adsabs.harvard.edu/abs/2020ApJ...889..166J}
}

@ARTICLE{Stern+04,
       author = {{Stern}, Daniel and {van Dokkum}, P.~G. and {Nugent}, Peter and {Sand}, D.~J. and {Ellis}, R.~S. and {Sullivan}, Mark and {Bloom}, J.~S. and {Frail}, D.~A. and {Kneib}, J. -P. and {Koopmans}, L.~V.~E. and {Treu}, Tommaso},
        title = "{Discovery of a Transient U-Band Dropout in a Lyman Break Survey: A Tidally Disrupted Star at z=3.3?}",
      journal = {\apj},
         year = 2004,
        month = sep,
       volume = {612},
       number = {2},
        pages = {690-697},
          doi = {10.1086/422744},
archivePrefix = {arXiv},
       eprint = {astro-ph/0405482},
 primaryClass = {astro-ph},
       adsurl = {https://ui.adsabs.harvard.edu/abs/2004ApJ...612..690S}
}

@ARTICLE{Gezari+06,
       author = {{Gezari}, S. and {Martin}, D.~C. and {Milliard}, B. and {Basa}, S. and {Halpern}, J.~P. and {Forster}, K. and {Friedman}, P.~G. and {Morrissey}, P. and {Neff}, S.~G. and {Schiminovich}, D. and {Seibert}, M. and {Small}, T. and {Wyder}, T.~K.},
        title = "{Ultraviolet Detection of the Tidal Disruption of a Star by a Supermassive Black Hole}",
      journal = {\apjl},
         year = 2006,
        month = dec,
       volume = {653},
       number = {1},
        pages = {L25-L28},
          doi = {10.1086/509918},
archivePrefix = {arXiv},
       eprint = {astro-ph/0612069},
 primaryClass = {astro-ph},
       adsurl = {https://ui.adsabs.harvard.edu/abs/2006ApJ...653L..25G}
}

@ARTICLE{Cenko+12,
       author = {{Cenko}, S. Bradley and {Krimm}, Hans A. and {Horesh}, Assaf and {Rau}, Arne and {Frail}, Dale A. and {Kennea}, Jamie A. and {Levan}, Andrew J. and {Holland}, Stephen T. and {Butler}, Nathaniel R. and {Quimby}, Robert M. and {Bloom}, Joshua S. and {Filippenko}, Alexei V. and {Gal-Yam}, Avishay and {Greiner}, Jochen and {Kulkarni}, S.~R. and {Ofek}, Eran O. and {Olivares E.}, Felipe and {Schady}, Patricia and {Silverman}, Jeffrey M. and {Tanvir}, Nial R. and {Xu}, Dong},
        title = "{Swift J2058.4+0516: Discovery of a Possible Second Relativistic Tidal Disruption Flare?}",
      journal = {\apj},
         year = 2012,
        month = jul,
       volume = {753},
       number = {1},
          eid = {77},
        pages = {77},
          doi = {10.1088/0004-637X/753/1/77},
archivePrefix = {arXiv},
       eprint = {1107.5307},
 primaryClass = {astro-ph.HE},
       adsurl = {https://ui.adsabs.harvard.edu/abs/2012ApJ...753...77C}
}

@ARTICLE{Ryu+20,
       author = {{Ryu}, Taeho and {Krolik}, Julian and {Piran}, Tsvi},
        title = "{Measuring Stellar and Black Hole Masses of Tidal Disruption Events}",
      journal = {\apj},
         year = 2020,
        month = nov,
       volume = {904},
       number = {1},
          eid = {73},
        pages = {73},
          doi = {10.3847/1538-4357/abbf4d},
archivePrefix = {arXiv},
       eprint = {2007.13765},
 primaryClass = {astro-ph.HE},
       adsurl = {https://ui.adsabs.harvard.edu/abs/2020ApJ...904...73R}
}

@ARTICLE{Papaloizou&Pringle84,
       author = {{Papaloizou}, J.~C.~B. and {Pringle}, J.~E.},
        title = "{The dynamical stability of differentially rotating discs with constant specific angular momentum}",
      journal = {\mnras},
         year = 1984,
        month = jun,
       volume = {208},
        pages = {721-750},
          doi = {10.1093/mnras/208.4.721},
       adsurl = {https://ui.adsabs.harvard.edu/abs/1984MNRAS.208..721P}
}

@ARTICLE{Stone+20,
       author = {{Stone}, James M. and {Tomida}, Kengo and {White}, Christopher J. and {Felker}, Kyle G.},
        title = "{The Athena++ Adaptive Mesh Refinement Framework: Design and Magnetohydrodynamic Solvers}",
      journal = {\apjs},
         year = 2020,
        month = jul,
       volume = {249},
       number = {1},
          eid = {4},
        pages = {4},
          doi = {10.3847/1538-4365/ab929b},
archivePrefix = {arXiv},
       eprint = {2005.06651},
 primaryClass = {astro-ph.IM},
       adsurl = {https://ui.adsabs.harvard.edu/abs/2020ApJS..249....4S}
}

@ARTICLE{Liu+22,
       author = {{Liu}, Xiao-Long and {Dou}, Li-Ming and {Chen}, Jin-Hong and {Shen}, Rong-Feng},
        title = "{The UV/Optical Peak and X-Ray Brightening in TDE Candidate AT 2019azh: A Case of Stream-Stream Collision and Delayed Accretion}",
      journal = {\apj},
         year = 2022,
        month = jan,
       volume = {925},
       number = {1},
          eid = {67},
        pages = {67},
          doi = {10.3847/1538-4357/ac33a9},
archivePrefix = {arXiv},
       eprint = {1912.06081},
 primaryClass = {astro-ph.HE},
       adsurl = {https://ui.adsabs.harvard.edu/abs/2022ApJ...925...67L}
}

@ARTICLE{Jiang+16,
       author = {{Jiang}, Ning and {Dou}, Liming and {Wang}, Tinggui and {Yang}, Chenwei and {Lyu}, Jianwei and {Zhou}, Hongyan},
        title = "{The WISE Detection of an Infrared Echo in Tidal Disruption Event ASASSN-14li}",
      journal = {\apjl},
         year = 2016,
        month = sep,
       volume = {828},
       number = {1},
          eid = {L14},
        pages = {L14},
          doi = {10.3847/2041-8205/828/1/L14},
archivePrefix = {arXiv},
       eprint = {1605.04640},
 primaryClass = {astro-ph.HE},
       adsurl = {https://ui.adsabs.harvard.edu/abs/2016ApJ...828L..14J}
}

@ARTICLE{Kozlowski+78,
       author = {{Kozlowski}, M. and {Jaroszynski}, M. and {Abramowicz}, M.~A.},
        title = "{The analytic theory of fluid disks orbiting the Kerr black hole.}",
      journal = {\aap},
         year = 1978,
        month = feb,
       volume = {63},
       number = {1-2},
        pages = {209-220},
       adsurl = {https://ui.adsabs.harvard.edu/abs/1978A&A....63..209K}
}

@ARTICLE{Davis&Gammie20,
       author = {{Davis}, Shane W. and {Gammie}, Charles F.},
        title = "{Covariant Radiative Transfer for Black Hole Spacetimes}",
      journal = {\apj},
         year = 2020,
        month = jan,
       volume = {888},
       number = {2},
          eid = {94},
        pages = {94},
          doi = {10.3847/1538-4357/ab5950},
archivePrefix = {arXiv},
       eprint = {1911.07950},
 primaryClass = {astro-ph.HE},
       adsurl = {https://ui.adsabs.harvard.edu/abs/2020ApJ...888...94D}
}

@ARTICLE{Caproni+23,
       author = {{Caproni}, Anderson and {Lanfranchi}, Gustavo A. and {Fria{\c{c}}a}, Am{\^a}ncio C.~S. and {Soares}, Jennifer F.},
        title = "{Boundary Conditions in Hydrodynamic Simulations of Isolated Galaxies and Their Impact on the Gas-loss Processes}",
      journal = {\apj},
         year = 2023,
        month = feb,
       volume = {944},
       number = {1},
          eid = {11},
        pages = {11},
          doi = {10.3847/1538-4357/acae85},
archivePrefix = {arXiv},
       eprint = {2302.04825},
 primaryClass = {astro-ph.GA},
       adsurl = {https://ui.adsabs.harvard.edu/abs/2023ApJ...944...11C}
}

@ARTICLE{Jiang+14b,
       author = {{Jiang}, Yan-Fei and {Stone}, James M. and {Davis}, Shane W.},
        title = "{An Algorithm for Radiation Magnetohydrodynamics Based on Solving the Time-dependent Transfer Equation}",
      journal = {\apjs},
         year = 2014,
        month = jul,
       volume = {213},
       number = {1},
          eid = {7},
        pages = {7},
          doi = {10.1088/0067-0049/213/1/7},
archivePrefix = {arXiv},
       eprint = {1403.6126},
 primaryClass = {astro-ph.IM},
       adsurl = {https://ui.adsabs.harvard.edu/abs/2014ApJS..213....7J}
}

@ARTICLE{Jiang21,
       author = {{Jiang}, Yan-Fei},
        title = "{An Implicit Finite Volume Scheme to Solve the Time-dependent Radiation Transport Equation Based on Discrete Ordinates}",
      journal = {\apjs},
         year = 2021,
        month = apr,
       volume = {253},
       number = {2},
          eid = {49},
        pages = {49},
          doi = {10.3847/1538-4365/abe303},
archivePrefix = {arXiv},
       eprint = {2102.02212},
 primaryClass = {astro-ph.IM},
       adsurl = {https://ui.adsabs.harvard.edu/abs/2021ApJS..253...49J}
}

@article{Charalampopoulos+21,
    author = "Charalampopoulos, P. and others",
    title = "{A detailed spectroscopic study of tidal disruption events}",
    eprint = "2109.00016",
    archivePrefix = "arXiv",
    primaryClass = "astro-ph.HE",
    doi = "10.1051/0004-6361/202142122",
    journal = "Astron. Astrophys.",
    volume = "659",
    pages = "A34",
    year = "2022"
}

@BOOK{Schwarzschild58,
       author = {{Schwarzschild}, Martin},
        title = "{Structure and evolution of the stars.}",
         year = 1958,
       adsurl = {https://ui.adsabs.harvard.edu/abs/1958ses..book.....S}
}

@ARTICLE{Shaviv98,
       author = {{Shaviv}, Nir J.},
        title = "{The Eddington Luminosity Limit for Multiphased Media}",
      journal = {\apjl},
         year = 1998,
        month = feb,
       volume = {494},
       number = {2},
        pages = {L193-L197},
          doi = {10.1086/311182},
       adsurl = {https://ui.adsabs.harvard.edu/abs/1998ApJ...494L.193S}
}

@ARTICLE{Shaviv00,
       author = {{Shaviv}, Nir J.},
        title = "{The Porous Atmosphere of {\ensuremath{\eta}} Carinae}",
      journal = {\apjl},
         year = 2000,
        month = apr,
       volume = {532},
       number = {2},
        pages = {L137-L140},
          doi = {10.1086/312585},
archivePrefix = {arXiv},
       eprint = {astro-ph/0002212},
 primaryClass = {astro-ph},
       adsurl = {https://ui.adsabs.harvard.edu/abs/2000ApJ...532L.137S}
}

@ARTICLE{Kiriakidis+93,
       author = {{Kiriakidis}, M. and {Fricke}, K.~J. and {Glatzel}, W.},
        title = "{The stability of massive stars and its dependence on metallicity and opacity.}",
      journal = {\mnras},
         year = 1993,
        month = sep,
       volume = {264},
        pages = {50-62},
          doi = {10.1093/mnras/264.1.50},
       adsurl = {https://ui.adsabs.harvard.edu/abs/1993MNRAS.264...50K}
}

@ARTICLE{Blaes&Socrates03,
       author = {{Blaes}, Omer and {Socrates}, Aristotle},
        title = "{Local Radiative Hydrodynamic and Magnetohydrodynamic Instabilities in Optically Thick Media}",
      journal = {\apj},
         year = 2003,
        month = oct,
       volume = {596},
       number = {1},
        pages = {509-537},
          doi = {10.1086/377637},
archivePrefix = {arXiv},
       eprint = {astro-ph/0304348},
 primaryClass = {astro-ph},
       adsurl = {https://ui.adsabs.harvard.edu/abs/2003ApJ...596..509B}
}

@INPROCEEDINGS{Owocki15,
       author = {{Owocki}, Stanley P.},
        title = "{Instabilities in the Envelopes and Winds of Very Massive Stars}",
    booktitle = {Very Massive Stars in the Local Universe},
         year = 2015,
       editor = {{Vink}, Jorick S.},
       series = {Astrophysics and Space Science Library},
       volume = {412},
        month = jan,
        pages = {113},
          doi = {10.1007/978-3-319-09596-7_5},
archivePrefix = {arXiv},
       eprint = {1403.6745},
 primaryClass = {astro-ph.SR},
       adsurl = {https://ui.adsabs.harvard.edu/abs/2015ASSL..412..113O}
}

@ARTICLE{Shaviv01,
       author = {{Shaviv}, Nir J.},
        title = "{The Nature of the Radiative Hydrodynamic Instabilities in Radiatively Supported Thomson Atmospheres}",
      journal = {\apj},
         year = 2001,
        month = mar,
       volume = {549},
       number = {2},
        pages = {1093-1110},
          doi = {10.1086/319428},
archivePrefix = {arXiv},
       eprint = {astro-ph/0010425},
 primaryClass = {astro-ph},
       adsurl = {https://ui.adsabs.harvard.edu/abs/2001ApJ...549.1093S}
}

@ARTICLE{Joss+73,
       author = {{Joss}, P.~C. and {Salpeter}, E.~E. and {Ostriker}, J.~P.},
        title = "{On the ``Critical Luminosity'' in Stellar Interiors and Stellar Surface Boundary Conditions}",
      journal = {\apj},
         year = 1973,
        month = apr,
       volume = {181},
        pages = {429-438},
          doi = {10.1086/152060},
       adsurl = {https://ui.adsabs.harvard.edu/abs/1973ApJ...181..429J}
}

@ARTICLE{vanMarle+08,
       author = {{van Marle}, A.~J. and {Owocki}, S.~P. and {Shaviv}, N.~J.},
        title = "{Numerical simulations of continuum-driven winds of super-Eddington stars}",
      journal = {\mnras},
         year = 2008,
        month = sep,
       volume = {389},
       number = {3},
        pages = {1353-1359},
          doi = {10.1111/j.1365-2966.2008.13648.x},
archivePrefix = {arXiv},
       eprint = {0806.4536},
 primaryClass = {astro-ph},
       adsurl = {https://ui.adsabs.harvard.edu/abs/2008MNRAS.389.1353V}
}

@BOOK{Kippenhahn&Weigert13,
       author = {{Kippenhahn}, Rudolf and {Weigert}, Alfred and {Weiss}, Achim},
        title = "{Stellar Structure and Evolution}",
         year = 2013,
          doi = {10.1007/978-3-642-30304-3},
       adsurl = {https://ui.adsabs.harvard.edu/abs/2013sse..book.....K}
}

@ARTICLE{Jiang+15,
       author = {{Jiang}, Yan-Fei and {Cantiello}, Matteo and {Bildsten}, Lars and {Quataert}, Eliot and {Blaes}, Omer},
        title = "{Local Radiation Hydrodynamic Simulations of Massive Star Envelopes at the Iron Opacity Peak}",
      journal = {\apj},
         year = 2015,
        month = nov,
       volume = {813},
       number = {1},
          eid = {74},
        pages = {74},
          doi = {10.1088/0004-637X/813/1/74},
archivePrefix = {arXiv},
       eprint = {1509.05417},
 primaryClass = {astro-ph.SR},
       adsurl = {https://ui.adsabs.harvard.edu/abs/2015ApJ...813...74J}
}

@ARTICLE{Kajava+20,
       author = {{Kajava}, Jari J.~E. and {Giustini}, Margherita and {Saxton}, Richard D. and {Miniutti}, Giovanni},
        title = "{Rapid late-time X-ray brightening of the tidal disruption event OGLE16aaa}",
      journal = {\aap},
         year = 2020,
        month = jul,
       volume = {639},
          eid = {A100},
        pages = {A100},
          doi = {10.1051/0004-6361/202038165},
archivePrefix = {arXiv},
       eprint = {2006.11179},
 primaryClass = {astro-ph.HE},
       adsurl = {https://ui.adsabs.harvard.edu/abs/2020A&A...639A.100K}
}

@ARTICLE{Yao+22,
       author = {{Yao}, Yuhan and {Lu}, Wenbin and {Guolo}, Muryel and {Pasham}, Dheeraj R. and {Gezari}, Suvi and {Gilfanov}, Marat and {Gendreau}, Keith C. and {Harrison}, Fiona and {Cenko}, S. Bradley and {Kulkarni}, S.~R. and {Miller}, Jon M. and {Walton}, Dominic J. and {Garc{\'\i}a}, Javier A. and {van Velzen}, Sjoert and {Alexander}, Kate D. and {Miller-Jones}, James C.~A. and {Nicholl}, Matt and {Hammerstein}, Erica and {Medvedev}, Pavel and {Stern}, Daniel and {Ravi}, Vikram and {Sunyaev}, R. and {Bloom}, Joshua S. and {Graham}, Matthew J. and {Kool}, Erik C. and {Mahabal}, Ashish A. and {Masci}, Frank J. and {Purdum}, Josiah and {Rusholme}, Ben and {Sharma}, Yashvi and {Smith}, Roger and {Sollerman}, Jesper},
        title = "{The Tidal Disruption Event AT2021ehb: Evidence of Relativistic Disk Reflection, and Rapid Evolution of the Disk-Corona System}",
      journal = {\apj},
         year = 2022,
        month = sep,
       volume = {937},
       number = {1},
          eid = {8},
        pages = {8},
          doi = {10.3847/1538-4357/ac898a},
archivePrefix = {arXiv},
       eprint = {2206.12713},
 primaryClass = {astro-ph.HE},
       adsurl = {https://ui.adsabs.harvard.edu/abs/2022ApJ...937....8Y}
}

@ARTICLE{Eyles-Ferris+25,
       author = {{Eyles-Ferris}, R.~A.~J. and {Starling}, R.~L.~C. and {O'Brien}, P.~T. and {Page}, K.~L. and {Evans}, P.~A.},
        title = "{Nine tidal disruption event candidates in eROSITA-DE DR1 discovered through supersoft X-ray selection}",
      journal = {\mnras},
         year = 2025,
        month = sep,
       volume = {542},
       number = {2},
        pages = {1654-1672},
          doi = {10.1093/mnras/staf1329},
archivePrefix = {arXiv},
       eprint = {2508.08389},
 primaryClass = {astro-ph.HE},
       adsurl = {https://ui.adsabs.harvard.edu/abs/2025MNRAS.542.1654E}
}

@ARTICLE{Alexander+20,
       author = {{Alexander}, Kate D. and {van Velzen}, Sjoert and {Horesh}, Assaf and {Zauderer}, B. Ashley},
        title = "{Radio Properties of Tidal Disruption Events}",
      journal = {\ssr},
         year = 2020,
        month = jun,
       volume = {216},
       number = {5},
          eid = {81},
        pages = {81},
          doi = {10.1007/s11214-020-00702-w},
archivePrefix = {arXiv},
       eprint = {2006.01159},
 primaryClass = {astro-ph.HE},
       adsurl = {https://ui.adsabs.harvard.edu/abs/2020SSRv..216...81A}
}

@ARTICLE{Bloom+11,
       author = {{Bloom}, Joshua S. and {Giannios}, Dimitrios and {Metzger}, Brian D. and {Cenko}, S. Bradley and {Perley}, Daniel A. and {Butler}, Nathaniel R. and {Tanvir}, Nial R. and {Levan}, Andrew J. and {O'Brien}, Paul T. and {Strubbe}, Linda E. and {De Colle}, Fabio and {Ramirez-Ruiz}, Enrico and {Lee}, William H. and {Nayakshin}, Sergei and {Quataert}, Eliot and {King}, Andrew R. and {Cucchiara}, Antonino and {Guillochon}, James and {Bower}, Geoffrey C. and {Fruchter}, Andrew S. and {Morgan}, Adam N. and {van der Horst}, Alexander J.},
        title = "{A Possible Relativistic Jetted Outburst from a Massive Black Hole Fed by a Tidally Disrupted Star}",
      journal = {Science},
         year = 2011,
        month = jul,
       volume = {333},
       number = {6039},
        pages = {203},
          doi = {10.1126/science.1207150},
archivePrefix = {arXiv},
       eprint = {1104.3257},
 primaryClass = {astro-ph.HE},
       adsurl = {https://ui.adsabs.harvard.edu/abs/2011Sci...333..203B}
}

@ARTICLE{Burrows+11,
       author = {{Burrows}, D.~N. and {Kennea}, J.~A. and {Ghisellini}, G. and {Mangano}, V. and {Zhang}, B. and {Page}, K.~L. and {Eracleous}, M. and {Romano}, P. and {Sakamoto}, T. and {Falcone}, A.~D. and {Osborne}, J.~P. and {Campana}, S. and {Beardmore}, A.~P. and {Breeveld}, A.~A. and {Chester}, M.~M. and {Corbet}, R. and {Covino}, S. and {Cummings}, J.~R. and {D'Avanzo}, P. and {D'Elia}, V. and {Esposito}, P. and {Evans}, P.~A. and {Fugazza}, D. and {Gelbord}, J.~M. and {Hiroi}, K. and {Holland}, S.~T. and {Huang}, K.~Y. and {Im}, M. and {Israel}, G. and {Jeon}, Y. and {Jeon}, Y. -B. and {Jun}, H.~D. and {Kawai}, N. and {Kim}, J.~H. and {Krimm}, H.~A. and {Marshall}, F.~E. and {P. M{\'e}sz{\'a}ros} and {Negoro}, H. and {Omodei}, N. and {Park}, W. -K. and {Perkins}, J.~S. and {Sugizaki}, M. and {Sung}, H. -I. and {Tagliaferri}, G. and {Troja}, E. and {Ueda}, Y. and {Urata}, Y. and {Usui}, R. and {Antonelli}, L.~A. and {Barthelmy}, S.~D. and {Cusumano}, G. and {Giommi}, P. and {Melandri}, A. and {Perri}, M. and {Racusin}, J.~L. and {Sbarufatti}, B. and {Siegel}, M.~H. and {Gehrels}, N.},
        title = "{Relativistic jet activity from the tidal disruption of a star by a massive black hole}",
      journal = {\nat},
         year = 2011,
        month = aug,
       volume = {476},
       number = {7361},
        pages = {421-424},
          doi = {10.1038/nature10374},
archivePrefix = {arXiv},
       eprint = {1104.4787},
 primaryClass = {astro-ph.HE},
       adsurl = {https://ui.adsabs.harvard.edu/abs/2011Natur.476..421B}
}

@ARTICLE{Horesh+21a,
       author = {{Horesh}, A. and {Cenko}, S.~B. and {Arcavi}, I.},
        title = "{Delayed radio flares from a tidal disruption event}",
      journal = {Nature Astronomy},
         year = 2021,
        month = may,
       volume = {5},
        pages = {491-497},
          doi = {10.1038/s41550-021-01300-8},
archivePrefix = {arXiv},
       eprint = {2102.11290},
 primaryClass = {astro-ph.HE},
       adsurl = {https://ui.adsabs.harvard.edu/abs/2021NatAs...5..491H}
}

@ARTICLE{Horesh+21b,
       author = {{Horesh}, Assaf and {Sfaradi}, Itai and {Fender}, Rob and {Green}, David A. and {Williams}, David R.~A. and {Bright}, Joe S.},
        title = "{Are Delayed Radio Flares Common in Tidal Disruption Events? The Case of the TDE iPTF 16fnl}",
      journal = {\apjl},
         year = 2021,
        month = oct,
       volume = {920},
       number = {1},
          eid = {L5},
        pages = {L5},
          doi = {10.3847/2041-8213/ac25fe},
archivePrefix = {arXiv},
       eprint = {2109.10921},
 primaryClass = {astro-ph.HE},
       adsurl = {https://ui.adsabs.harvard.edu/abs/2021ApJ...920L...5H}
}

@ARTICLE{Sfaradi+22,
       author = {{Sfaradi}, Itai and {Horesh}, Assaf and {Fender}, Rob and {Green}, David A. and {Williams}, David R.~A. and {Bright}, Joe and {Schulze}, Steve},
        title = "{A Late-time Radio Flare Following a Possible Transition in Accretion State in the Tidal Disruption Event AT 2019azh}",
      journal = {\apj},
         year = 2022,
        month = jul,
       volume = {933},
       number = {2},
          eid = {176},
        pages = {176},
          doi = {10.3847/1538-4357/ac74bc},
archivePrefix = {arXiv},
       eprint = {2202.00026},
 primaryClass = {astro-ph.HE},
       adsurl = {https://ui.adsabs.harvard.edu/abs/2022ApJ...933..176S}
}

@ARTICLE{Cendes+22,
       author = {{Cendes}, Y. and {Berger}, E. and {Alexander}, K.~D. and {Gomez}, S. and {Hajela}, A. and {Chornock}, R. and {Laskar}, T. and {Margutti}, R. and {Metzger}, B. and {Bietenholz}, M.~F. and {Brethauer}, D. and {Wieringa}, M.~H.},
        title = "{A Mildly Relativistic Outflow Launched Two Years after Disruption in Tidal Disruption Event AT2018hyz}",
      journal = {\apj},
         year = 2022,
        month = oct,
       volume = {938},
       number = {1},
          eid = {28},
        pages = {28},
          doi = {10.3847/1538-4357/ac88d0},
archivePrefix = {arXiv},
       eprint = {2206.14297},
 primaryClass = {astro-ph.HE},
       adsurl = {https://ui.adsabs.harvard.edu/abs/2022ApJ...938...28C}
}

@ARTICLE{Murase+20,
       author = {{Murase}, Kohta and {Kimura}, Shigeo S. and {Zhang}, B. Theodore and {Oikonomou}, Foteini and {Petropoulou}, Maria},
        title = "{High-energy Neutrino and Gamma-Ray Emission from Tidal Disruption Events}",
      journal = {\apj},
         year = 2020,
        month = oct,
       volume = {902},
       number = {2},
          eid = {108},
        pages = {108},
          doi = {10.3847/1538-4357/abb3c0},
archivePrefix = {arXiv},
       eprint = {2005.08937},
 primaryClass = {astro-ph.HE},
       adsurl = {https://ui.adsabs.harvard.edu/abs/2020ApJ...902..108M}
}

@ARTICLE{Stein+21,
       author = {{Stein}, Robert and {van Velzen}, Sjoert and {Kowalski}, Marek and {Franckowiak}, Anna and {Gezari}, Suvi and {Miller-Jones}, James C.~A. and {Frederick}, Sara and {Sfaradi}, Itai and {Bietenholz}, Michael F. and {Horesh}, Assaf and {Fender}, Rob and {Garrappa}, Simone and {Ahumada}, Tom{\'a}s and {Andreoni}, Igor and {Belicki}, Justin and {Bellm}, Eric C. and {B{\"o}ttcher}, Markus and {Brinnel}, Valery and {Burruss}, Rick and {Cenko}, S. Bradley and {Coughlin}, Michael W. and {Cunningham}, Virginia and {Drake}, Andrew and {Farrar}, Glennys R. and {Feeney}, Michael and {Foley}, Ryan J. and {Gal-Yam}, Avishay and {Golkhou}, V. Zach and {Goobar}, Ariel and {Graham}, Matthew J. and {Hammerstein}, Erica and {Helou}, George and {Hung}, Tiara and {Kasliwal}, Mansi M. and {Kilpatrick}, Charles D. and {Kong}, Albert K.~H. and {Kupfer}, Thomas and {Laher}, Russ R. and {Mahabal}, Ashish A. and {Masci}, Frank J. and {Necker}, Jannis and {Nordin}, Jakob and {Perley}, Daniel A. and {Rigault}, Mickael and {Reusch}, Simeon and {Rodriguez}, Hector and {Rojas-Bravo}, C{\'e}sar and {Rusholme}, Ben and {Shupe}, David L. and {Singer}, Leo P. and {Sollerman}, Jesper and {Soumagnac}, Maayane T. and {Stern}, Daniel and {Taggart}, Kirsty and {van Santen}, Jakob and {Ward}, Charlotte and {Woudt}, Patrick and {Yao}, Yuhan},
        title = "{A tidal disruption event coincident with a high-energy neutrino}",
      journal = {Nature Astronomy},
         year = 2021,
        month = feb,
       volume = {5},
        pages = {510-518},
          doi = {10.1038/s41550-020-01295-8},
archivePrefix = {arXiv},
       eprint = {2005.05340},
 primaryClass = {astro-ph.HE},
       adsurl = {https://ui.adsabs.harvard.edu/abs/2021NatAs...5..510S}
}

@ARTICLE{vanVelzen+24,
       author = {{van Velzen}, Sjoert and {Stein}, Robert and {Gilfanov}, Marat and {Kowalski}, Marek and {Hayasaki}, Kimitake and {Reusch}, Simeon and {Yao}, Yuhan and {Garrappa}, Simone and {Franckowiak}, Anna and {Gezari}, Suvi and {Nordin}, Jakob and {Fremling}, Christoffer and {Sharma}, Yashvi and {Yan}, Lin and {Kool}, Erik C. and {Stern}, Daniel and {Veres}, Patrik M. and {Sollerman}, Jesper and {Medvedev}, Pavel and {Sunyaev}, Rashid and {Bellm}, Eric C. and {Dekany}, Richard G. and {Duev}, Dimitri A. and {Graham}, Matthew J. and {Kasliwal}, Mansi M. and {Kulkarni}, Shrinivas R. and {Laher}, Russ R. and {Riddle}, Reed L. and {Rusholme}, Ben},
        title = "{Establishing accretion flares from supermassive black holes as a source of high-energy neutrinos}",
      journal = {\mnras},
         year = 2024,
        month = apr,
       volume = {529},
       number = {3},
        pages = {2559-2576},
          doi = {10.1093/mnras/stae610},
archivePrefix = {arXiv},
       eprint = {2111.09391},
 primaryClass = {astro-ph.HE},
       adsurl = {https://ui.adsabs.harvard.edu/abs/2024MNRAS.529.2559V}
}

@ARTICLE{Reusch+22,
       author = {{Reusch}, Simeon and {Stein}, Robert and {Kowalski}, Marek and {van Velzen}, Sjoert and {Franckowiak}, Anna and {Lunardini}, Cecilia and {Murase}, Kohta and {Winter}, Walter and {Miller-Jones}, James C.~A. and {Kasliwal}, Mansi M. and {Gilfanov}, Marat and {Garrappa}, Simone and {Paliya}, Vaidehi S. and {Ahumada}, Tom{\'a}s and {Anand}, Shreya and {Barbarino}, Cristina and {Bellm}, Eric C. and {Brinnel}, Val{\'e}ry and {Buson}, Sara and {Cenko}, S. Bradley and {Coughlin}, Michael W. and {De}, Kishalay and {Dekany}, Richard and {Frederick}, Sara and {Gal-Yam}, Avishay and {Gezari}, Suvi and {Giroletti}, Marcello and {Graham}, Matthew J. and {Karambelkar}, Viraj and {Kimura}, Shigeo S. and {Kong}, Albert K.~H. and {Kool}, Erik C. and {Laher}, Russ R. and {Medvedev}, Pavel and {Necker}, Jannis and {Nordin}, Jakob and {Perley}, Daniel A. and {Rigault}, Mickael and {Rusholme}, Ben and {Schulze}, Steve and {Schweyer}, Tassilo and {Singer}, Leo P. and {Sollerman}, Jesper and {Strotjohann}, Nora Linn and {Sunyaev}, Rashid and {van Santen}, Jakob and {Walters}, Richard and {Zhang}, B. Theodore and {Zimmerman}, Erez},
        title = "{Candidate Tidal Disruption Event AT2019fdr Coincident with a High-Energy Neutrino}",
      journal = {\prl},
         year = 2022,
        month = jun,
       volume = {128},
       number = {22},
          eid = {221101},
        pages = {221101},
          doi = {10.1103/PhysRevLett.128.221101},
archivePrefix = {arXiv},
       eprint = {2111.09390},
 primaryClass = {astro-ph.HE},
       adsurl = {https://ui.adsabs.harvard.edu/abs/2022PhRvL.128v1101R}
}

\end{CJK*}

\end{document}